\documentclass[useAMS,usenatbib]{mn2e}
\usepackage{graphicx}
\usepackage{amssymb, amsmath}
\usepackage{times}
\usepackage{color}
\usepackage{lscape}
\usepackage{multirow}
\usepackage[section]{placeins}
\usepackage{setspace}
\usepackage{url}

\usepackage{lineno}
\newcommand{\hmpc}{\,h^{-1}\mathrm{Mpc}}
\newcommand{\khmpc}{\,h\,{\rm Mpc}^{-1}}

\newcommand{\lya}{Ly$\alpha$}
\newcommand{\hubunits}{\,{\rm km}\,{\rm s}^{-1}\,{\rm Mpc}^{-1}}
\newcommand{\zrec}{z_{\rm rec}}
\newcommand{\DM}{D_M}
\newcommand{\DV}{D_V}
\newcommand{\DA}{D_A}
\newcommand{\DHub}{D_H}
\newcommand\rd{r_{d}}
\newcommand\rdfid{r_{d,{\rm fid}}}

\newcommand{\apj}{ApJ}
\newcommand{\aj}{AJ}
\newcommand{\apjs}{ApJ Supp}
\newcommand{\apjl}{ApJ}
\newcommand{\mnras}{MNRAS}
\newcommand{\araa}{ARA\&A}

\newcommand{\upd}[1]{#1}    

\def\aj{\rm{AJ}}                   % Astronomical Journal
\def\araa{\rm{ARA\&A}}             % Annual Review of Astron and Astrophys
\def\apj{\rm{ApJ}}                 % Astrophysical Journal
\def\apjl{\rm{ApJ}}                % Astrophysical Journal, Letters
\def\apjs{\rm{ApJS}}               % Astrophysical Journal, Supplement
\def\mnras{\rm{MNRAS}}             % Monthly Notices of the RAS

\begin{document}
% \linenumbers

% --- title --- %
\title[Cosmological Analysis of BOSS galaxies] {The clustering of galaxies in the completed SDSS-III Baryon Oscillation Spectroscopic Survey: cosmological analysis of the DR12 galaxy sample}

% --- authors --- %
\author[S. Alam et al.]{\parbox{\textwidth}{
\Large 
Shadab Alam$^{1,2}$,
Metin Ata$^{3}$,
Stephen Bailey$^{4}$,
Florian Beutler$^{4}$,
Dmitry Bizyaev$^{5,6}$,
Jonathan A. Blazek$^{7}$,
Adam S. Bolton$^{8,9}$,
Joel R. Brownstein$^{8}$,
Angela Burden$^{10}$,
Chia-Hsun Chuang$^{11,3}$,
Johan Comparat$^{11,12}$,
Antonio J. Cuesta$^{13}$,
Kyle S. Dawson$^{8}$,
Daniel J. Eisenstein$^{14}$,
Stephanie Escoffier$^{15}$,
H\'ector Gil-Mar\'in$^{16,17}$,
Jan Niklas Grieb$^{18,19}$,
Nick Hand$^{20}$,
Shirley Ho$^{1,2}$,
Karen Kinemuchi$^{5}$,
David Kirkby$^{21}$,
Francisco Kitaura$^{3,4,20}$,
Elena Malanushenko$^{5}$,
Viktor Malanushenko$^{5}$,
Claudia Maraston$^{22}$,
Cameron K. McBride$^{14}$,
Robert C. Nichol$^{22}$,
Matthew D. Olmstead$^{23}$,
Daniel Oravetz$^{5}$,
Nikhil Padmanabhan$^{10}$,
Nathalie Palanque-Delabrouille$^{24}$,
Kaike Pan$^{5}$,
Marcos Pellejero-Ibanez$^{25,26}$,
Will J. Percival$^{22}$,
Patrick Petitjean$^{27}$,
Francisco Prada$^{11,28,29}$,
Adrian M. Price-Whelan$^{30}$,
Beth A. Reid$^{4,31,32}$,
Sergio A. Rodr\'iguez-Torres$^{11,28,12}$,
Natalie A. Roe$^{4}$,
Ashley J. Ross$^{7,22}$,
Nicholas P. Ross$^{33}$,
Graziano Rossi$^{34}$,
Jose Alberto Rubi\~no-Mart\'in$^{25,26}$,
Ariel G. S\'anchez$^{19}$,
Shun Saito$^{35,36}$,
Salvador Salazar-Albornoz$^{18,19}$,
Lado Samushia$^{37}$,
Siddharth Satpathy$^{1,2}$,
Claudia G. Sc\'occola$^{11,38,39}$,
David J. Schlegel\thanks{BOSS PI: djschlegel@lbl.gov}$^{4}$,
Donald P. Schneider$^{40,41}$,
Hee-Jong Seo$^{42}$,
Audrey Simmons$^{5}$,
An\v{z}e Slosar$^{43}$,
Michael A. Strauss$^{30}$,
Molly E. C. Swanson$^{14}$,
Daniel Thomas$^{22}$,
Jeremy L. Tinker$^{44}$,
Rita Tojeiro$^{45}$,
Mariana Vargas Maga\~na$^{1,2,46}$,
Jose Alberto Vazquez$^{43}$,
Licia Verde$^{13,47,48,49}$,
David A. Wake$^{50,51}$,
Yuting Wang$^{52,22}$,
David H. Weinberg$^{53,7}$,
Martin White$^{4,32}$,
W. Michael Wood-Vasey$^{54}$,
Christophe Y\`eche$^{24}$,
Idit Zehavi$^{55}$,
Zhongxu Zhai$^{44}$,
Gong-Bo Zhao$^{52,22}$
 } 
\vspace*{-20pt}\\ 
}

\date{\today} 
\pagerange{\pageref{firstpage}--\pageref{lastpage}} \pubyear{2016}
\maketitle
\label{firstpage}

%\clearpage
% --- abstract --- %
\begin{abstract}
We present cosmological results from the final galaxy clustering data
set of the Baryon Oscillation Spectroscopic Survey, part of the Sloan
Digital Sky Survey III.  Our combined galaxy sample comprises 1.2
million massive galaxies over an effective area of 9329~deg$^2$ and
volume of 18.7 Gpc$^3$, divided into three partially overlapping
redshift slices centred at effective redshifts 0.38, 0.51, and 0.61.
We measure the angular diameter distance $\DM$ and Hubble parameter
$H$ from the baryon acoustic oscillation (BAO) method after applying
reconstruction to reduce non-linear effects on the BAO feature.  Using
the anisotropic clustering of the pre-reconstruction density field, we
measure the product $\DM H$ from the Alcock-Paczynski (AP) effect and
the growth of structure, quantified by $f\sigma_8(z)$, from
redshift-space distortions (RSD).  We combine individual measurements
presented in seven companion papers into a set of consensus values and
likelihoods, obtaining constraints that are tighter and more robust
than those from any one method; in particular, the AP measurement from
sub-BAO scales sharpens constraints from post-reconstruction BAO by
breaking degeneracy between $\DM$ and $H$. Combined with Planck 2015
cosmic microwave background measurements, our distance scale
measurements simultaneously imply curvature $\Omega_K=0.0003\pm0.0026$
and a dark energy equation of state parameter $w=-1.01\pm0.06$, in
strong affirmation of the spatially flat cold dark matter model with a
cosmological constant ($\Lambda$CDM). Our RSD measurements of
$f\sigma_8$, at 6 per cent precision, are similarly consistent with
this model. When combined with supernova Ia data, we find
$H_0=67.3\pm1.0\hubunits$ even for our most general dark energy model,
in tension with some direct measurements. Adding extra relativistic
species as a degree of freedom loosens the constraint only slightly,
to $H_0=67.8\pm1.2 \hubunits$. Assuming flat $\Lambda$CDM we find
$\Omega_m=0.310\pm0.005$ and $H_0=67.6\pm0.5\hubunits$, and we find a
95\% upper limit of 0.16 eV/$c^2$ on the neutrino mass sum.

\end{abstract}

\begin{keywords}
  cosmology: observations, distance scale, large-scale structure
\end{keywords}

%\clearpage
\begin{scriptsize}
\begin{spacing}{1.0}
\noindent
 $^{1}$ Department of Physics, Carnegie Mellon University, 5000 Forbes Avenue, Pittsburgh, PA 15213, USA\vspace*{0pt} \\ 
 $^{2}$ The McWilliams Center for Cosmology, Carnegie Mellon University, 5000 Forbes Ave., Pittsburgh, PA 15213, USA\vspace*{0pt} \\ 
 $^{3}$ Leibniz-Institut f\"{u}r Astrophysik Potsdam (AIP), An der Sternwarte 16, D-14482 Potsdam, Germany\vspace*{0pt} \\ 
 $^{4}$ Lawrence Berkeley National Laboratory, 1 Cyclotron Road, Berkeley, CA 94720, USA\vspace*{0pt} \\ 
 $^{5}$ Apache Point Observatory and New Mexico State University, P.O. Box 59, Sunspot, NM 88349, USA\vspace*{0pt} \\ 
 $^{6}$ Sternberg Astronomical Institute, Moscow State University, Universitetski pr. 13, 119992 Moscow, Russia\vspace*{0pt} \\ 
 $^{7}$ Center for Cosmology and Astro-Particle Physics, Ohio State University, Columbus, Ohio, USA\vspace*{0pt} \\ 
 $^{8}$ Department Physics and Astronomy, University of Utah, 115 S 1400 E, Salt Lake City, UT 84112, USA\vspace*{0pt} \\ 
 $^{9}$ National Optical Astronomy Observatory, 950 N Cherry Ave, Tucson, AZ 85719, USA\vspace*{0pt} \\ 
 $^{10}$ Department of Physics, Yale University, 260 Whitney Ave, New Haven, CT 06520, USA\vspace*{0pt} \\ 
 $^{11}$ Instituto de F\'isica Te\'orica (UAM/CSIC), Universidad Aut\'onoma de Madrid, Cantoblanco, E-28049 Madrid, Spain\vspace*{0pt} \\ 
 $^{12}$ Departamento de F\'isica Te\'orica M8, Universidad Aut\'onoma de Madrid, E-28049 Cantoblanco, Madrid, Spain\vspace*{0pt} \\ 
 $^{13}$ Institut de Ci\`encies del Cosmos (ICCUB), Universitat de Barcelona (IEEC-UB), Mart\'i i Franqu\`es 1, E08028 Barcelona, Spain\vspace*{0pt} \\ 
 $^{14}$ Harvard-Smithsonian Center for Astrophysics, 60 Garden St., Cambridge, MA 02138, USA\vspace*{0pt} \\ 
 $^{15}$ CPPM, Aix-Marseille Universit\'e, CNRS/IN2P3, CPPM UMR 7346, 13288, Marseille, France\vspace*{0pt} \\ 
 $^{16}$ Sorbonne Universit\'es, Institut Lagrange de Paris (ILP), 98 bis Boulevard Arago, 75014 Paris, France\vspace*{0pt} \\ 
 $^{17}$ Laboratoire de Physique Nucl\'eaire et de Hautes Energies, Universit\'e Pierre et Marie Curie, 4 Place Jussieu, 75005 Paris, France\vspace*{0pt} \\ 
 $^{18}$ Universit\"ats-Sternwarte M\"unchen, Scheinerstrasse 1, 81679 Munich, Germany\vspace*{0pt} \\ 
 $^{19}$ Max-Planck-Institut f\"ur Extraterrestrische Physik, Postfach 1312, Giessenbachstr., 85748 Garching, Germany\vspace*{0pt} \\ 
 $^{20}$ Department of Astronomy, University of California at Berkeley, Berkeley, CA 94720, USA\vspace*{0pt} \\ 
 $^{21}$ Department of Physics and Astronomy, UC Irvine, 4129 Frederick Reines Hall, Irvine, CA 92697, USA\vspace*{0pt} \\ 
 $^{22}$ Institute of Cosmology \& Gravitation, Dennis Sciama Building, University of Portsmouth, Portsmouth, PO1 3FX, UK\vspace*{0pt} \\ 
 $^{23}$ Department of Chemistry and Physics, King's College, 133 North River St, Wilkes Barre, PA 18711, USA\vspace*{0pt} \\ 
 $^{24}$ CEA, Centre de Saclay, IRFU/SPP, F-91191 Gif-sur-Yvette, France\vspace*{0pt} \\ 
 $^{25}$ Instituto de Astrof\'isica de Canarias (IAC), C/V\'ia L\'actea, s/n, E-38200, La Laguna, Tenerife, Spain\vspace*{0pt} \\ 
 $^{26}$ Dpto. Astrof\'isica, Universidad de La Laguna (ULL), E-38206 La Laguna, Tenerife, Spain\vspace*{0pt} \\ 
 $^{27}$ Institut d'Astrophysique de Paris, Universit\'e Paris 6 et CNRS, 98bis Boulevard Arago, 75014 Paris, France\vspace*{0pt} \\ 
 $^{28}$ Campus of International Excellence UAM+CSIC, Cantoblanco, E-28049 Madrid, Spain\vspace*{0pt} \\ 
 $^{29}$ Instituto de Astrof\'isica de Andaluc\'ia (CSIC), E-18080 Granada, Spain\vspace*{0pt} \\ 
 $^{30}$ Department of Astrophysical Sciences, Princeton University, Ivy Lane, Princeton, NJ 08544, USA\vspace*{0pt} \\ 
 $^{31}$ Hubble Fellow\vspace*{0pt} \\ 
 $^{32}$ Department of Physics, University of California, 366 LeConte Hall, Berkeley, CA 94720, USA\vspace*{0pt} \\ 
 $^{33}$ Institute for Astronomy, University of Edinburgh, Royal Observatory, Edinburgh, EH9 3HJ, UK\vspace*{0pt} \\ 
 $^{34}$ Department of Astronomy and Space Science, Sejong University, Seoul 143-747, Korea\vspace*{0pt} \\ 
 $^{35}$ Kavli Institute for the Physics and Mathematics of the Universe (WPI), The University of Tokyo Institutes for Advanced Study, The University of Tokyo, Kashiwa, Chiba 277-8583, Japan\vspace*{0pt} \\ 
 $^{36}$ Max Planck Institut f\"ur Astrophysik, Karl-Schwarzschild-Stra{\ss}e 1, D-85740 Garching bei M\"unchen, Germany\vspace*{0pt} \\ 
 $^{37}$ Department of Physics, Kansas State University, 116 Cardwell Hall, Manhattan, KS 66506, USA\vspace*{0pt} \\ 
 $^{38}$ Facultad de Ciencias Astron\'omicas y Geof\'isicas - Universidad Nacional de La Plata. Paseo del Bosque S/N, (1900) La Plata, Argentina\vspace*{0pt} \\ 
 $^{39}$ CONICET, Rivadavia 1917, (1033) Buenos Aires, Argentina\vspace*{0pt} \\ 
 $^{40}$ Department of Astronomy and Astrophysics, The Pennsylvania State University, University Park, PA 16802, USA\vspace*{0pt} \\ 
 $^{41}$ Institute for Gravitation and the Cosmos, The Pennsylvania State University, University Park, PA 16802, USA\vspace*{0pt} \\ 
 $^{42}$ Department of Physics and Astronomy, Ohio University, 251B Clippinger Labs, Athens, OH 45701, USA\vspace*{0pt} \\ 
 $^{43}$ Brookhaven National Laboratory, Bldg 510, Upton, New York 11973, USA\vspace*{0pt} \\ 
 $^{44}$ Center for Cosmology and Particle Physics, New York University, New York, NY 10003, USA\vspace*{0pt} \\ 
 $^{45}$ School of Physics and Astronomy, University of St Andrews, St Andrews, KY16 9SS, UK\vspace*{0pt} \\ 
 $^{46}$ Instituto de F\'isica, Universidad Nacional Aut\'onoma de M\'exico, Apdo. Postal 20-364, M\'exico\vspace*{0pt} \\ 
 $^{47}$ ICREA (Instituci\'o Catalana de Recerca i Estudis Avan\c{c}ats) Passeig Llu\'is Companys 23, E-08010 Barcelona, Spain\vspace*{0pt} \\ 
 $^{48}$ Radcliffe Institute for Advanced Study \& ITC, Harvard-Smithsonian Center for Astrophysics, Harvard University, MA 02138, USA\vspace*{0pt} \\ 
 $^{49}$ Institute of Theoretical Astrophysics, University of Oslo, 0315 Oslo, Norway\vspace*{0pt} \\ 
 $^{50}$ Department of Astronomy, University of Wisconsin-Madison, 475 N. Charter Street, Madison, WI, 53706, USA\vspace*{0pt} \\ 
 $^{51}$ Department of Physical Sciences, The Open University, Milton Keynes, MK7 6AA, UK\vspace*{0pt} \\ 
 $^{52}$ National Astronomy Observatories, Chinese Academy of Science, Beijing, 100012, P.R. China\vspace*{0pt} \\ 
 $^{53}$ Department of Astronomy, Ohio State University, Columbus, Ohio, USA\vspace*{0pt} \\ 
 $^{54}$ PITT PACC, Department of Physics and Astronomy, University of Pittsburgh, 3941 O'Hara Street, Pittsburgh, PA 15260, USA\vspace*{0pt} \\ 
 $^{55}$ Department of Astronomy, Case Western Reserve University, Cleveland, OH 44106, USA\vspace*{0pt} \\ 

\end{spacing}
\end{scriptsize}

\section{Introduction}
\label{sec:intro}

Observations and theoretical studies over the past four decades
have led to the emergence of
a standard cosmological model, $\Lambda$CDM, based on a spatially
flat universe, cold dark matter, a cosmological constant that drives
accelerated expansion at late times, and structure seeded by quantum
fluctuations during an epoch of inflation at very early times.
The goals of ``precision cosmology'' are to test the underlying
assumptions of this model and to measure its parameters with sufficient
precision to yield new physical insights, such as the mass scale
of neutrinos, the presence of unknown relativistic species, 
possible small departures from flatness, and the physics of
inflation or alternative scenarios of the early universe.
Observations on galactic and sub-galactic scales can test the hypothesis
that dark matter is weakly interacting and cold (in the sense that
its primordial velocity dispersion was too small to affect structure
formation).  The biggest question of contemporary cosmology
is the origin of cosmic
acceleration: does it arise from a constant vacuum energy as assumed
in $\Lambda$CDM, or from another form of dark energy that varies in
time and space, or from a breakdown of General Relativity (GR) on
cosmological scales?  This question can be addressed by precisely
measuring the cosmic expansion history over a wide span
of redshift and by comparing measurements of the growth of matter
clustering to the predictions of $\Lambda$CDM+GR.

This paper presents cosmological results from the final galaxy 
clustering data set of the Baryon Oscillation Spectroscopic Survey 
(BOSS; \citealt{Dawson13}), conducted as part of the 
Sloan Digital Sky Survey III (SDSS-III; \citealt{Eis11}).  
As the name suggests, the defining goal of BOSS is to measure the
cosmic expansion history by means of baryon acoustic oscillations (BAO),
which imprint a characteristic scale detectable in the clustering of
galaxies and of intergalactic \lya\ forest absorption.  BOSS is
the premier current data set for measurements of large scale galaxy
clustering, which can also be used to constrain cosmological parameters
through the full shape of the galaxy power spectrum and the 
anisotropy induced by redshift-space distortions (RSD).  As discussed
further below, this paper draws on results from a number of supporting
papers, which present analyses of BAO, RSD, and full shape constraints
using a variety of measurement and modelling techniques and provide
the infrastructure to derive statistical uncertainties and test
for systematic effects.  Here we synthesize these results into
``consensus'' cosmological constraints from BOSS galaxy clustering,
in combination with a variety of external data sets.
The galaxy data set that underpins these measurements comes
from SDSS Data Release 12 (DR12; \citealt{DR12})
and the large scale structure catalogue with the additional information 
(masks, completeness, etc.) required for clustering measurements
appears in \cite{ReidEtAl15}.

The first direct evidence for cosmic acceleration came from surveys
of Type Ia supernovae (SNe) in the late 1990s \citep{Riess98,Perl99}.
This evidence had immediate impact in part because studies of 
cosmic microwave background (CMB) anisotropy and large scale structure
(LSS) already favoured $\Lambda$CDM as an economical explanation for
observed cosmic structure (see, e.g., 
\citealt{Efstathiou90,Krauss95,Ostriker95}).
The case for $\Lambda$CDM 
sharpened quickly with balloon-based CMB measurements that found the
first acoustic peak at the angular location
predicted for a flat universe (\citealt{DeBernardis00,Hanany00}; see
\citealt{Netterfield97} for earlier ground-based results pointing
in this direction).
Today the web of evidence for cosmic acceleration is extremely strong,
and nearly all observations remain consistent with a cosmological
constant form of dark energy.  CMB measurements from the
Wilkinson Microwave Anisotropy Probe (WMAP; \citealt{Bennett13}),
ground-based experiments such as the Atacama Cosmology Telescope
\citep{Das14} and the South Pole Telescope \citep{George15},
and, especially, the Planck satellite \citep{Planck2015I} now
provide strong constraints on the cosmic matter and radiation density,
the angular diameter distance to the surface of last scattering,
and the shape and amplitude of the matter power spectrum at the
recombination epoch $\zrec \approx 1090$. These measurements also probe lower 
redshift matter clustering through gravitational lensing and the
integrated Sachs-Wolfe (ISW; \citealt{Sachs67}) effect.
Within $\Lambda$CDM, CMB data alone are sufficient to provide tight
parameter constraints, but these weaken considerably when 
non-zero curvature or more flexible forms of dark energy are allowed
(Planck Collaboration XIII. 2015, hereafter Planck2015).
Supernova measurements of the expansion history
have improved dramatically thanks to large ground-based surveys that
span the redshift range $0.2 < z < 0.8$, improved local calibrator
samples, {\it Hubble Space Telescope} searches that extend the
Hubble diagram to $z \approx 1.5$, and major efforts by independent
groups to place different data sets on a common scale and to identify
and mitigate sources of systematic error
(see \citealt{Suzuki12}; \citealt{JLA}; and references therein).
BAO measurements, now spanning $z=0.1-0.8$ and $z \approx 2.5$, complement
the SN measurements by providing an absolute distance scale, direct
measurement of the expansion rate $H(z)$, and robustness to systematic
errors (see discussion and references below).
Direct ``distance ladder'' measurements of $H_0$ constrain the present
day expansion rate, providing the longest lever arm against the CMB
\citep{riess11,Riess16,Freedman12}.
RSD and weak gravitational lensing measurements provide complementary
probes of structure growth that have somewhat different parameter sensitivity
and very different systematics.  Consistency of RSD and weak lensing can 
also test modified gravity models that predict different effective potentials 
governing light-bending and acceleration of non-relativistic
tracers.  At present, these structure growth measurements are substantially
less precise than expansion history measurements ($\sim 5-10\%$ vs. 
$\sim 1-2\%$), so they serve primarily to test departures from GR and
constrain neutrino masses rather than measure dark energy parameters.
This situation is likely to change in next-generation experiments.
Observational probes of dark energy are reviewed by, e.g.,
\cite{Albrecht06}, \cite{Frieman08}, \cite{Blanchard10},
\cite{Astier12}, and more comprehensively by \cite{Weinberg13}.
Reviews focused more on theories of dark energy and modified gravity
include \cite{Copeland06}, \cite{Jain10}, and \cite{Joyce16}.
Reviews focused on future observational facilities include
\cite{LSST09}, \cite{Kim15}, \cite{Huterer15}, and \cite{Amendola16}.

While acoustic oscillations were already incorporated in early
theoretical calculations of CMB anisotropies \citep{PeeblesYu70,SZ70},
interest in using the BAO feature as a ``standard ruler'' in galaxy 
clustering grew after the discovery of cosmic acceleration
\citep{Eisenstein98,Blake03,Seo03}.  The physics of BAO and
contemporary methods of BAO analysis are reviewed at length in
Ch.\ 4 of \cite{Weinberg13}, and details specific to our analyses
appear in the supporting papers listed below.  In brief, pressure waves
in the pre-recombination universe imprint a characteristic scale on
late-time matter clustering at the radius of the sound horizon,
\begin{equation}
\rd = \int_{z_d}^\infty {c_s(z) \over H(z)}dz~,
\label{eqn:rd}
\end{equation}
evaluated at the drag epoch $z_d$, shortly after recombination,
when photons and baryons decouple (see \citealt{Aubourg} for 
more precise discussion).
This scale appears as a localized peak in the correlation function
or a damped series of oscillations in the power spectrum.
Assuming standard matter and radiation content,
the Planck 2015 measurements of the matter and baryon density
determine the sound horizon to 0.2\%. An anisotropic BAO analysis that measures the BAO feature in
the line-of-sight and transverse directions can separately
measure $H(z)$ and the comoving angular diameter distance
$\DM(z)$, which is related to the physical angular diameter distance by
$\DM(z) = (1+z)\DA(z)$ \citep{Padmanabhan08}.
Adjustments in cosmological parameters or 
changes to the pre-recombination energy density (e.g.,
from extra relativistic species)
can alter $\rd$, so BAO measurements really constrain
the combinations $\DM(z)/\rd$, $H(z)\rd$.
An angle-averaged galaxy BAO measurement
constrains a combination that is approximately
\begin{equation}
\DV(z) = \left[cz \DM^2(z)/H(z)\right]^{1/3}~.
\label{eqn:dv}
\end{equation}
An anisotropic BAO analysis automatically incorporates the
so-called Alcock-Paczynski (\citeyear{AP}; AP) test, which
uses the requirement of statistical isotropy to constrain 
the parameter combination $H(z)\DM(z)$.

The localized three-dimensional nature of the BAO feature makes BAO measurements
robust to most observational systematics (see \citealt{RossEtAl12,Ross16}),
which tend to introduce only smooth distortions in clustering
measurements.  Similarly, non-linear evolution and galaxy bias are
expected to produce smooth rather than localized distortions of
clustering.  Our BAO analysis methods introduce parametrized templates
to marginalize over smooth distortions of observational or
astrophysical origin, and results are insensitive to details of
these templates and to many other analysis details
\citep{VargasMaganaEtAl14,VargasMagana16}.
Non-linear evolution broadens the BAO peak in the correlation
function (or damps high-$k$ oscillations in the power spectrum),
and simulations and perturbation theory calculations indicate that 
non-linear evolution and galaxy bias can shift the location of the
BAO peak at a level of $0.2-0.5\%$
\citep{Eis07b,PadWhite09,Seo10,Mehta11,Sherwin12}.
Measurements of the BAO scale using samples
with considerable differences in galaxy bias that share the same volume
have obtained results consistent with such small shifts 
\citep{Ross14,Beutler16}.
A key element of recent BAO analyses is reconstruction, which
attempts to reverse non-linear effects so as to sharpen the
BAO peak and thereby restore measurement precision 
\citep{Eisenstein07,PadmanabhanEtAl12,Burden15,Schmittfull15}.
Simulation tests and perturbation theory calculations
show that reconstruction also removes the
small shifts induced by non-linearity and galaxy bias, to
a level of $\approx 0.1\%$ or better 
\citep{Padmanabhan09,Noh09,Seo10,Mehta11,Tassev12,White15}.
The combination of precision, complementarity to SNe, and 
robustness to systematics has made BAO a pillar of contemporary
cosmology.

Early analyses of the power spectrum of the 2-Degree Field Galaxy
Redshift Survey (2dFGRS; \citealt{Colless03}) showed strong
hints of baryonic features \citep{Percival01}, but the first
clear detections of BAO came in 2005 with analyses of the
final 2dFGRS data set \citep{Cole05} and the SDSS DR3 data
set \citep{Eis05}.  These detections were already sufficient
to yield $3-4\%$ distance scale constraints.  The SDSS measurement
was based on the luminous red galaxy (LRG) sample, constructed to
provide sparse but relatively uniform sampling over a large volume
\citep{Eisenstein01}.  Subsequent milestones in BAO measurement
include: isotropic BAO analyses of the final (DR7) SDSS-I/II LRG and main galaxy
redshift surveys \citep{Per07}; 
detection of BAO in clustering of SDSS galaxies with photometric redshifts
\citep{Padmanabhan07};
analyses of anisotropic BAO signals in SDSS-I/II
\citep{Okumura08,Gaz09,Chuang12,Chuang13a,Chuang13b};
the first BAO measurements at $z > 0.5$ from the WiggleZ survey
\citep{Blake11}; a low redshift ($z\approx 0.1$) BAO measurement
from the 6-degree Field Galaxy Survey (6dFGS; \citealt{Beutler11});
improved measurements from applying reconstruction to the SDSS LRG
survey \citep{PadmanabhanEtAl12} and main galaxy survey
(MGS; \citealt{Ross15MGS}); BAO measurements from the
BOSS DR9 and DR11 galaxy redshift surveys
\citep{Anderson2012,Anderson2014a,Anderson2014b,TojeiroEtAl14};
and BAO measurements at $z\approx 2.5$ in the BOSS \lya\ forest
using auto-correlations in DR9
\citep{Busca13,Slosar13} and both auto-correlations and
quasar-\lya\ cross-correlations in DR11 \citep{Font14,Delubac}.
The BOSS DR11 measurements achieve distance scale precision of
2.0\% at $z=0.32$, 1.0\% at $z=0.57$, and
$\approx 2\%$ at $z = 2.5$ (where the best constrained combination
is $\DM^{0.3}H^{-0.7}$ rather than $\DV$).
\cite{Aubourg} present cosmological constraints and model
tests derived from these measurements in concert with other data,
and they provide a high-level discussion of the interplay between
BAO measurements and complementary probes.
Section~\ref{sec:cosmology} of this paper updates these constraints
and model tests to our final DR12 galaxy clustering results.
The DR12 \lya\ forest BAO measurements are in process and will
be reported in future work 
(J.\ Bautista et al., in prep.).

The linear theory description of RSD is three decades old
\citep{Kaiser87}, but progress on high-precision RSD constraints
has been slow because a variety of non-linear effects influence
RSD signals even out to very large scales
\citep{Cole94,Scoccimarro04,Tinker06}.  
RSD constraints thus require both
large survey volumes and analytic or numerical models for
non-linear evolution and galaxy bias.  Milestones in large scale
RSD analysis include measurements from the 1.2-Jy \citep{Cole95}
and PSCz \citep{Tadros99} IRAS redshift surveys,
the Stromlo-APM redshift survey \citep{Loveday96},
the 2dFGRS \citep{Peacock01,Hawkins03,Percival04}, 
the VVDS \citep{Guzzo08}, VIPERS \citep{delaTorre2013}, the SDSS LRG sample
\citep{Okumura08,Chuang13a,Chuang13b,Oka14} and main galaxy redshift survey
\citep{Howlett2015}, and the 6dFGS \citep{Beutler12}
and WiggleZ \citep{Blake12} surveys.
RSD measurements from earlier BOSS data releases, using
a variety of technical approaches, include
\cite{Reid12,Reid14,Tojeiro12,Chuang13a,Samushia13,Samushia14,Sanchez13,
Sanchez14,Beutler14,Gil15RSD,AlamRSDDR112015}.
Modern RSD analyses usually frame their results in terms 
of constraints on $f(z)\sigma_8(z)$, where $\sigma_8(z)$ describes
the normalization of the linear theory matter power spectrum
at redshift $z$ (via the rms fluctuation in $8\hmpc$ spheres)
and 
\begin{equation}
f(z) \equiv {d\ln G \over d\ln a}
\label{eqn:fz}
\end{equation} is the logarithmic growth rate of the linear
fluctuation amplitude $G(t)$ with respect to expansion
factor $a(t) = (1+z)^{-1}$ (see \citealt{Percival09};
\citealt{Song09}; \S 7.2 of \citealt{Weinberg13} and
references therein).  The papers above adopt a variety
of approaches to RSD measurement and, crucially, to modelling
non-linear effects.  There is frequently a trade-off between
decreasing statistical errors and increasing theoretical
systematics as one probes to smaller scales.  There is also
partial degeneracy between clustering caused by peculiar 
velocities and the geometric distortion from the AP effect.
Analyses that reach to BAO scales, or that include BAO as an
external constraint, can achieve better $f\sigma_8$ constraints
because the BAO themselves constrain the AP distortion.
Conversely, AP constraints from anisotropic clustering analysis on sub-BAO
scales can help break the degeneracy between $\DM(z)$ and $H(z)$ in
BAO.  Thus, the potential gains from combining
BAO analyses with analyses of the full shape of the galaxy power
spectrum or correlation function are large.

This paper derives cosmological constraints from the combination
of BAO-only measurements that incorporate reconstruction and
full shape (FS) measurements of galaxy clustering without
reconstruction.  FS measurements do not have the precision gains
available from reconstruction at the BAO scale, and their
interpretation relies more heavily on non-linear modelling.  However, FS
analyses take advantage of the rich information on cosmological
parameters encoded in the broad band power spectrum, they use broad
band information to improve measurement of the AP effect, and,
most importantly for purposes of this paper, they yield constraints
on structure growth through RSD.  The input measurements for
our analysis are summarized in this paper and detailed in
seven supporting papers (Table~\ref{tab:supportingpapers}). 
The BAO scale is measured using the anisotropic two-point
correlation function in \cite{Ross16} and \cite{VargasMagana16}
and using the anisotropic power spectrum in \cite{BeutlerBAO16}.
The full shape of the anisotropic two-point correlation function
is computed and analysed using multipoles in \cite{Satpathy16} and
using $\mu$-wedges in \cite{Sanchez16}. The equivalent measurements
in Fourier space are made using power-spectrum multipoles in
\cite{BeutlerRSD16} and $\mu$-wedges in \cite{Grieb16}. 
Other key supporting papers are 
\cite{ReidEtAl15}, who describe the LSS catalogues used for
all of these analyses,
\cite{kitaura_etal:2016}, who describe the MultiDark-Patchy mock catalogues
used to test analysis methods and derive covariance matrices,
\cite{tinker_etal:2016}, who present high-resolution mock catalogues
and use them to test the RSD performance of our FS methods, and
\cite{SanchezStat16}, who describe and test our statistical 
methodology for combining results from these analyses.
The resulting final consensus likelihoods are publicly
available\footnote{\url{
    https://sdss3.org/science/boss_publications.php}.  The MCMC chains used to 
infer cosmological parameters will be made available after acceptance
of the paper.}.

\begin{table*}
\centering
\caption{Supporting papers providing input to this analysis, based
on the galaxy correlation function $\xi({\bf s})$ or power spectrum
$P({\bf k})$.  BAO-only analyses use post-reconstruction galaxy
distributions, while full shape (FS/RSD) analyses use pre-reconstruction 
distributions.  The last four papers provide technical underpinnings
for our analysis.}
\begin{tabular}{ll}
\hline\hline
\cite{Ross16}         & BAO, 
  $\xi({\bf s})$ multipoles, observational systematics\\
\cite{VargasMagana16} & BAO, 
  $\xi({\bf s})$ multipoles, modelling systematics\\
\cite{BeutlerBAO16}   & BAO, $P({\bf k})$ multipoles\\
\cite{Satpathy16}     & FS/RSD, $\xi({\bf s})$ multipoles\\
\cite{BeutlerRSD16}   & FS/RSD, $P({\bf k})$ multipoles\\
\cite{Sanchez16}      & FS/RSD, $\xi({\bf s})$ $\mu$-wedges\\
\cite{Grieb16}        & FS/RSD, $P({\bf k})$ $\mu$-wedges\\
\hline
\cite{ReidEtAl15}        & LSS catalogues\\
\cite{kitaura_etal:2016} & MD-Patchy mock catalogues\\
\cite{tinker_etal:2016}  & High-resolution mock catalogues, FS/RSD tests\\
\cite{SanchezStat16}     & Combined likelihoods methodology\\
\hline
\hline
\end{tabular}
\label{tab:supportingpapers}
\end{table*}

While each of these analyses is individually a major endeavour,
this multi-faceted approach has two key virtues.
First, we obtain results from several groups working 
semi-independently with a variety of analysis tools and
modelling assumptions, allowing powerful cross checks for
errors or for systematic effects that might influence
one method more than another.  Second, even though they
are applied to the same data set, these methods extract
information in different ways that are not entirely
redundant, even within the BAO-only or FS subsets.
We evaluate the covariance of their results using mock catalogues,
and even though the covariances are often strong, the
combined precision is higher than that of any individual input
\citep{SanchezStat16}.
Even a 10\% gain of precision is equivalent to a 20\% increase
of data volume, or a full year of BOSS observations.

In addition to these papers providing direct input to our consensus
analysis, a number of other BOSS Collaboration papers investigate
cosmological constraints from DR12 galaxy clustering using different 
approaches.  \cite{Cuesta16} and \cite{Gil15BAO} measure BAO in configuration
space and Fourier space, respectively, using the DR12 LOWZ and CMASS 
galaxy samples instead of the optimally binned combined sample 
(see \S\ref{sec:data}).  \cite{Gil15RSD} carry out a Fourier space
RSD analysis on these samples.
\cite{Slepian16BAO} present a $\sim$4.5$\sigma$ detection of BAO in the
3-point correlation function of BOSS CMASS galaxies.
\cite{Slepian16RV}, 
following a method suggested by \cite{Yoo11},
use the CMASS 3-point correlation function
to constrain the impact of baryon-dark matter relative velocities 
\citep{TH10} on galaxy clustering, 
setting a 0.3\% rms limit of a shift of the BAO distance scale
from this coupling.
\cite{Chuang2016} use DR12 clustering as a ``single-probe'' constraint
on $H(z)$, $\DM(z)$, $f\sigma_8$, and $\Omega_m h^2$, adopting only
broad priors in place of external data.
\cite{Pellejero-Ibanez16} add Planck CMB data to this analysis to derive
``double-probe'' constraints.
\cite{Wang16} and \cite{Zhao16} extract ``tomographic'' constraints from the 
BOSS combined sample adopting redshift-binning that is much finer than
used here, in configuration space and Fourier space, respectively.
\cite{Salazar16} derive constraints from the angular
auto-correlations and cross-correlations of BOSS galaxies divided
into redshift shells.

Our analyses make use of a fiducial cosmological model to convert
redshifts to comoving distances before calculating the clustering
signal. Thus the configuration-space and Fourier-space clustering
statistics we present are slightly distorted from their true comoving values
to the extent that the fiducial cosmological model is not exactly correct.
We allow for this distortion when comparing models with the data,
so our results are not biased by this step, even though we do not
recompute the correlation function and power spectrum from the galaxy
data for each set of cosmological parameters that we consider.
One can think of this use of a fiducial model as a form of 
``data-compression'', summarizing clustering by statistics
that can be modelled in an unbiased way by including the conversion
of length scales in the model predictions.
The fiducial cosmological model
used in this paper is a flat $\Lambda$CDM model with the following
parameters: matter density $\Omega_m = 0.31$, Hubble constant $h
\equiv H_0/(100\hubunits) = 0.676$, baryon density $\Omega_bh^2
= 0.022$, fluctuation amplitude $\sigma_8 = 0.8$, and spectral tilt
$n_\mathrm{s} = 0.97$. 
These parameters are generally within $1\sigma$ of the best-fit 
Planck2015 values (the CMB value of $\sigma_8$ is sensitive
to the choice of polarization data).
The sound horizon for this fiducial model is $r_{d,{\rm fid}}=147.78$ Mpc,
and convenient scalings of $\rd$ with cosmological parameters
can be found in \cite{Aubourg}.  
We quote constraints on distances in Mpc with a scaling factor,
e.g., $\DM(z) \times (\rdfid/\rd)$, so that the numbers
we provide are independent of the fiducial model choice.
Our inferences of $f(z)\sigma_8(z)$ and
the Alcock-Paczynski parameter $F_{\rm AP}(z)$ are likewise
independent of the choice of fiducial model.

The current paper is organised as follows: in Section~\ref{sec:data} we
summarise the SDSS data and define the BOSS combined sample.
Section~\ref{sec:methodology} summarises our general methodology
and introduces some relevant formalism. Our mock catalogues for the estimation
of the covariance matrices are presented in Section~\ref{sec:mocks}.
The BAO scale is measured in Section~\ref{sec:measuring_BAO} whereas
Section~\ref{sec:full_shape} presents AP and growth rate measurements using the full-shape of the two-point clustering statistics.
Our error analysis, including tests on high-fidelity mocks and
systematic error budget, is presented in Section~\ref{sec:errors}.
We combine our measurements and likelihoods in
Section~\ref{sec:combining_measurements}, where we present our final
consensus constraints and likelihoods. Finally, we use the latter
to infer cosmological parameters in Section~\ref{sec:cosmology}.
We conclude in Section~\ref{sec:discuss}.

\section{The Data}
\label{sec:data}
\subsection{SDSS-III data}

The Sloan Digital Sky Survey \citep{York00} observed more than one quarter of the sky using the 2.5-meter Sloan Telescope \citep{Gunn06} in Apache Point, New Mexico. Photometry in five passbands was obtained using a drift-scanning mosaic CCD camera \citep{Gunn08}, to a depth of 22.5 magnitudes in the $r-$band. Details on the camera, photometry and photometric pipeline can be found in 
\cite{Fukugita96}, \cite{Lupton01}, \cite{Smith02}, 
\cite{Pier03}, \cite{Padmanabhan08}, and \cite{Doi10}. 
All the photometry was re-processed and released in the eighth data release \citep{DR8}. 
Since 2008, the Baryon Oscillation Spectroscopic Survey 
(BOSS; \citealt{Dawson13}) of SDSS-III \citep{Eis11} has
collected optical spectra for over 1.5 million targets, 
distributed across a footprint of nearly 10,000 deg$^2$. 
Using double-armed spectrographs, significantly upgraded from
those used for SDSS-I and II,
BOSS obtained medium-resolution spectra ($R \approx 1500 $ to 2600) in the wavelength range from 3600 to 10000 \AA\ through 2-arcsecond fibres. 
\cite{Smee13} provide 
a detailed description of the spectrographs, and
\cite{Bolton12} describe the spectroscopic data reduction pipeline
and redshift determination.
Discussions of survey design, spectroscopic target selection, and
their implications for large scale structure analysis can be
found in \cite{Dawson13} and \cite{ReidEtAl15}.

\subsection{Catalogue creation}

\begin{table}
\begin{center}
\begin{tabular}{|l|c|c|c|c|}
\hline \hline
 & & $N_{gals}$ & $V_{\rm eff}$ (Gpc$^3$) & $V$ (Gpc$^3$) \\ \hline
 \multirow{3}{*}{$0.2 < z < 0.5$} & NGC & 429182 & 2.7 & 4.7\\
 		 & SGC & 174819& 1.0 & 1.7 \\ 
		 & {\bf Total} & {\bf 604001} & {\bf 3.7} & {\bf 6.4} \\ \hline
 \multirow{3}{*}{$0.4 < z < 0.6$} & NGC & 500872 & 3.1 & 5.3\\
 				 & SGC & 185498& 1.1   & 2.0\\ 
				  & {\bf Total} & {\bf 686370} & {\bf 4.2} & {\bf 7.3} \\ \hline
\multirow{3}{*}{$0.5 < z < 0.75$} & NGC & 435741& 3.0  & 9.0 \\
 				  & SGC & 158262& 1.1  & 3.3 \\
				   & {\bf Total} & {\bf 594003} & {\bf 4.1} & {\bf 12.3} \\ 
\hline \hline
\end{tabular}
\end{center}
\caption{Number of galaxies and effective volume for the combined
sample in each of the three redshift bins used in this paper. The
number of galaxies quoted is the total number of galaxies used in
the large-scale clustering catalogue, constructed as described in
\protect\cite{ReidEtAl15}. Please see their Table 2 for further
details. The effective volume is computed according to their Eq.\
52 with $P_0 = 10000 h^{-3}$Mpc$^3$ and includes the effects of
sector completeness and veto mask. Also included is the total volume within
each redshift bin. The expected BAO uncertainty scales closely with $\sqrt{V_{\rm eff}}$, which would equal the total volume given an infinite sampling density.
It is quoted here in Gpc$^3$ for our fiducial model value of
$h=0.676$.
}
\label{tab:dr12_combined}
\end{table}

\begin{figure*}
 \includegraphics[width=87mm]{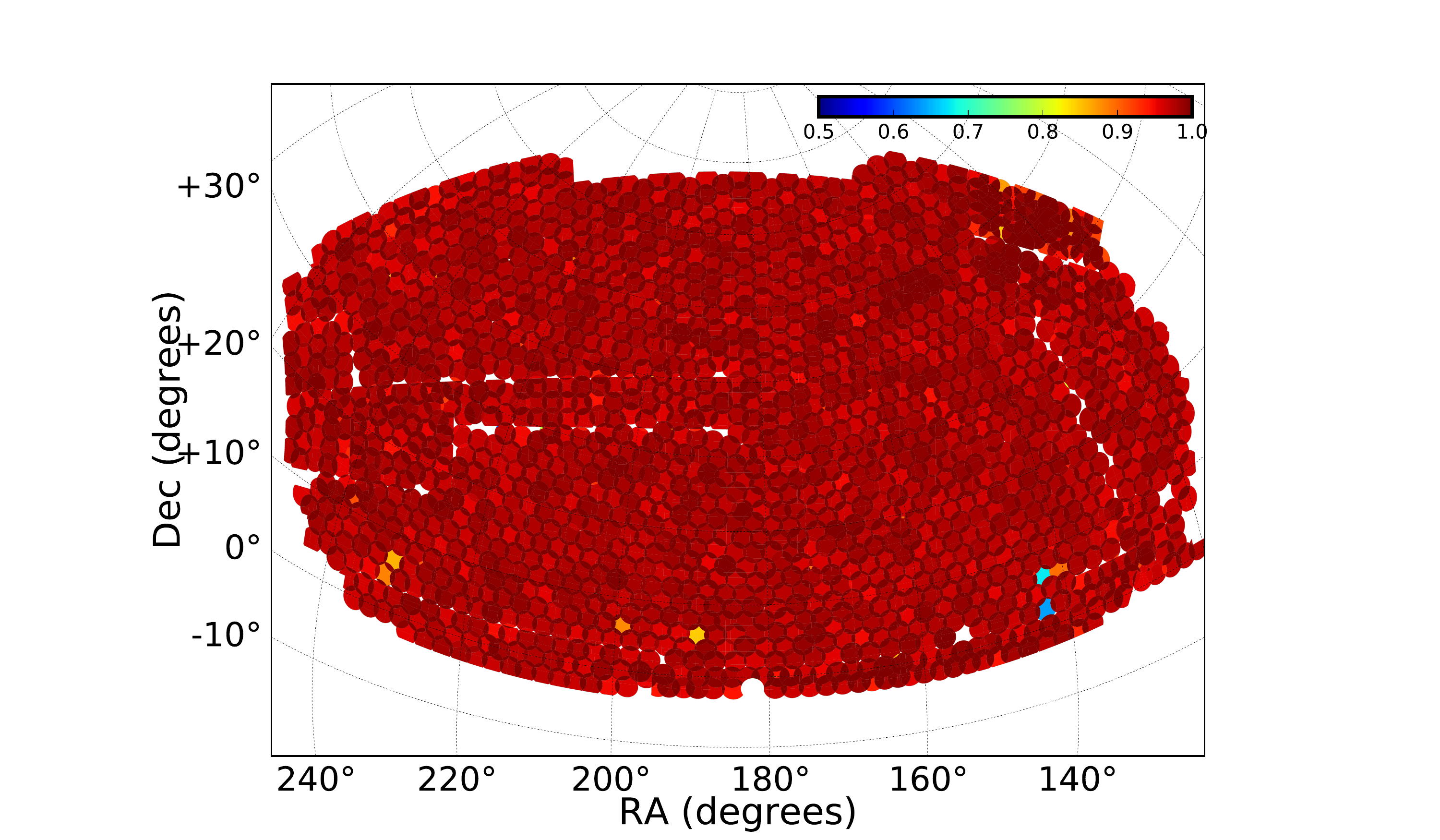}
  \includegraphics[width=87mm]{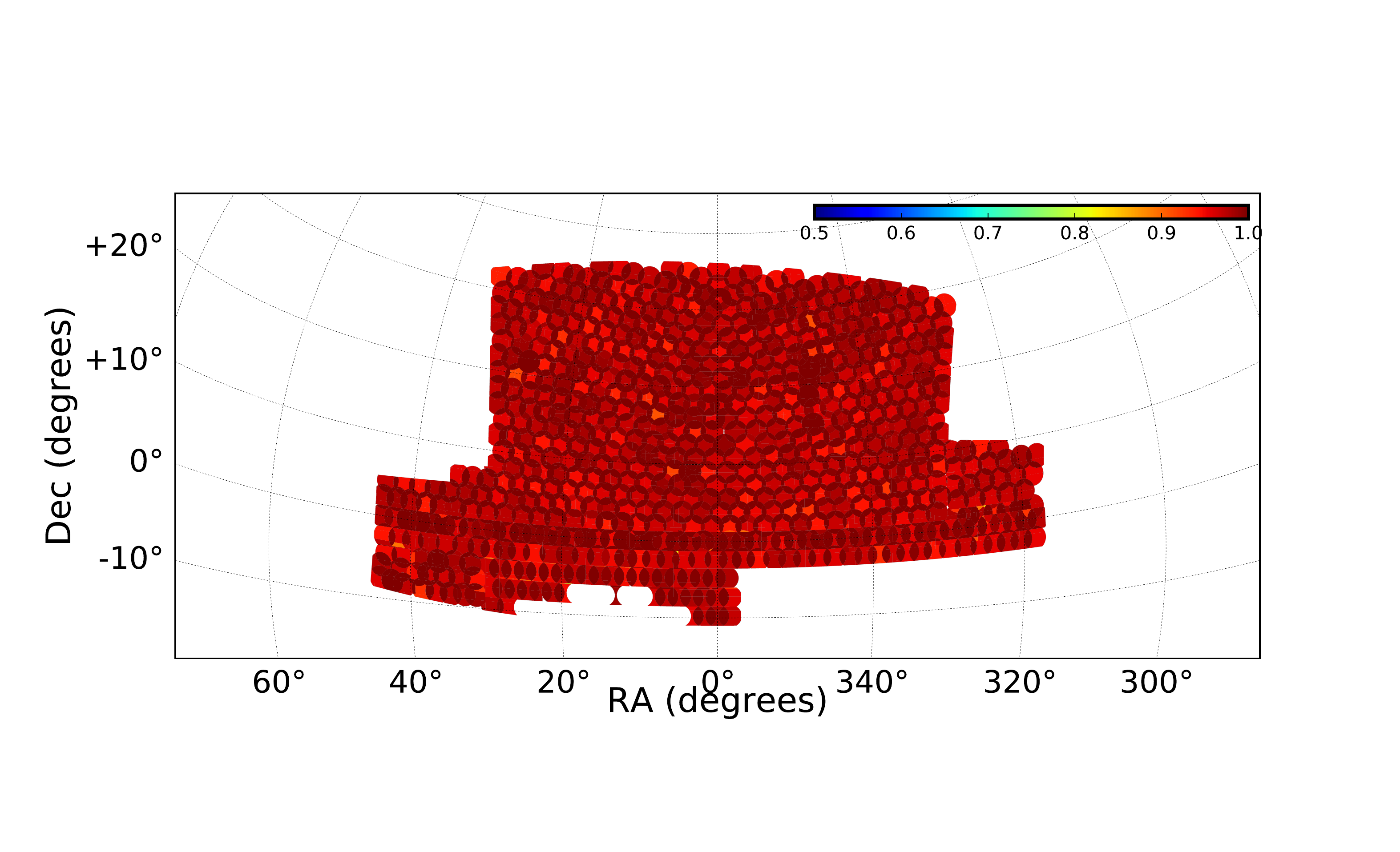}
 \caption{The footprint of the subsamples corresponding to the Northern and Southern galactic caps of the BOSS DR12 combined sample.
  The circles indicate the different pointings of the telescope and their colour corresponds to the sector completeness. The total area in the combined sample footprint, weighted by completeness, is 10,087 deg$^2$. Of these, 759 deg$^2$ are excluded by a series of veto masks, leaving a total effective area of 
  9329 deg$^2$. See \citealt{ReidEtAl15} for further details on completeness calculation and veto masks.}
\label{fig:footprint}
\end{figure*}

\begin{figure}
 \includegraphics[width=85mm]{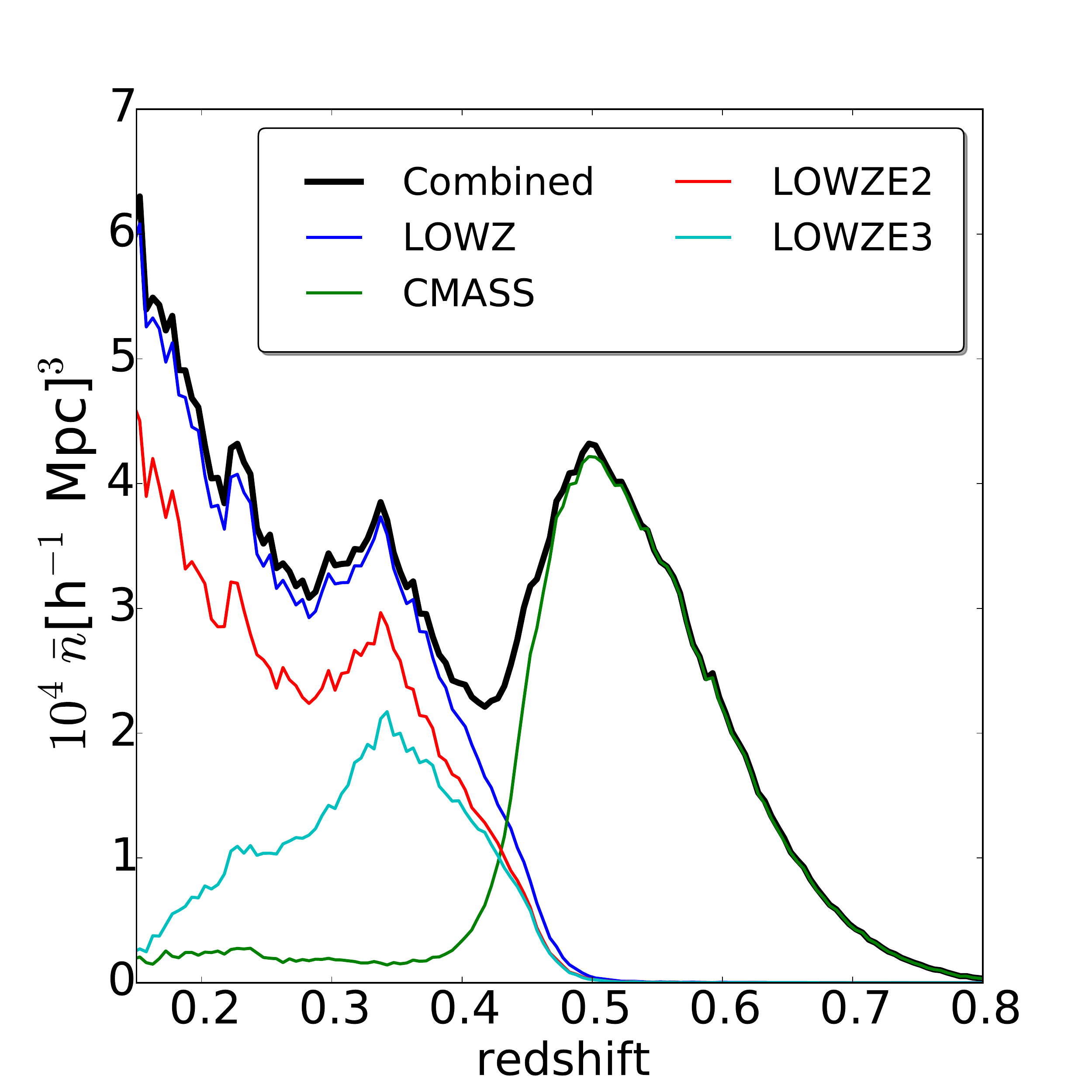}
\caption{Number density of all four target classes assuming our
fiducial cosmology with $\Omega_m = 0.31$, along with the sum of the
CMASS and LOWZ number densities (black). }
\label{fig:nz}
\end{figure}

The creation of the large-scale structure
catalogues from the BOSS spectroscopic observations is detailed in \cite{ReidEtAl15}. In brief, we consider the survey footprint, veto masks and survey-related systematics (such as fibre collisions and redshift failures) in order to construct data and random catalogues for the DR12 BOSS galaxies. The veto masks exclude 6.6\% (9.3\%) of the area within the north (south) galactic cap footprint, mostly due to regions of non-photometric quality but we also consider plate centerposts, collision priorities, bright stars, bright objects, Galactic extinction and seeing. The DR12 footprint is shown in Fig.~\ref{fig:footprint} 
and Table~\ref{tab:dr12_combined} summarises our sample, 
which spans a completeness-weighted effective area of 9329 deg$^2$ (after removing the vetoed area).
The total un-vetoed area with completeness $c>0.7$ is 9486 deg$^2$.

BOSS utilizes two target selection algorithms: LOWZ was designed
to target luminous red galaxies up to $z\approx0.4$, while CMASS was
designed to target massive galaxies from $0.4 < z < 0.7$. The spatial
number density of these samples can be seen in Fig.~\ref{fig:nz}.
In previous papers, we analyzed these two samples separately, splitting at $z=0.43$
and omitting a small fraction of galaxies in the tails of both redshift distributions
as well as the information from cross-correlations between the two samples.
For the current analysis, we instead construct a {\it combined
sample} that we describe in Section~\ref{sec:combined_sample}. 
With the combined map, we more optimally divide the observed volume 
into three partially overlapping redshift slices.
As in
\cite{Anderson2014b}, the CMASS galaxies are weighted to correct
for dependencies between target density and both stellar density
and seeing. The definitions and motivations for these weights are
described in \cite{ReidEtAl15} and \cite{Ross16}.  Clustering analyses of the
DR12 LOWZ and CMASS samples, using two-point statistics, can be
found in \cite{Cuesta16} and \cite{Gil15BAO}.

In addition to the LOWZ and CMASS samples, we use data from two early (i.e., while the final selection was being settled on) LOWZ selections, each of which are subsets of the final LOWZ selection. These are defined in \cite{ReidEtAl15} and denoted `LOWZE2' (total area of 144 deg$^2$) and `LOWZE3' (total area of 834 deg$^2$). Together with the LOWZ sample, these three samples occupy the same footprint as the CMASS sample. As detailed in \cite{Ross16}, the `LOWZE3' sample requires a weight to correct for a dependency with seeing. The LOWZ and LOWZE2 samples require no correction for systematic dependencies, as these were found to be negligible. We thus have four BOSS selections that we can use to construct a combined sample. This combined sample uses all of the CMASS, LOWZ, LOWZE2, and LOWZE3 galaxies with $0.2 < z < 0.75$ and allows us to define redshift slices of equal volume, thereby optimising our signal over the whole sample (see Section~\ref{sec:combined_sample}).

\subsection{The Combined BOSS Sample}\label{sec:combined_sample}

In this section, we motivate the methods we use to combine the four BOSS samples into one combined sample.

In principle, when combining galaxy populations with different
clustering amplitudes, it would be optimal to apply a weight to each
sample to account for these differences 
\citep{PVP}. \cite{Ross16} present measurements of the
redshift-space correlation function for each of the four BOSS
selections. Section 5.1 of that paper shows that the clustering
amplitudes of each selection match to within 20 per cent and that
combining the selections together where they overlap in redshift
has no discernible systematic effect. Given the small difference
in clustering amplitudes, a weighting scheme would improve the
results by a negligible factor while imparting considerable
additional complexity. We therefore choose to weight each sample equally when
combining the catalogues. Each galaxy in this combined sample
is then weighted by the redshift-dependent FKP weight
\citep{Feldman94}.

The clustering amplitude of different selections {\it within} the CMASS sample varies considerably more than the individual target selections (LOWZ/LOWZE2/LOWZE3/CMASS): the difference in clustering amplitude between the reddest and bluest galaxies within CMASS is a factor of two \citep{Ross14,Favole15,Patej16}. However, even when optimally weighting for this difference, the forecasted improvement in the statistical power of BOSS is 2.5 percent and our attempts to employ such a weighting in mock samples were unable to obtain even this improvement. Therefore, we have chosen to not introduce this additional complexity into our analysis.

We define the overall redshift range to consider for BOSS galaxies as $0.2 < z < 0.75$. Below $z=0.2$, the sample is affected by the bright limit of $r > 16$, and the BAO 
scale has been measured for $z<0.2$ 
galaxies in the SDSS-I/II main galaxy redshift survey
\citep{Strauss02} by \cite{Ross15MGS}. 
The upper limit of 0.75 is higher than in our previous analyses as we 
find no systematic concerns associated with using the $z > 0.7$ data, but the number density has decreased to 
10$^{-5}h^{3}$Mpc$^{-3}$ at $z = 0.75$ (a factor of 40 below its peak at $z\approx 0.5$; see Fig.~\ref{fig:nz}) and 
any additional data at higher redshift offer negligible 
improvement in the statistical power of the BOSS sample.

We defined the redshift bins used in this analysis based on an ensemble of 100 mock catalogues of the 
combined BOSS sample in the range $0.2<z<0.75$. We tested several binning schemes 
by means of anisotropic BAO measurements on these mock catalogues. For each configuration, we ran an MCMC analysis 
using the mean value and errors from the BAO measurements, combining them with synthetic CMB measurements 
(distance priors) corresponding to the same cosmology of these mock catalogues. We chose the binning
that provides the strongest constraints on the dark energy equation-of-state parameter $w_{\rm{DE}}$. 
It consists of two independent redshift bins of nearly equal effective volume for $0.2<z<0.5$ and $0.5<z<0.75$. 
In order to ensure we have counted every pair of BOSS galaxies, we also define an overlapping redshift bin of nearly 
the same volume as the other two, covering the redshift range $0.4<z<0.6$. 
Using our mock catalogues, with the original LOWZ and CMASS 
redshift binning we obtain a 3.5\%  (9.6\%) precision measurement of the
transverse (line-of-sight) BAO scale in the LOWZ sample and a
1.8\% (4.3\%) precision measurement for the CMASS sample. 
With our chosen binning for the combined sample, we instead obtain
transverse (line-of-sight) precision of 2.5\%  (6.3\%) 
in our low redshift bin and 
2.3\% (5.6\%) in our high redshift bin ,
comparable for the two samples by design.
Our results in \S~\ref{sec:comparison_to_previous} are consistent with
these expected changes of precision relative to the LOWZ and CMASS samples.
Measurements in the overlapping redshift bin are of course covariant
with those in the two independent bins, and we take this covariance
(estimated from mock catalogues) into account
when deriving cosmological constraints.
See Table~\ref{tab:dr12_combined} for a summary of the combined sample.

\subsection{The NGC and SGC sub-samples}

The DR12 combined sample is observed across the two Galactic
hemispheres, referred to as the Northern and Southern
galactic caps (NGC and SGC, respectively).  As these two regions do not overlap, they are
prone to slight offsets in their photometric calibration.
As described in appendix~\ref{app:NGC_vs_SGC}, we find good evidence
that the NGC and SGC subsamples probe slightly different galaxy
populations in the low-redshift part of the combined sample, and
that this difference is consistent with an offset in photometric calibration
between the NGC and the SGC (first reported by \citealt{Schlafly:2010dz}).
Having established the reason for the observed difference in
clustering amplitude, we decide not to re-target the SGC but rather
to simply allow sufficient freedom when fitting models to the
clustering statistics in each galactic cap, as to allow for this
slight change in galaxy population.  In particular, the different Fourier-space 
statistics are modelled with different nuisance parameters in
the two hemispheres, as appropriate for each method. Using fits of
the MD-Patchy mocks, we find that this approach brings no penalty in uncertainty
of fitted parameters. We refer the reader to the individual companion
papers for details on how this issue was tackled in each case.

\section{Methodology}
\label{sec:methodology}

\subsection{Clustering measurements}

We study the clustering properties of the BOSS combined sample by
means of anisotropic two-point statistics in configuration and
Fourier space. Rather than studying the full two-dimensional
correlation function and power spectrum, we use the information
contained in their first few Legendre multipoles or in the clustering
wedges statistic \citep{Kazin2012}.

In configuration space, the Legendre multipoles $\xi_{\ell}(s)$ are given by
\begin{equation}
\xi_\ell(s)\equiv \frac{2\ell+1}{2}\int^1_{-1} L_\ell(\mu)\xi(\mu,s)\,{\rm d}\mu,
\label{eq:multipoles_xi}
\end{equation}
where $\xi(\mu,s)$ is the two-dimensional correlation function, $L_\ell$ is the Legendre polynomial or order $\ell$, and $\mu$ is the cosine of the angle 
between the separation vector $\mathbf{s}$ and the line-of-sight
direction. The power spectrum multipoles $P_{\ell}(k)$ are defined in an analogous way in terms of the 
two-dimensional power spectrum $P(\mu,k)$
\begin{equation}
P_\ell(k)\equiv \frac{2\ell+1}{2}\int^1_{-1} L_\ell(\mu)P(\mu,k)\,{\rm d}\mu,
\label{eq:multipoles_pk}
\end{equation}
and are related to the configuration-space $\xi_{\ell}(s)$ by
\begin{equation}
\xi_{\ell}(s)\equiv \frac{i^{\ell}}{2\pi^2}\int^{\infty}_{0} P_{\ell}(k) j_{\ell}(ks)\,k^2{\rm d}k,
\label{eq:pl2xil}
\end{equation}
where $j_{\ell}(x)$ is the spherical Bessel function of order $\ell$. We use the information from the 
monopole, quadrupole and hexadecapole moments ($\ell = 0$, 2 and 4), which are a full description of the $\mu$ dependence
of $\xi(s,\mu)$ in the linear regime and in the distant observer approximation.

The configuration- and Fourier-space wedges, $\xi_{\mu_1}^{\mu_2}(s)$ and $P_{\mu_1}^{\mu_2}(k)$ correspond 
to the average of the two-dimensional correlation function and power spectrum over the interval 
$\Delta\mu=\mu_{2}-\mu_{1}$, that is
\begin{equation}
\xi_{\mu_1}^{\mu_2}(s)\equiv \frac{1}{\Delta \mu}\int^{\mu_2}_{\mu_1}{\xi(\mu,s)}\,{{\rm d}\mu}
\label{eq:wedges_xi}
\end{equation}
and
\begin{equation}
P_{\mu_1}^{\mu_2}(k)\equiv \frac{1}{\Delta \mu}\int^{\mu_2}_{\mu_1}P(\mu,k)\,{{\rm d}\mu}.
\label{eq:wedges_P}
\end{equation}
Here we define three clustering wedges by splitting the $\mu$ range from 0 to 1 into three equal-width intervals.
We denote these measurements by $\xi_{3{\rm w}}(s)$ and $P_{3{\rm w}}(k)$. 

The information content of the multipoles and the wedges is highly covariant, as they are related by 
\begin{equation}
 \label{eq:xi_w_from_ell}
 \xi_{\mu_1}^{\mu_2}(s) = \sum_\ell \xi_\ell(s) \, {\bar L}_\ell,
\end{equation}
where ${\bar L}_\ell$ is the average of the Legendre polynomial of order $\ell$ over the $\mu$-range of the 
wedge,
\begin{equation}
 \label{eq:Lpbar}
 {\bar L}_\ell \equiv \frac{1}{\Delta \mu} \int_\mu^{\mu + \Delta \mu} L_\ell(\mu)\, {\rm d} \mu.
\end{equation}
More details on the estimation of these statistics using data from the 
BOSS combined sample can be found in the supporting papers 
listed in 
Table~\ref{tab:supportingpapers}.

\subsection{Parametrizing the Distance Scale}

The BAO scale is measured anisotropically in redshift-space in both 
the two-point correlation function and the power spectrum. 
We measure the shift of the BAO peak position with respect to 
its position in a fiducial cosmology, which directly gives 
the Hubble expansion rate, $H(z)$, and the comoving
angular diameter distance, $\DM(z)$, relative to the 
sound horizon at the drag epoch, $\rd$ (eq.~\ref{eqn:rd}).
We define the dimensionless ratios
\begin{equation}
\alpha_{\perp} = \frac{\DM(z)\rdfid}{\DM^{\rm fid}(z)\rd},
\qquad 
\alpha_{\parallel} = \frac{H^{\rm fid}(z)\rdfid}{H(z)\rd},
\end{equation}
to describe shifts perpendicular and parallel to the line of sight. 
The anisotropy of galaxy clustering
is also often parametrized using an isotropically-averaged 
shift $\alpha$ and a warping factor $\epsilon$ with
\begin{equation}
\alpha = \alpha_\perp ^{2/3} \alpha_{\parallel}^{1/3}, \qquad
\epsilon + 1 = \left( \frac{\alpha_{\parallel}}{\alpha_\perp}\right)^{1/3}.
\label{eqn:alpha_epsilon_def}
\end{equation}
Converting equation~(\ref{eqn:alpha_epsilon_def}) to more physical
quantities, we can define a spherically-averaged distance 
$\DV(z)$ and an anisotropy parameter 
(often referred to as the Alcock-Paczynski parameter) $F_{\rm AP}(z)$ as
\begin{equation}
\DV(z) = \left( \DM^2(z) \frac{cz}{H(z)}\right)^{1/3}~,
\end{equation}
\begin{equation}
F_{\rm AP}(z) = \DM(z)H(z)/c~.
\end{equation}

Although these quantities are trivially interchangeable, 
we will adopt in each section the most natural 
parametrization. In particular, 
we quote our measurements in physical units: $D_{\rm A}(z)$, $H(z)$, $D_{\rm V}(z)$, 
$F_{\rm AP}(z)$ and $D_{\rm M}(z)$.
We generally use $\alpha_\perp$ and $\alpha_{\parallel}$ 
when referring to studies and checks on our mock catalogues
and $\alpha$ and $\epsilon$ when describing our systematic error budget.
In our fiducial cosmological model, $\rdfid=147.78$ Mpc, and 
convenient approximations for the scaling of $r_d$ with cosmological
parameters (including neutrino mass) can be found in \cite{Aubourg}.
Within $\Lambda$CDM, the uncertainty in $r_d$ given Planck CMB
constraints is 0.2\%, substantially smaller than our statistical errors.
However, changes to the pre-recombination energy density, such as
additional relativistic species or early dark energy, can change
$r_d$ by altering the age-redshift relation at early epochs.

\section{Mock catalogues and the covariance matrix}
\label{sec:mocks}
\label{sec:mocks_cov}

We use
mock galaxy catalogues to estimate the covariance matrix of our
clustering measurements and to extensively test our methods. For this
work, we utilized two distinct methods of mock galaxy creation:
MultiDark-Patchy (hereafter MD-Patchy;
\citealt{kitaura_etal:2016}) and Quick Particle Mesh (QPM;
\citealt{white_etal:2014}). MD-Patchy simulates the growth of density
perturbations through a combination of second-order Lagrange
perturbation theory and a stochastic halo biasing scheme calibrated on
high-resolution N-body simulations. QPM uses low-resolution particle
mesh simulations to evolve the density field, then selects particles
from the density field such that they match the one- and two-point
statistics of dark matter halos. Both mock algorithms then use halo
occupation methods to construct galaxy density fields that match the
observed redshift-space clustering of BOSS galaxies as a function of
redshift. Each mock matches both the angular selection function of the
survey, including fibre collisions, and the observed redshift
distribution $n(z)$. A total of 1000 MD-Patchy mocks and 1000 QPM
mocks were utilized in this analysis.

Analyses of
previous data releases utilized mocks created from the PTHalos method
\citep{Manera2015}. Comparison of the QPM and PTHalos in the context
of our BAO analysis can be found in \cite{VargasMaganaEtAl15}, and a
comparison of MD-Patchy to PTHalos, as well as other leading methods for generating mock catalogues, can be found in
\cite{Chuang2015}. \cite{VargasMagana16} directly compared the values
and errors found in pre-reconstructed BAO analysis between PTHalos and
QPM for the large-scale structure sample of the SDSS 11th data release (DR11). They found that the
derived quantities, $\alpha$ and $\epsilon$, and their uncertainties,
were consistent between the two methods.

Our reconstruction and BAO fitting procedures, as well as tests of the
clustering measurements, have been applied to both sets 
\citep{VargasMagana16, Ross16}. For details on the use of the mocks
in the full shape analyses, see the respective papers
(Table~\ref{tab:supportingpapers}) for each individual analysis.
Having two sets of mock simulations allows us to test the
dependence of our errors on the mock-making technique. QPM and
MD-Patchy differ in their methods of creating an evolved density
field, as well as their underlying cosmology. As a conservative
choice, our final error bars on all measurements, both BAO and RSD,
are taken from the MD-Patchy covariance matrix because the errors
obtained using the MD-Patchy covariance matrix are roughly $10-15$ per cent
larger, and the clustering of the MD-Patchy mocks is a better match to
the observed data. The larger derived error bar from the MD-Patchy
covariance matrix is obtained both when fitting the observations and when
fitting mock data.

For each clustering measurement, we use the distribution of the measured
quantity measured from the mocks to estimate the covariance matrix
used in all fittings. For the MD-Patchy mocks for the DR12 combined
sample, distinct boxes were used for the NGC and SGC footprints, 
a change in practice compared to our analyses of previous data releases.
Thus, the covariance
matrices for the NGC and the SGC are each estimated from 997 total
mocks\footnote{Three MD-Patchy mocks were removed from the final
  analysis due to unusual, non-Gaussian, clustering properties that
  were likely due to errors in the simulations.}. The full procedure
for estimating the covariance matrices is described in
\cite{Percival2014}, which includes the uncertainties in the
covariance matrix when derived from a finite sample of simulations.

\section{Post-reconstruction BAO measurements}\label{sec:measuring_BAO}

\subsection{Reconstruction} \label{sec:reconstruction}

\begin{figure*}
\begin{minipage}{7in}
\includegraphics[width=7in]{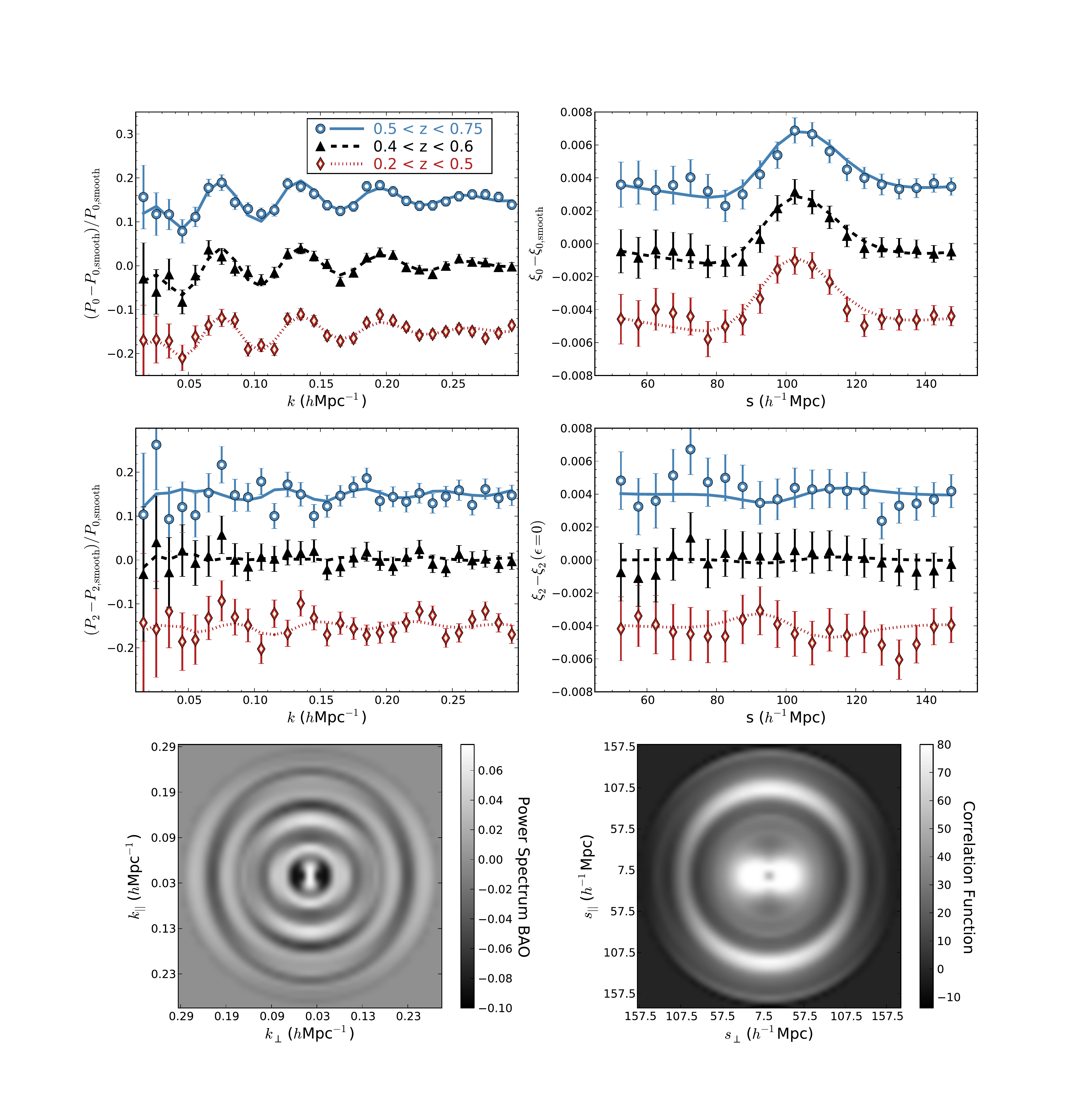}
\caption{BAO signals in the measured post-reconstruction 
power spectrum (left panels) and correlation function (right panels)
and predictions of the best-fit BAO models (curves).
To isolate the BAO in the monopole (top panels), predictions of a 
smooth model with the best-fit cosmological parameters
but no BAO feature have been subtracted, and the same
smooth model has been divided out in the power spectrum panel.
For clarity,
vertical offsets of $\pm 0.15$ (power spectrum) and $\pm 0.004$
(correlation function) have been added to the points and curves
for the high- and low-redshift bins, while the intermediate redshift
bin is unshifted.
For the quadrupole (middle panels), we subtract the quadrupole of
the smooth model power spectrum, and for the correlation function we
subtract the quadrupole of a model that has the same parameters
as the best-fit but with $\epsilon=0$.
If reconstruction were perfect and the fiducial model were exactly 
correct, the curves and points in these panels would be flat;
oscillations in the model curves indicate best-fit $\epsilon \neq 0$.
The bottom panels show the measurements for the $0.4 < z < 0.6$ 
redshift bin decomposed into the component of the separations transverse to and 
along the line of sight, 
based on $x(p,\mu) = x_0(p)+L_2(\mu)x_2(p)$, 
where $x$ represents either $s^2$ multiplied by the correlation function 
or the BAO component power spectrum displayed in the upper panels, 
$p$ represents either the separation or the Fourier mode, 
$L_2$ is the 2nd order Legendre polynomial, 
$p_{||} = \mu p$, and $p_{\perp} = \sqrt{p^2-\mu^2p^2}$.
}
  \label{fig:pkxirecBAOfit}
  \end{minipage}
\end{figure*}

\begin{figure*}
    \centering 
    \includegraphics[width=0.33\textwidth]{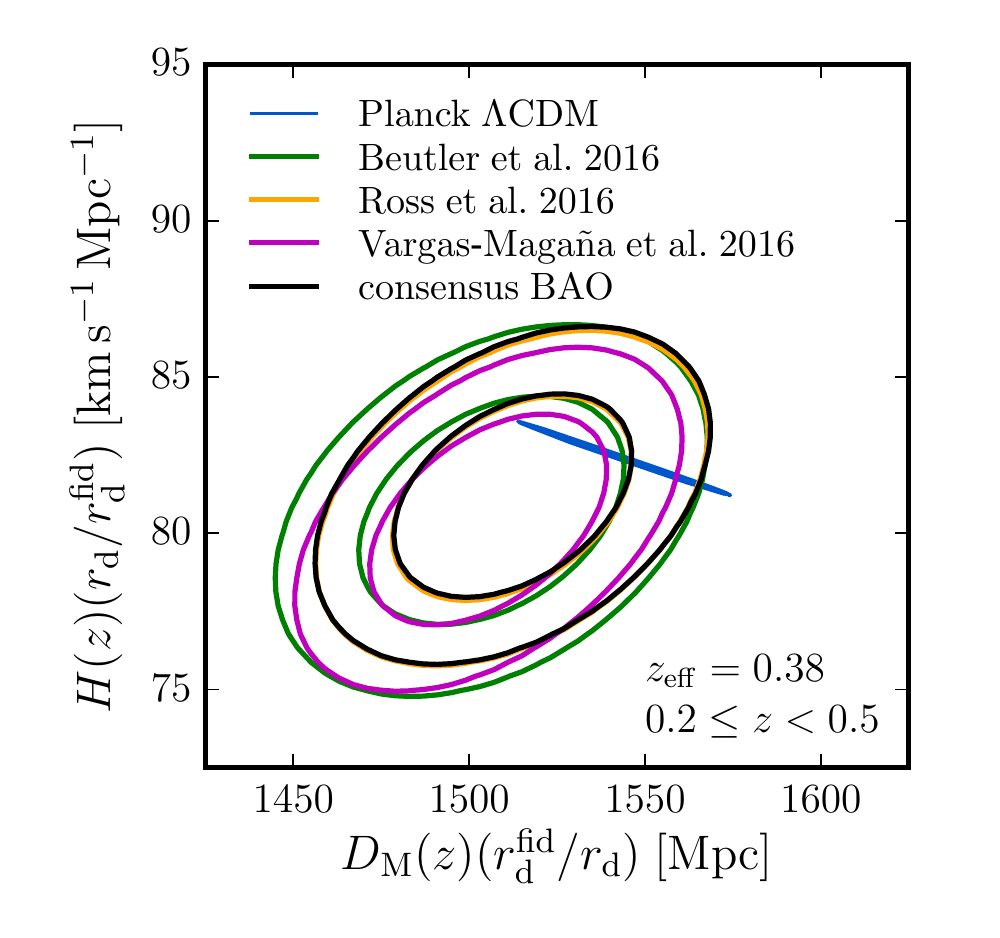}
    \includegraphics[width=0.33\textwidth]{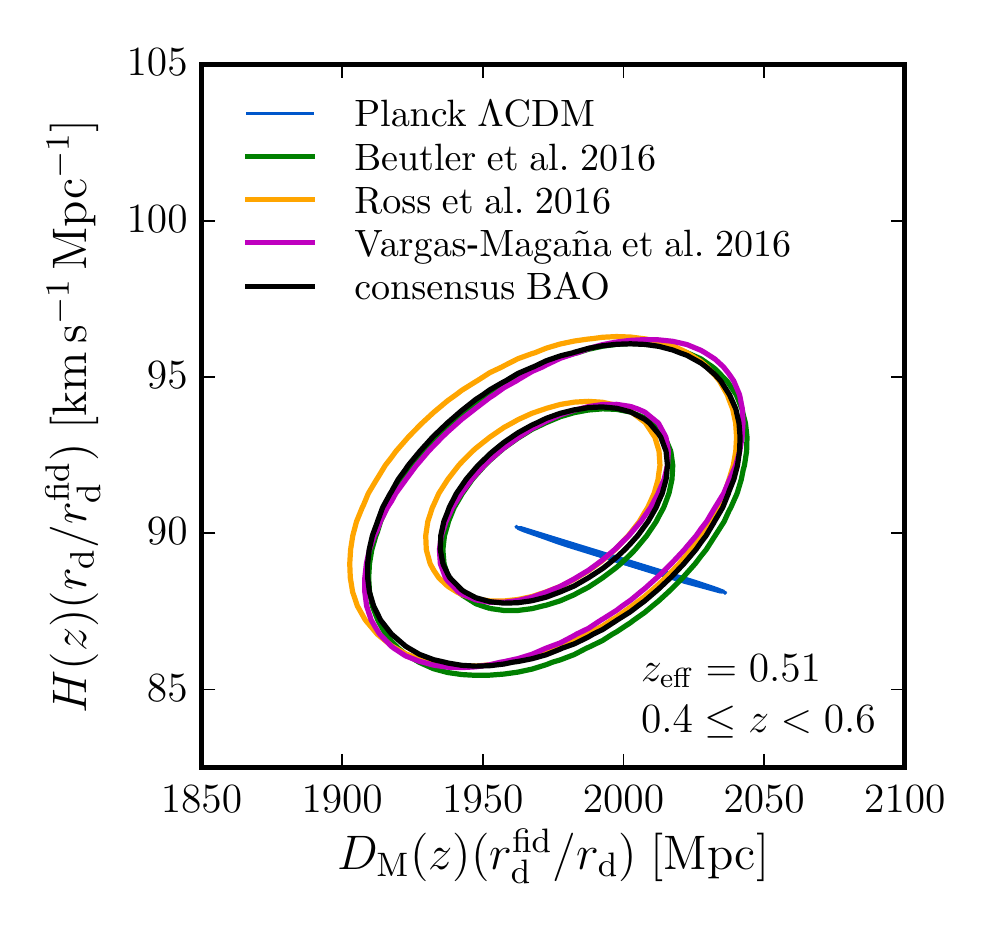}
    \includegraphics[width=0.33\textwidth]{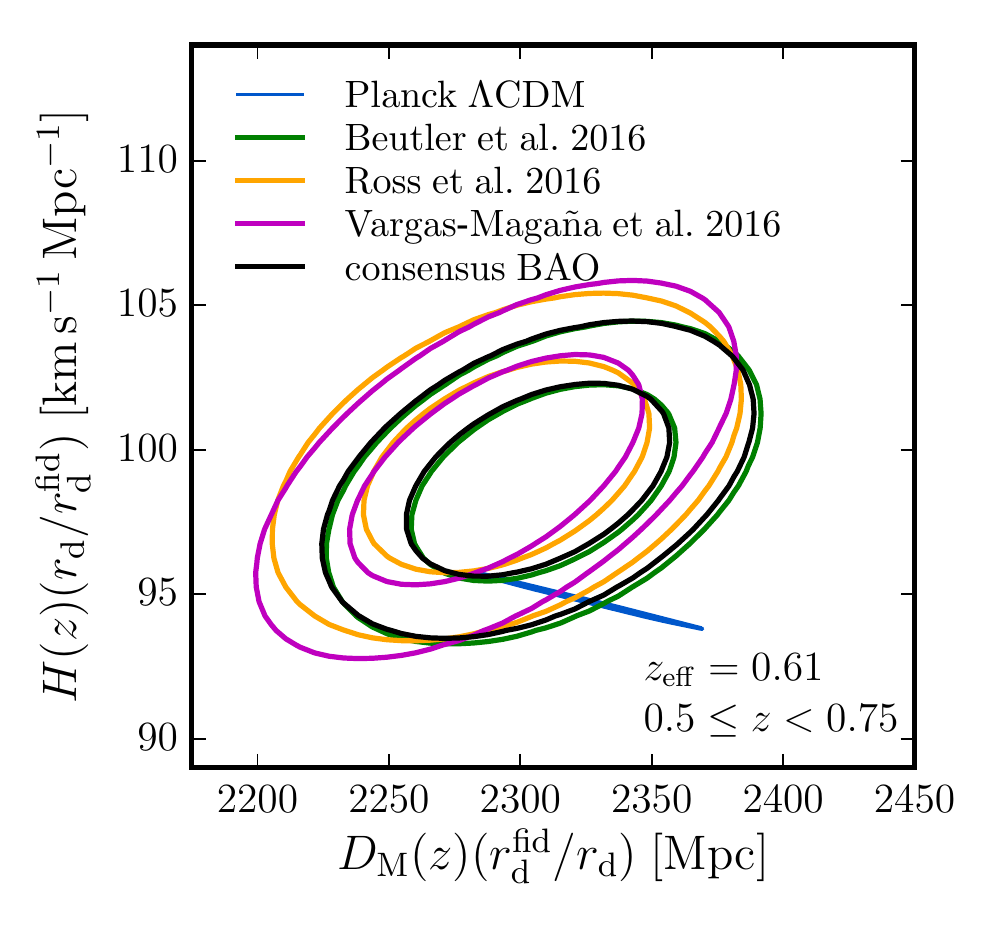}
    \caption{
    Two-dimensional 68 and 95 per cent marginalized constraints on 
	$\DM(z)\times(\rdfid/\rd)$ and
    $H(z)\times (\rd/\rdfid)$
    obtained by fitting the BAO signal in the post-reconstruction monopole and quadrupole in configuration and 
    Fourier space. The black solid lines represent the combination of these results into a set of consensus 
    BAO-only constraints, as described in Section~\ref{sec:consensus_results}.      
    The blue solid lines correspond to the constraints inferred from the Planck CMB temperature and
    polarization measurements under the assumption of a $\Lambda$CDM model. 
    }
  \label{fig:2d_bao}
\end{figure*}

\begin{table*}
\begin{center}
\caption{Summary table of post-reconstruction BAO-only constraints on 
$\DM\times\left(r_{\rm d,fid}/r_{\rm d}\right)$ 
and $H\times\left(r_{\rm d}/r_{\rm d,fid}\right)$}
\begin{tabular}{ccccc}
\hline \hline
 Measurement & redshift &  Beutler et al. (b) & Vargas-Maga{\~n}a et al. &  Ross et al.  \\ 
                       &                &  $P(k)$ & $\xi(s)$ & $\xi(s) $   \\ \hline
$\DM\times\left(\rdfid/\rd\right)$ [Mpc] & $z=0.38 $ & $1507 \pm 25$ &$1507 \pm 22$ &$1512 \pm 23$    \\ 
$\DM\times\left(\rdfid/\rd\right)$ [Mpc] & $z=0.51 $ &  $1976 \pm 29 $&$1975 \pm 27$ & $1971 \pm 27$ \\ 
$\DM\times\left(\rdfid/\rd\right)$ [Mpc] & $z=0.61 $ &$2307 \pm 35$ & $ 2291 \pm 37$& $2296 \pm 37$  \\ 
$H\times\left(\rd/\rdfid\right)\,[{\rm km}\,{\rm s}^{-1}{\rm Mpc}^{-1}]$ & $z=0.38 $ & $80.7 \pm 2.4$ & $80.4 \pm 2.4$ & $81.1 \pm 2.2$  \\
$H\times\left(\rd/\rdfid\right)\,[{\rm km}\,{\rm s}^{-1}{\rm Mpc}^{-1}]$ & $z=0.51 $ & $90.8 \pm 2.2$ & $91.0 \pm 2.1$ & $91.1 \pm 2.1$  \\ 
$H\times\left(\rd/\rdfid\right)\,[{\rm km}\,{\rm s}^{-1}{\rm Mpc}^{-1}]$ & $z=0.61 $ & $98.8 \pm 2.3$ & $99.3 \pm 2.5$ & $99.4 \pm 2.2$ \\ 
\hline \hline
\label{tab:combined_BAO}
\end{tabular}
\end{center}
\end{table*}

Following our previous work, we approximately
reconstruct the linear density field in order to increase the
significance and precision of our measurement of the BAO peak
position. The reconstruction algorithm we use is described in
\cite{PadmanabhanEtAl12}. This algorithm takes two input parameters:
the growth rate parameter $f(z)$ (eq.~\ref{eqn:fz}) to correct
for redshift-space distortion effects and the galaxy bias parameter
$b$ to convert between the galaxy density field and the matter density
field. In the case of BOSS galaxies we find that the galaxy bias
is a rather shallow function of redshift except for the very
high redshift end, so a single value is assumed for the three
redshift bins in our analysis. Furthermore, for a $\Lambda$CDM
cosmology the value of $f$ is not strongly redshift-dependent either,
varying by $\sim$10 per cent in the range $0.20<z<0.75$.
Variations of this size in the input parameters have been proven
not to affect in any significant way the post-reconstruction BAO
measurements \citep{Anderson2012}. This allows us to run reconstruction
on the full survey volume (as opposed to running the code on individual
redshift bins) and thus take into account the contributions from
larger scales to bulk flows. The values of the input parameters we
used correspond to the value of the growth rate at $z=0.5$ for our
fiducial cosmology, $f=0.757$, and a galaxy bias of $b=2.2$ for the
QPM and MD-Patchy mocks. For the data, we assumed a bias of $b=1.85$.
The finite-difference grid is $512^3$ (each cell being roughly 
6$h^{-1}$Mpc on a side), and we use a Gaussian kernel
of 15$h^{-1}$Mpc to smooth the density field, a choice found
to provide conservative error bars in BAO fitting
\citep{VargasMaganaEtAl15, Burden15}.

Fig. \ref{fig:pkxirecBAOfit} displays the post-reconstruction BAO
feature in the combined sample data. Each panel uses different means
to isolate the BAO information.  The upper panels represent the BAO
information in the monopole of the clustering measurements; this
information provides the spherically averaged BAO distance constraint.
For the power spectrum, we display $(P_0 - P_{0,{\rm
    smooth}})/P_{0,{\rm smooth}}$ and for the correlation function
$\xi_0 - \xi_{0,{\rm smooth}}$, where the subscript `smooth' denotes
the best-fit model but substituting a template with no BAO feature for
the nominal BAO template.  One can observe that the
spherically-averaged BAO feature is of nearly equal strength in each
redshift bin, as expected given their similar effective volumes.  Note
that data points in $\xi_0(s)$ are strongly correlated, while those in
$P_0(k)$ are more nearly independent.  Qualitatively, our ability to
measure the isotropic BAO scale comes down to our ability to centroid
the BAO peaks in $\xi_0(s)$ or to determine the phases of the
oscillations in $P_0(k)$.  Best-fit models are slightly offset
horizontally because the best-fit values of $\alpha$ are slightly
different in the low, middle, and high-redshift bins.  (Vertical
offsets are added for visual clarity.)

The middle panels of Fig. \ref{fig:pkxirecBAOfit} illustrate the
information provided by the quadrupole of the clustering measurements,
which constrains the anisotropy parameter $\epsilon$ (or,
equivalently, $F_{\rm AP}$).  The nature of the BAO signature is more
subtle here, since if reconstruction perfectly removed redshift-space
distortions and the fiducial cosmological were exactly correct then
clustering would be isotropic and the quadrupole would vanish.  For
the power spectrum, we display $(P_2 - P_{2,{\rm smooth}})/P_{0,{\rm
    smooth}}$ and for the correlation function $\xi_2 -
\xi_{2}(\epsilon=0)$, where $\xi_{2}(\epsilon=0)$ is computed using
the same parameters as the best-fit model but with $\epsilon=0$.  For
the $0.4 < z < 0.6$ redshift bin, $\epsilon$ is close to zero (the
significance is 0.3$\sigma$ for both the power spectrum and the
correlation function; see Table 3 in \citealt{BeutlerBAO16}), and thus
no clear feature is observed in the data or the model.  In the low and
high redshift bins, $\epsilon$ is marginally significant ($\sim
1\sigma$ for both) and of opposite signs.  Thus, the data points and
best-fit curves show weak features that are opposite in sign in the
two redshift bins.

The bottom panel of Fig. \ref{fig:pkxirecBAOfit} displays the BAO
ring(s) in the $0.4 < z < 0.6$ redshift bin, as reconstructed from the
monopole and quadrupole, thereby filtering the higher-order multipoles
that are treated as noise in our analysis. The results are decomposed
into the component of the separations transverse to and along the line
of sight, based on $x(p,\mu) = x_0(p)+L_2(\mu)x_2(p)$, where $x$
represents either $s^2$ multiplied by the correlation function or
$(P_{\ell}-P_{\ell,{\rm smooth}})/P_{0,{\rm smooth}}(k)$ for the power
spectrum, $p$ represents either the separation, $s$, or the Fourier
mode, $k$, $L_2$ is the 2nd order Legendre polynomial, $p_{||} = \mu
p$, and $p_{\perp} = \sqrt{p^2-\mu^2p^2}$. Plotted in this fashion,
the radius at which the BAO feature(s) represents the
spherically-averaged BAO measurement and the degree to which the
ring(s) is(are) circular represents the AP test as applied to BAO
measurements.

\subsection{Measuring the BAO scale}

Our companion papers \citet{Ross16}, \citet{VargasMagana16} and
\cite{BeutlerBAO16} use the BAO signal in the post-reconstruction
monopole and quadrupole, in configuration space and Fourier space, to
constrain the geometric parameter combinations $\DM(z)/\rd$ and $H(z)
\rd$.  We now present a brief summary of these analyses and refer the
reader to those papers for more details.

\cite{Ross16} and \cite{VargasMagana16} measure the anisotropic
redshift-space two-point correlation function.  Both methods rely on
templates for $\xi_0$ and $\xi_2$, which have BAO features that are
altered as function of the relative change in $\DM(z)$ and $H(z)$ away
from the values assumed in the fiducial templates (which are
constructed using the fiducial cosmology). These templates are allowed
to vary in amplitude and are combined with third-order polynomials,
for both $\xi_0$ and $\xi_2$, that marginalize over any shape
information. This methodology follows that of 
\cite{Xu13,Anderson2014a} and \cite{Anderson2014b}. Small differences between 
\cite{Ross16} and \cite{VargasMagana16} exist in the modelling of the fiducial templates and the
choices for associated nuisance parameters. The choices in
\cite{Ross16} are motivated by the discussion in \cite{Seo15} and \cite{Ross152D}, and
they carry out detailed investigations to show that observational
systematics have minimal impact on the BAO measurement.
\cite{VargasMagana16} use as their fiducial methodology the templates and
choices used in previous works
\citep{Cuesta16,Anderson2014a,Anderson2014b} enabling direct
comparison of the results with those previous papers. In addition, \cite{VargasMagana16} perform
a detailed investigation of possible sources of theoretical
systematics in anisotropic BAO measurements in configuration space,
examining the various steps of the analysis and studying the potential
systematics associated with each step. This work extends the previous
effort in \cite{VargasMaganaEtAl14}, which focused on systematic
uncertainties associated with fitting methodology, to more general
aspects such as the estimators, covariance matrices, and use of higher
order multipoles in the analysis.

\cite{BeutlerBAO16} extract the BAO information from the power
spectrum. The analysis uses power spectrum bins of $\Delta
k=0.01\khmpc$ and makes use of scales up to $k_{\rm max} =
0.3\khmpc$. The covariance matrix used in this analysis has been
derived from the MD-Patchy mocks described in
Section~\ref{sec:mocks}. The reduced $\chi^2$ for all redshift bins is
close to 1.

The two-dimensional 68 and 95 per cent confidence levels (CL) on
$\DM(z)/r_{\rm d}$ and $H(z)r_{\rm d}$ recovered from these fits are
shown in Fig.~\ref{fig:2d_bao}, where we have scaled our measurements
by the sound horizon scale in our fiducial cosmology, $\rdfid =
147.78\,{\rm Mpc}$, to express them in the usual units of Mpc and
${\rm km}\,{\rm s}^{-1}{\rm Mpc}^{-1}$.  The corresponding
one-dimensional constraints are summarised in
Table~\ref{tab:combined_BAO}.  The results inferred from the three
methods are in excellent agreement.  As expected, given the small
differences in methodology and data, the results of
\cite{Ross16} and \cite{VargasMagana16} are very similar.  Tests on
the results obtained from mock samples show that the results are
correlated to a degree that combining them affords no improvement in
the statistical uncertainty of the measurements.  Differences between
the results are at most 0.5$\sigma$ and are typically considerably
smaller; these differences are consistent with expectations (further details of tests on mock samples are presented in
Section~\ref{sec:mockcomp}).  Thus, for simplicity, we select only the
measurements and likelihoods presented in \cite{Ross16} to combine
with the power-spectrum BAO results and full-shape measurements. The
consensus values are computed and discussed in
Section~\ref{sec:combining_measurements}.

\section{Pre-reconstruction full-shape measurements}\label{sec:full_shape}

\begin{figure*}
\begin{minipage}{7in}
\includegraphics[width=0.33\textwidth]{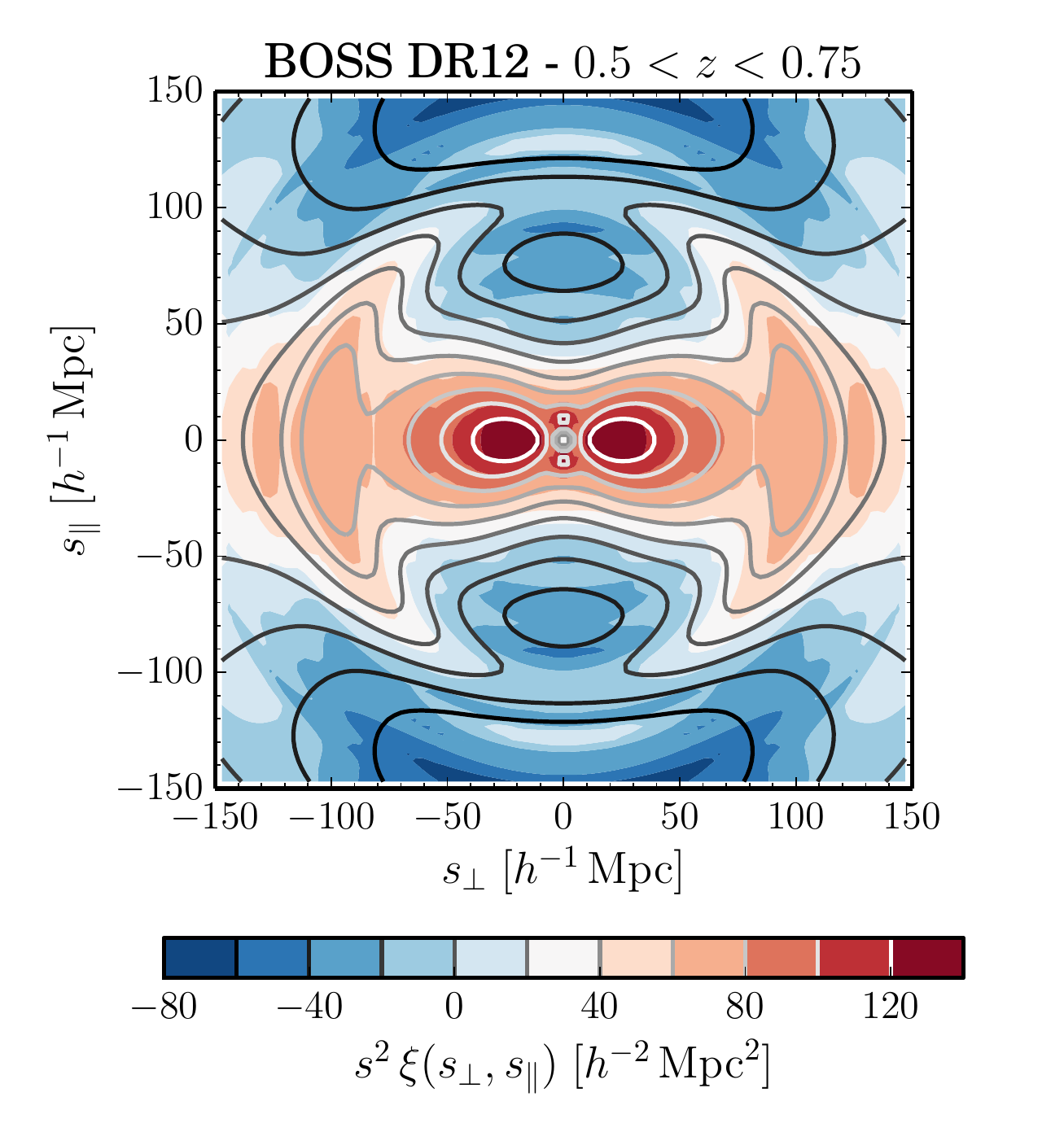}
\includegraphics[width=0.33\textwidth]{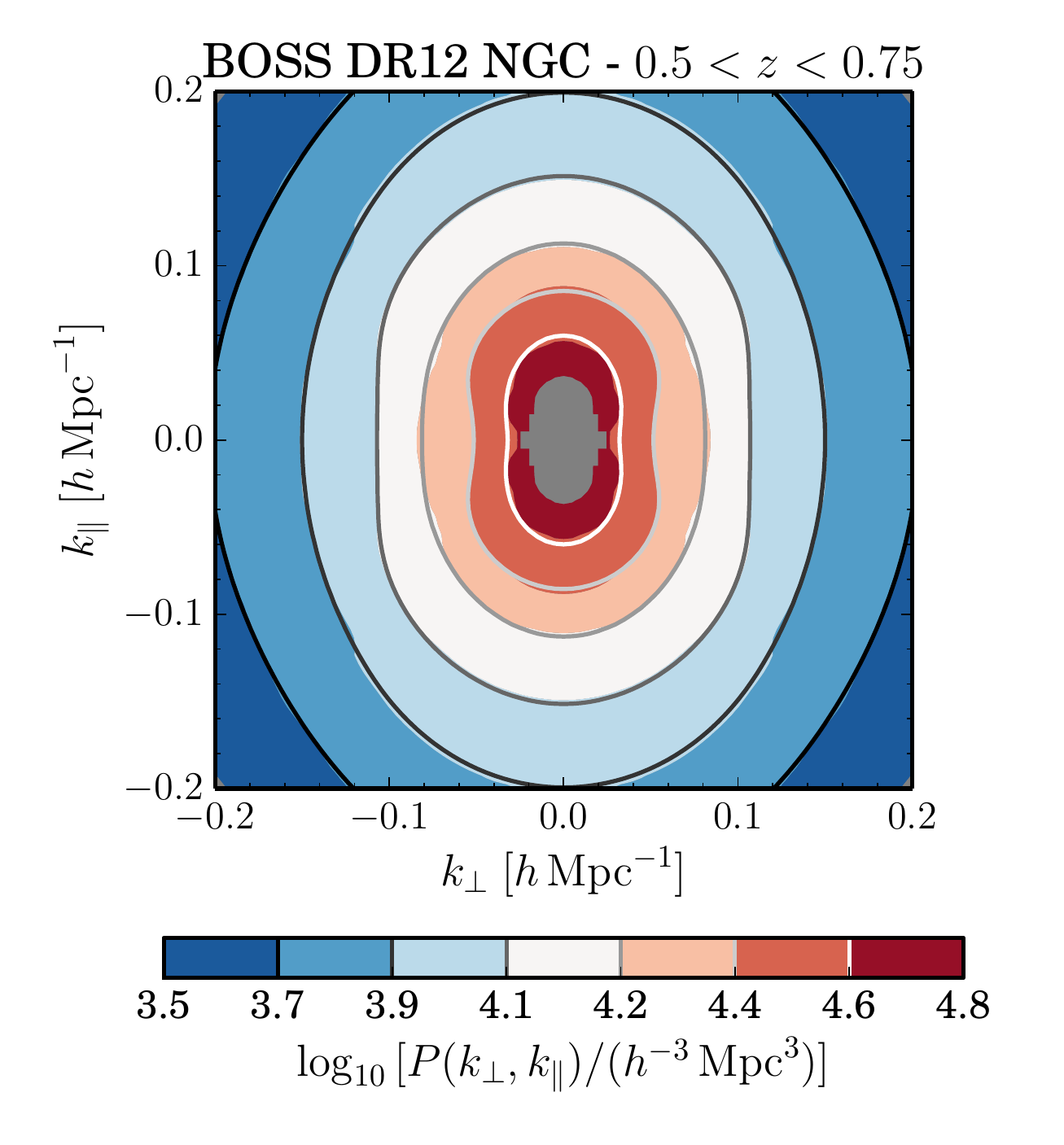}
\includegraphics[width=0.33\textwidth]{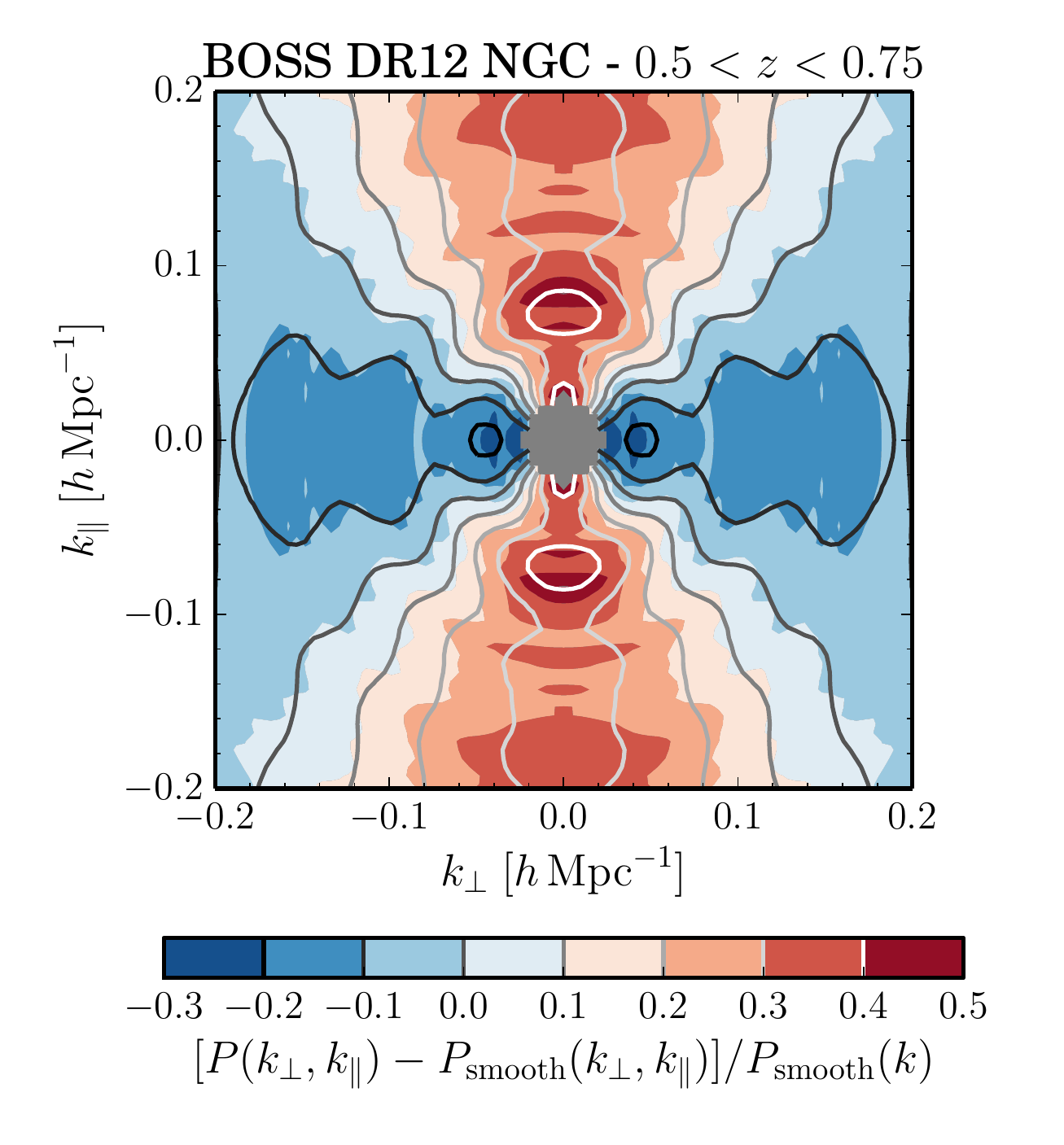}
  \caption{The measured pre-reconstruction correlation function (left) and power spectrum (middle) in the directions  
  perpendicular and parallel to the line of sight, 
  shown for the NGC only in the redshift range $0.50<z<0.75$. 
  In each panel, the color
  scale shows the data and the contours show the prediction of the
  best-fit model. 
  The anisotropy of the contours seen in both plots reflects a combination of RSD and the AP effect, and holds most of the information used to separately constrain 
  $\DM(z)/\rd$, $H(z)\rd$, and $f\sigma_8$. 
  The BAO ring can be seen in two dimensions on the 
  correlation function plot. To more clearly show the anisotropic BAO ring in the power spectrum, the right panel plots
  the two-dimensional power-spectrum divided by the best-fit smooth component.
  The wiggles seen in this panel are analogous to the oscillations seen in 
  the top left panel of Fig~\ref{fig:pkxirecBAOfit}.
  }
  \label{fig:2DPkxi}
  \end{minipage}
\end{figure*}

\begin{table*}
\centering
\caption{Summary table of pre-reconstruction full-shape constraints 
on the parameter combinations 
$\DM\times\left(\rdfid/\rd\right)$,
$H\times \left(\rd/\rdfid\right)$, and $f\sigma_8(z)$ 
derived in the supporting papers for each of our three overlapping 
redshift bins}
\begin{tabular}{cccccc}
\hline\hline
 Measurement & redshift &  Satpathy et al. & Beutler et al. (b) &  Grieb et al.  & S{\'a}nchez et al. \\
                       &                &  $\xi(s) $ multipoles & $P(k)$  multipoles & $P(k)$ wedges &$\xi(s) $ wedges  \\ \hline
$\DM\times\left(\rdfid/\rd\right)$ [Mpc] & $z=0.38 $ &$1476 \pm 33$  & $1549 \pm 41$ & $1525 \pm 25$ & $1501 \pm 27$ \\ 
$\DM\times\left(\rdfid/\rd\right)$ [Mpc] & $z=0.51 $ & $1985 \pm 41$ & $2015 \pm 53$ & $1990 \pm 32$ & $2010 \pm 30$ \\
$\DM\times\left(\rdfid/\rd\right)$ [Mpc] & $z=0.61 $ &$2287 \pm 54$  & $2270 \pm 57$ & $2281 \pm 43$ & $2286 \pm 37$ \\
$H\times\left(\rd/\rdfid\right)\,[{\rm km}\,{\rm s}^{-1}{\rm Mpc}^{-1}]$ & $z=0.38 $ & $79.3 \pm 3.3$ &$82.5 \pm 3.2$ & $81.2 \pm 2.3$ & $82.5 \pm 2.4$ \\
$H\times\left(\rd/\rdfid\right)\,[{\rm km}\,{\rm s}^{-1}{\rm Mpc}^{-1}]$ & $z=0.51 $ & $88.3 \pm 4.1$ &$88.4 \pm 4.1$ & $87.0 \pm 2.4$ & $90.2 \pm 2.5$ \\
$H\times\left(\rd/\rdfid\right)\,[{\rm km}\,{\rm s}^{-1}{\rm Mpc}^{-1}]$ & $z=0.61 $ & $99.5 \pm 4.4$ &$97.0 \pm 4.0$ & $94.9 \pm 2.5$ & $97.3 \pm 2.7$ \\
$f\sigma_8$    & $z=0.38 $ &$0.430 \pm 0.054$   &$0.479 \pm 0.054$ & $0.498 \pm 0.045$ & $0.468 \pm 0.053$ \\
$f\sigma_8$    & $z=0.51 $ & $0.452 \pm 0.058$  &$0.454 \pm 0.051$  &$0.448 \pm 0.038$ & $0.470 \pm 0.042$ \\
$f\sigma_8$    & $z=0.61 $ & $0.456 \pm 0.052$  &$0.409 \pm 0.044$ & $0.409 \pm 0.041$ & $0.440 \pm 0.039$ \\\hline\hline
\end{tabular}
\label{tab:all_RSD}
\end{table*}

\begin{figure*}
    \centering 
    \includegraphics[width=0.92\textwidth]{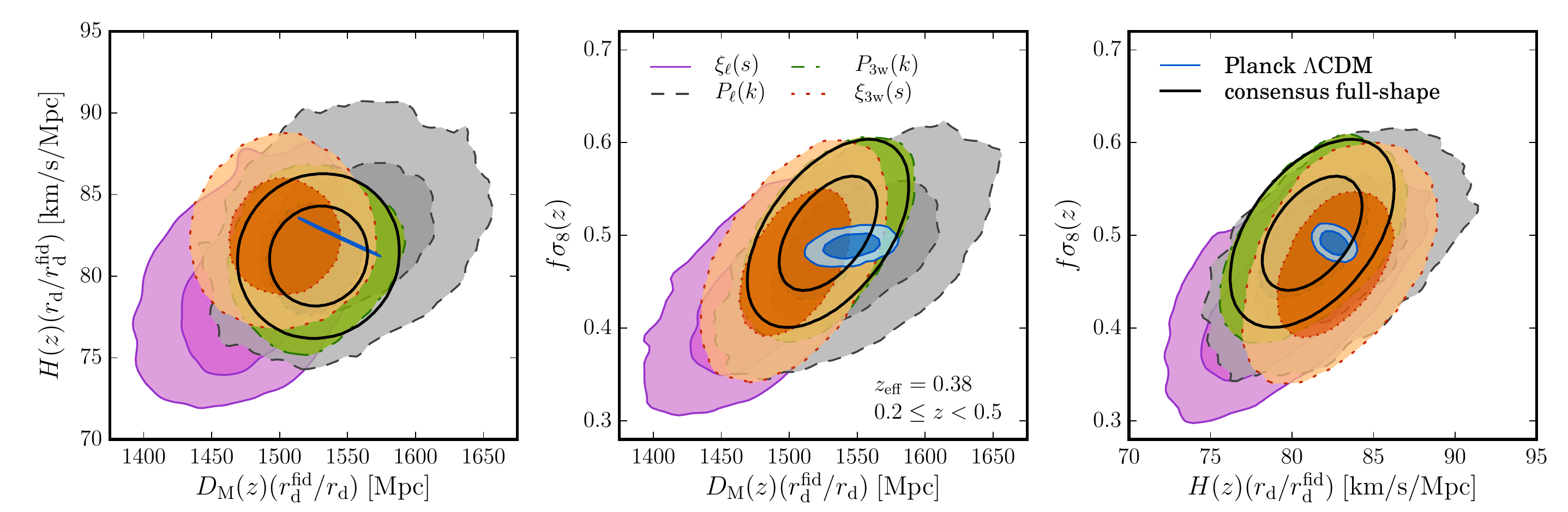}
    \includegraphics[width=0.92\textwidth]{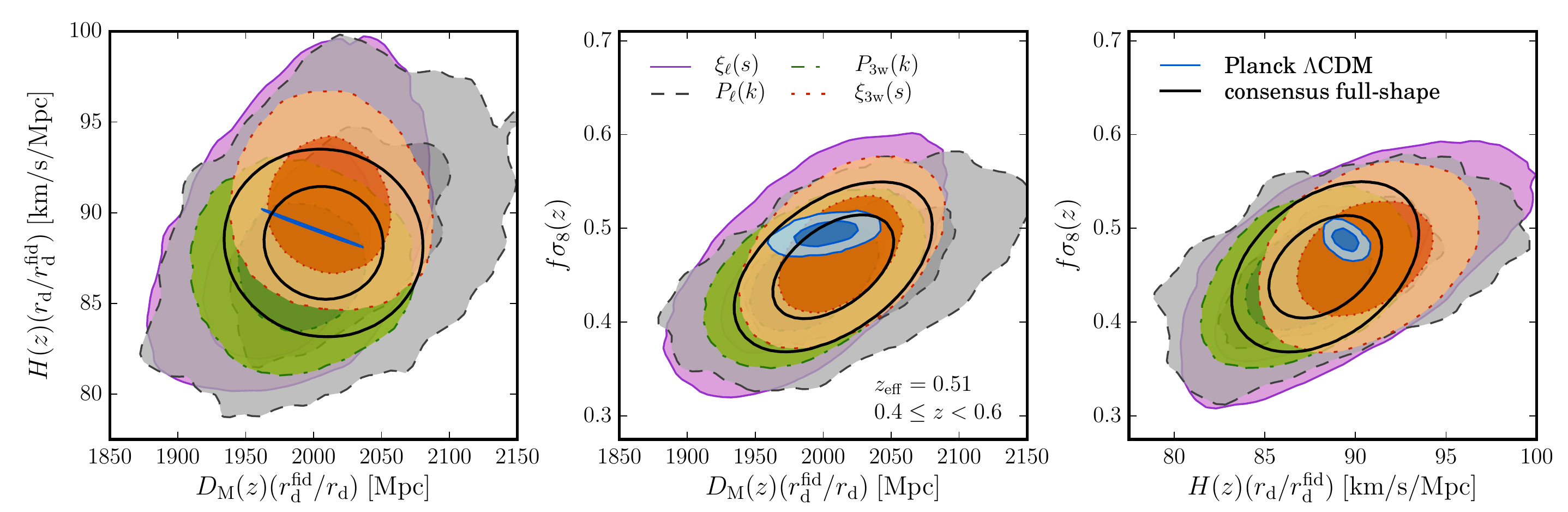}
    \includegraphics[width=0.92\textwidth]{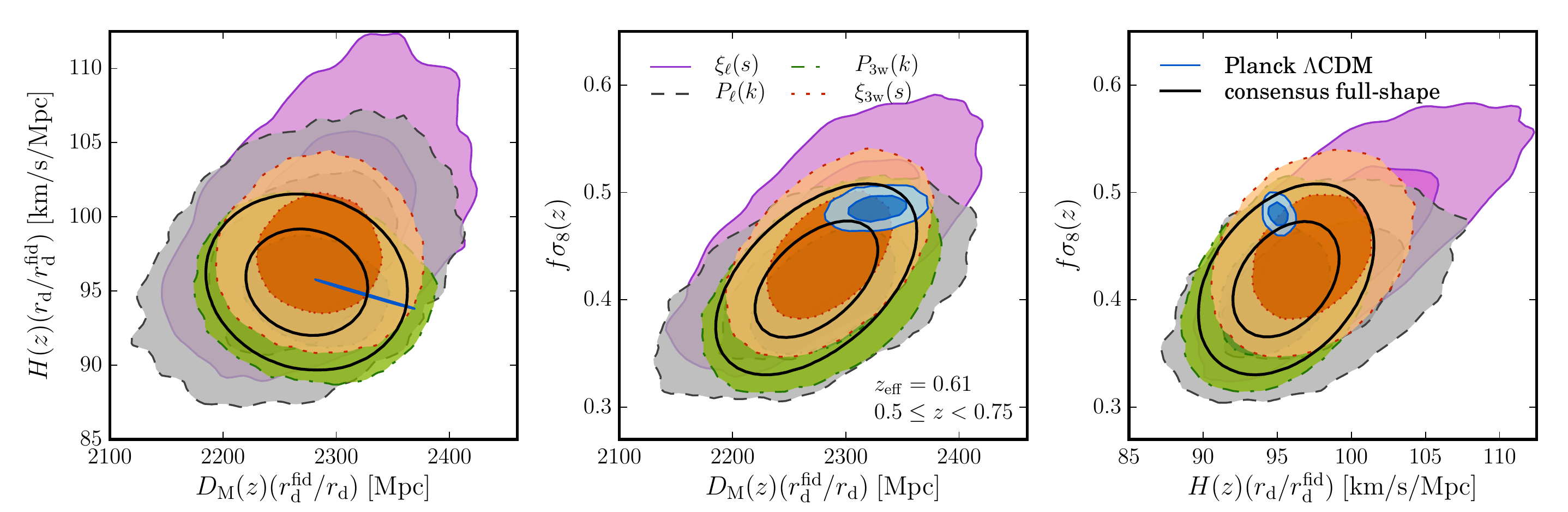}
    \caption{
    Two-dimensional 68 and 95 per cent marginalized constraints 
	on $\DM(z)\times \left(\rdfid/\rd\right)$,
    $H(z) \times \left(\rd/\rdfid\right)$,
	and $f(z)\sigma_8(z)$ 
    obtained from our pre-reconstruction full-shape fits to multipoles and wedges in configuration and Fourier space.  
    The solid lines represent the combination of these results into a set of consensus constraints, as described in
    Section~\ref{sec:consensus_results}. The blue solid lines correspond to the constraints inferred from the Planck CMB
    measurements under the assumption of a $\Lambda$CDM model. 
	Top, middle, and bottom rows show our low, intermediate, and 
	high redshift bins.
	Comparison to and combination with post-reconstruction BAO measurements
	appears in Fig.~\ref{fig:combined_contours} below.
    }
  \label{fig:2d_rsd}
\end{figure*}

Fig.~\ref{fig:2DPkxi} 
shows the two-dimensional correlation function, $\xi(s_{\perp},s_{\parallel})$ (left panel), and power spectrum,
$P(k_{\perp},k_{\parallel})$ (middle and right panels), of the NGC BOSS combined galaxy sample, in the redshift range 
$0.5 < z < 0.75$. The figures for other redshifts and the SGC would look similar. 
The full shape of these measurements encodes additional information beyond that of the BAO feature. 
If we had access to the real-space positions of the galaxies and in the absence of AP distortions, the contours 
of these functions would correspond to perfect circles. However, the RSD caused by the peculiar velocities of the 
galaxies distort these contours, compressing (stretching) them along the line-of-sight direction in 
configuration (Fourier) space. These anisotropies encode information on the growth rate of cosmic structures, 
which can be used to constrain the parameter combination $f\sigma_8(z)$, where $f \equiv d\ln D /d\ln a $.

On large scales most of the information contained in $\xi(s_{\perp},s_{\parallel})$ and $P(k_{\perp},k_{\parallel})$
can be compressed into a small number of one-dimensional projections such as the their first few Legendre 
multipoles \citep[e.g.][]{PadWhite08}, or the clustering wedges statistic \citep{Kazin2012}.
Each of the four supporting papers 
\citep{Satpathy16, Sanchez16, BeutlerRSD16,Grieb16}
uses the information of either multipoles or wedges in $\mu$, in configuration or Fourier Space, 
employing different approaches to model
the clustering statistics in the non-linear regime. 
The four methods were tested in high-fidelity mocks, via a blind challenge that we describe in
Section~\ref{sec:mock_challenge} and that will later inform our systematic error budget. 
These measurements simultaneously capture the impact of the expansion rate, AP-effect and growth rate on 
the distribution of galaxies, allowing us to determine the parameter combinations $D_{\rm M}(z)/r_{\rm d}$, 
$H(z)r_{\rm d}$ (or some combination thereof) and $f\sigma_8(z)$.
Here we give a brief description of these analyses and refer the reader to those papers for more details 
on the measurements, modelling, fitting procedures and tests with mocks, as well as figures showing each of
the measurements individually.

\citet{Satpathy16} analyses the monopole and quadrupole of the two-point correlation function. The covariance matrix 
of these measurements is estimated using 997 MD-Patchy mock catalogues. The multipoles are modelled using Convolution Lagrangian 
Perturbation Theory (CLPT) and the Gaussian Streaming model (GSM) \citep{Carlson13,Wang14}. This model has been tested for both dark matter 
and biased tracers using N-body simulations \citep{Wang14} and has been tested for various observational 
and theoretical systematic errors \citep{AlamRSDDR112015}. \citet{Satpathy16} fit scales between 25 and 
$150\,h^{-1}{\rm Mpc}$ with bin width of $5\,h^{-1}{\rm Mpc}$ and extract the cosmological and growth parameters
with a Markov Chain Monte Carlo (MCMC) algorithm using \textsc{CosmoMC} \citep{cosmomc}.

\cite{Sanchez16} extract cosmological information from the full shape of three clustering wedges in 
configuration space, defined by dividing the $\mu$ range from 0 to 1 into three equal-width intervals,
whose covariance matrix was obtained from a set of 2045 MD-Patchy mock catalogues.
This analysis is based on a new description of the effects of the non-linear evolution of density
fluctuations (gRPT, Blas et al. in prep.), bias and RSD that is applied to the BOSS measurements for scales
$s$ between 20 and $160\,h^{-1}{\rm Mpc}$ with a bin width of $5\,h^{-1}{\rm Mpc}$. 
\cite{Sanchez16} perform extensive tests of this model using the large-volume Minerva N-body simulations
\citep{Grieb15} to show that it can extract cosmological information from three clustering wedges 
without introducing any significant systematic errors.

\citet{BeutlerRSD16} analyse the anisotropic power spectrum using the estimator suggested in 
\citet{Bianchi2015} and \citet{Scoccimarro2015}, which employs Fast Fourier Transforms to measure all 
relevant higher order multipoles. The analysis uses power spectrum bins of $\Delta k=0.01\khmpc$ and makes use of 
scales up to $k_{\rm max} = 0.15h\,\mathrm{Mpc}^{-1}$ for the monopole and quadrupole and 
$k_{\rm max} = 0.1h\,\mathrm{Mpc}^{-1}$ for the hexadecapole. These measurements are then compared to a model 
based on renormalized perturbation theory \citep{TaruyaNishimichiSaito2010}. This model has been extensively 
tested with N-body simulations in configuration space
\citep[e.g.][]{delaTorreGuzzo2012} and Fourier space 
\citep[e.g.][]{Beutler12}. The covariance matrix used in this analysis has been derived from 
$2048$ Multidark-Patchy mock catalogues (The NGC uses only 2045 mock catalogues) and the reduced 
$\chi^2$ for all redshift bins is close to 1. 

The methodology in \citet{Grieb16} extends the application of the clustering wedges statistic to Fourier space.
In order to make use of new estimators based on fast Fourier transforms
\citep[FFT;][]{Bianchi2015,Scoccimarro2015}, their analaysis uses the power
spectrum clustering wedges, filtering out the information of Legendre multipoles $\ell>4$.
This information is combined to three power spectrum wedges, measured in wavenumber bins of
$\Delta k = 0.005 \, h\,\mathrm{Mpc}^{-1}$, up to the mildly non-linear regime, $k < 0.2 \, h\,\mathrm{Mpc}^{-1}$.
The full shape of these measurements is fitted with theoretical predictions based on the same underlying model
of non-linearities, bias and RSD as in \citet{Sanchez16}. Thus, these two complementary analyses represent
the first time that the same model is applied in configuration and Fourier space fits. The methodology has been
validated using the Minerva simulations and mock catalogues and found to give unbiased cosmological constraints.
Besides the covariance matrix, which is derived from 2045 MD-Patchy mock catalogues, this analysis depends on a
framework for the wedge window function, which was developed based on the recipe for the power spectrum multipoles
of \citet{Beutler14}. The power spectrum wedges of the NGC and SGC sub-samples in the low-redshift bin are
modelled with two different bias, RSD, and shot noise parameters, while the intermediate and high redshift bins
are fitted with the same nuisance parameters for the two sub-samples.

The constraints on $\DM(z)/\rd$, $H(z)\rd$, and $f(z)\sigma_8(z)$
produced by each of the four individual methods are 
shown in Fig.~\ref{fig:2d_rsd} where, as before, we 
have rescaled our measurements by the sound horizon scale 
in our fiducial cosmology. 
The corresponding one-dimensional constraints are summarised in 
Table~\ref{tab:all_RSD}.  
The agreement between the results inferred from the different 
clustering statistics and analysis methodologies is good, 
and the scatter between methods is consistent with what we observe 
in mocks (see Section~\ref{sec:mock_challenge} 
and Fig.~\ref{fig:fs8_nseries_diff}).
In all cases the $\mu$-wedges analyses give significantly
tighter constraints than the multipole analyses, in both
configuration space and Fourier space. The consensus constraints,
described in \S\ref{sec:consensus_results} below, are slightly
tighter than those of the individual wedge analyses.
At all three redshifts and for all three quantities,
mapping distance, expansion rate, and the growth of structure,
the 68\% confidence contour for the consensus results overlaps
the 68\% confidence contour derived from Planck 2015 data
assuming a $\Lambda$CDM cosmology.
We illustrate the combination of these full shape results
with the post-reconstruction BAO results in 
Fig.~\ref{fig:combined_contours} below.

\section{Systematic Error Estimation}
\label{sec:errors}
\subsection{Tests using MD-Patchy mock catalogues}
\label{sec:mockcomp}

Here we examine the consistency and correlation among the three
methods applied to obtain post-reconstruction BAO measurements
(described in Section \ref{sec:measuring_BAO}) and the four methods
applied to obtain pre-reconstruction full-shape constraints (described
in Section \ref{sec:full_shape}). We do these comparisons using the
results obtained from the fits to the MD-Patchy mock catalogues, which
enabled at least 996 comparisons in all cases. These comparisons are
used to inform the final systematic uncertainties, to be described in
Section \ref{sec:sys_err}, that we apply to our measurements.

Table \ref{tab:Pkxicompmocks} compares the mean BAO results for the
post-reconstruction correlation function and the power spectrum
measurements and then the mean result when the individual results are
combined as described in Section \ref{sec:consensus_method}. The
standard deviations are improved by the combination, as they are not
perfectly correlated. The correlations between the correlation
function and power spectrum results range between 0.88 and 0.90. The
correlations between the two correlation function results are such
that the optimal combination does not affect the standard deviation at
the quoted precision. The results differ by an average of 0.002 for
both $\alpha$ and for $\epsilon$, with the $P(k)$ results
having greater $\epsilon$ and lesser $\alpha$. These differences are
smaller than the systematic modelling uncertainties will we adopt for each parameter
(0.003 and 0.005), described in Section \ref{sec:sys_err}.

The detailed tests presented in \cite{BeutlerBAO16} and \cite{VargasMagana16}
suggest no reason to believe any of the results should be biased
relative to the others and they can be thus combined to produce the
consensus results. The consensus results are obtained as described in
\cite{SanchezStat16}. The $P(k)$ results are slightly more precise and the
consensus results are thus weighted towards these results. The consensus $\alpha$
results are biased by at most 0.001 and the mean bias is 0.000. The consensus
$\epsilon$ results are each biased by 0.002; this is 0.15$\sigma$ and is substantially
smaller than the systematic uncertainty we adopt for $\epsilon$.

\begin{figure*}
\begin{minipage}{6in}
\includegraphics[width=6in]{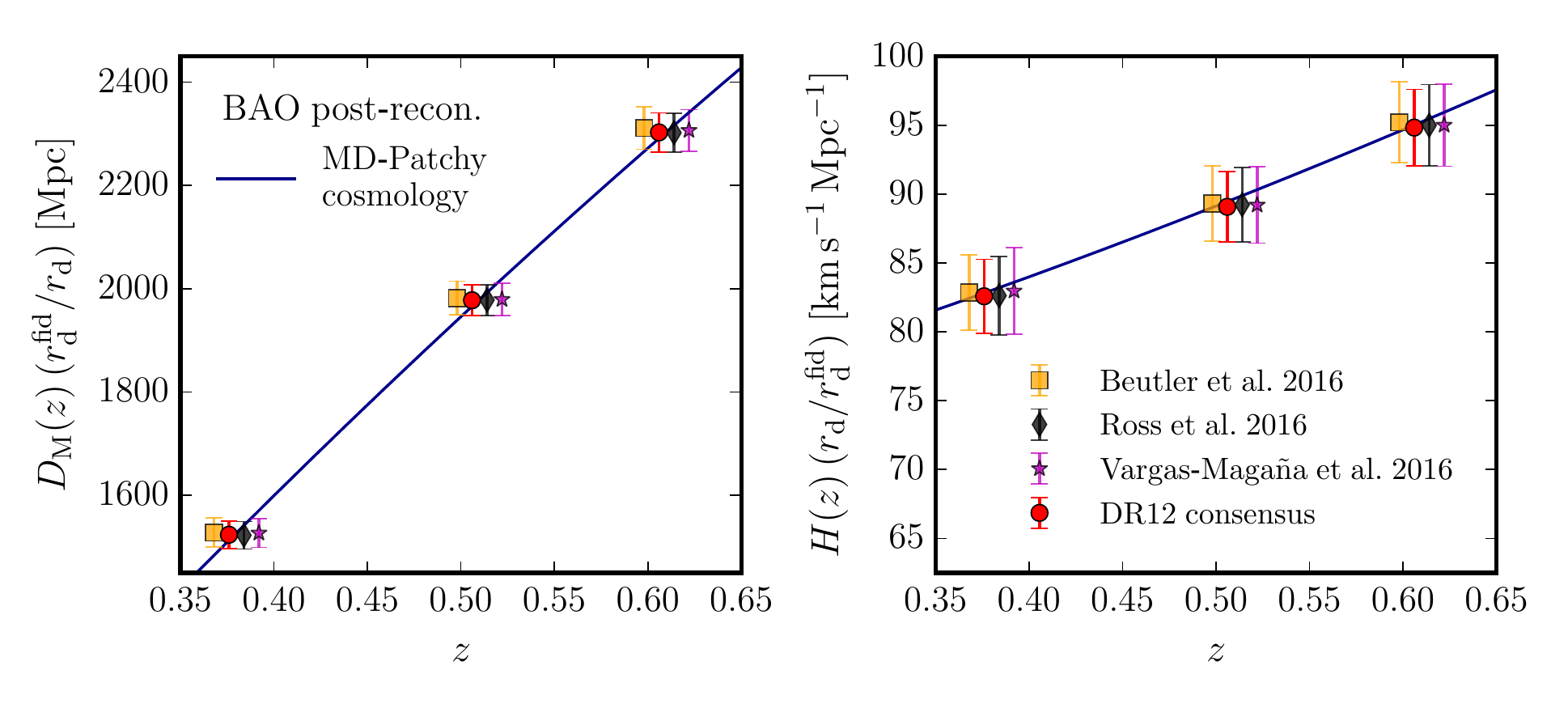}
  \caption{The mean distance and Hubble factor recovered from post-reconstruction MD-Patchy mocks (points) 
  compared to the expected results for the input cosmology of 
  MD-Patchy (curves).  Error bars on the points show the 
  mock-to-mock dispersion (the standard deviation of the mean
  would be smaller by $\approx 1000^{1/2} \approx 30$).
  Red circles display the results from the combination of correlation
  function and power spectrum measurements and are plotted at the
  measurement redshift; other points are offset in redshift for
  visual clarity.
  }
  \label{fig:Patchyfits_BAO}
  \end{minipage}
\end{figure*}

Fig. \ref{fig:Patchyfits_BAO} is an illustration of the results
presented in Table \ref{tab:Pkxicompmocks}, with the $\alpha$ and $\epsilon$ values
converted to $D_{\rm M}$ and $H$ values. Visually, it is clear that differences
between the results using each methodology are negligibly small and that
the consensus results match those expected for the MD-Patchy cosmology.

\begin{table}
\centering
\caption{Post-reconstruction combined sample 2D BAO fits for $\xi(s)$ and $P(k)$ in MD-Patchy mock samples. The $\Delta$ values are the mean with the expected value subtracted. $S$ denotes standard deviation. `R' denotes results from Ross et al. (2016) and `V' denotes results from Vargas-Maga{\~n}a et al. (2016). The $P(k)$ results are from Beutler et al. (2016b).}
\begin{tabular}{ccccc}
\hline
\hline
 sample & $\Delta\langle \alpha\rangle$ & $S_{\alpha}$ & $\Delta\langle\epsilon\rangle$  & $S_{\epsilon}$\\
\hline
{\bf $0.2 < z < 0.5$:}\\
consensus & --0.001 & 0.012 & 0.002 & 0.014\\
$\xi$ R & 0.000 & 0.013 & 0.001 & 0.015\\
$\xi$ V & --0.001 & 0.014 & 0.001 & 0.016\\
$P(k)$ & --0.001 & 0.013 & 0.003 & 0.015\\
\hline
{\bf $0.4 < z < 0.6$:}\\
consensus & 0.001 & 0.011 & 0.002 & 0.013\\
$\xi$ R & 0.001 & 0.012 & 0.001 & 0.014\\
$\xi$ V & 0.000 & 0.012 & 0.002 & 0.013\\
$P(k)$ & 0.000 & 0.012 & 0.002 & 0.014\\
\hline
{\bf $0.5 < z < 0.75$:}\\
consensus & 0.000 & 0.011 & 0.002 & 0.013\\
$\xi$ R & 0.002 & 0.012 & --0.001 & 0.015\\
$\xi$ V & 0.000 & 0.012 & 0.001 & 0.014\\
$P(k)$ & --0.001 & 0.012 & 0.001 & 0.015\\
\hline
\label{tab:Pkxicompmocks}
\end{tabular}
\end{table}

Table \ref{tab:RSDcompmocks} presents results 
for the four methods that apply pre-reconstruction
full-shape fits to the DR12 BOSS galaxy data,
including results for $f\sigma_8$. Their combination, obtained as
described in Section \ref{sec:consensus_method}, produces the
`combined' results, and this combination reduces the standard
deviations of the recovered results, taking advantage of 
partial complementarity (i.e., not complete correlations)
among the methods. See \cite{SanchezStat16} for further
details.

\begin{figure*}
\begin{minipage}{7in}
\includegraphics[width=7in]{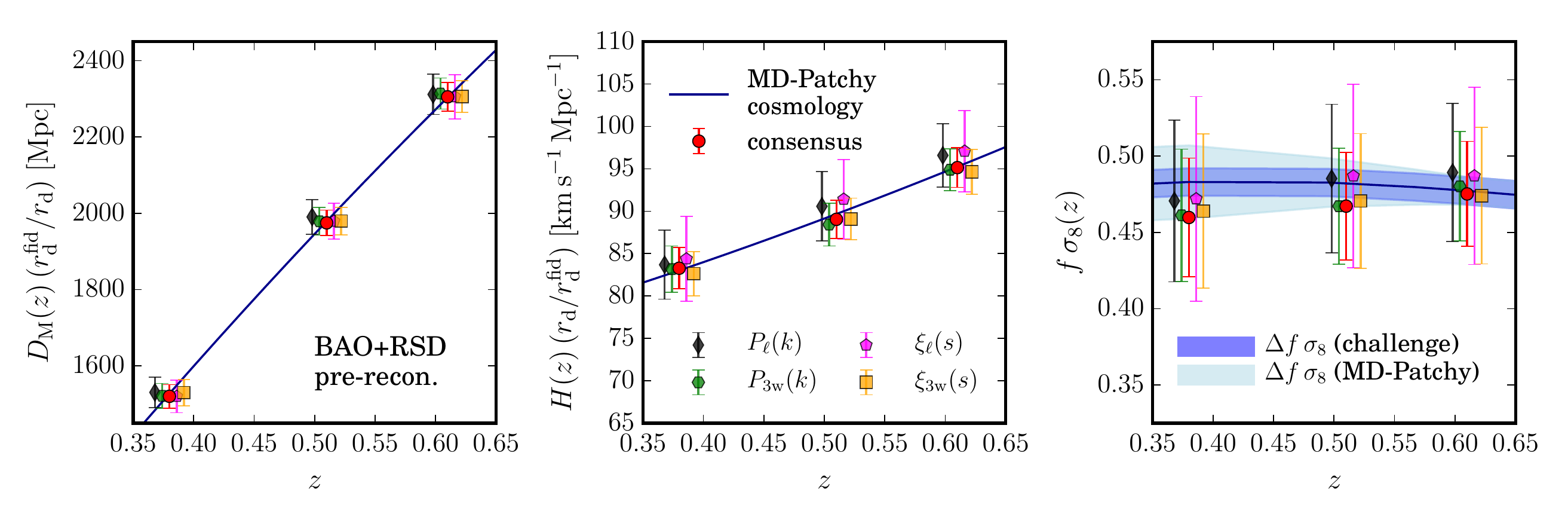}
\caption{The mean distance, Hubble parameter, and $f\sigma_8$
  recovered from pre-reconstruction MD-Patchy mocks (points) compared
  to the expected results for the input cosmology of MD-Patchy
  (curves).  Error bars show the mock-to-mock dispersion.  Red circles
  display the results recovered by combining the four methods and are
  plotted at the correct redshift.  Other points are offset in
  redshift for visual clarity.  The two blue bands indicate the systematic
  error on $f\sigma_8$ estimated from as the maximum of the biases
  found from the MD-Patchy mocks and the high resolution boxes of
  section~\ref{sec:mock_challenge}.
  }
  \label{fig:Patchyfits}
  \end{minipage}
\end{figure*}

The biases in the combined $\alpha$ and $\epsilon$ values 
from these pre-reconstruction analyses
are below $0.3\sigma$ compared to the mock-to-mock dispersion,
and they are smaller than the systematic uncertainty defined in \S\ref{sec:sys_err}.
The biases on the recovered $f\sigma_8$,
0.024, 0.016, and 0.003 in the three redshift bins,
are up to $0.6\sigma$ compared to the mock-to-mock dispersion.
Fig. \ref{fig:Patchyfits} compares the results of our RSD
fitting methods to the natural cosmology of the MD-Patchy mocks.  When
setting the systematic error for $f\sigma_8$, we compare the bias inferred from
the MD-Patchy mocks to the systematic error obtained using high
resolution N-body simulations described in the next subsection.
We use the maximum of these two numbers in each redshift bin as the 
systematic error.

\begin{table*}
\centering
\caption{Pre-reconstruction combined sample full-shape fits in MD-Patchy mock samples. The $\Delta$ values are the mean with the expected value subtracted. $S$ denotes standard deviation. }
\begin{tabular}{ccccccc}
\hline
\hline
  sample & $\Delta\langle \alpha\rangle$ & $S_{\alpha}$ & $\Delta\langle\epsilon\rangle$  & $S_{\epsilon}$ & $\Delta f \sigma_8$ & $S_{f\sigma_8}$ \\
 \hline
{\bf $0.2 < z < 0.5$:} \\
 consensus   &   --0.003 &     0.018 &     0.000 &     0.011 &    --0.024 &     0.038 \\
$\xi_{3w}$   &     0.004 &     0.020 &     0.001 &     0.012 &    --0.019 &     0.050 \\
  $P_\ell$   &   --0.001 &     0.021 &   --0.004 &     0.020 &    --0.012 &     0.053 \\
  $P_{3w}$   &   --0.002 &     0.019 &     0.000 &     0.013 &    --0.022 &     0.043 \\
$\xi_\ell$   &   --0.008 &     0.020 &   --0.004 &     0.025 &    --0.011 &     0.067 \\
\hline
{\bf $0.4 < z < 0.6$:} \\
 consensus   &     0.002 &     0.015 &     0.003 &     0.009 &   --0.016 &     0.035 \\
$\xi_{3w}$   &     0.003 &     0.017 &     0.002 &     0.010 &   --0.012 &     0.044 \\
  $P_\ell$   &     0.001 &     0.019 &   --0.005 &     0.018 &     0.002 &     0.049 \\
  $P_{3w}$   &     0.005 &     0.017 &     0.005 &     0.010 &   --0.016 &     0.038 \\
$\xi_\ell$   &   --0.006 &     0.017 &   --0.007 &     0.022 &     0.004 &     0.060 \\
\hline
{\bf $0.5 < z < 0.75$:} \\
 consensus   &     0.002 &     0.015 &     0.001 &     0.009 &   --0.003 &     0.034 \\
$\xi_{3w}$   &     0.004 &     0.016 &     0.002 &     0.011 &   --0.004 &     0.045 \\
  $P_\ell$   &   --0.001 &     0.017 &   --0.005 &     0.017 &     0.011 &     0.045 \\
  $P_{3w}$   &     0.006 &     0.016 &     0.000 &     0.009 &     0.002 &     0.036 \\
$\xi_\ell$   &   --0.005 &     0.016 &   --0.006 &     0.022 &     0.009 &     0.058 \\
\hline
\label{tab:RSDcompmocks}
\end{tabular}
\end{table*}

\subsection{RSD Tests using High Resolution Mocks}
\label{sec:mock_challenge}

In addition to the large suite of covariance mocks, we also utilize a
small series of high-fidelity mocks to test the accuracy and precision
of our multiple RSD methods. Details of this ``mock challenge'' are
presented in \cite{tinker_etal:2016}. There are two complementary sets
of mocks. The first is a homogeneous set in which all mocks have the
same underlying galaxy bias model built upon on the same cosmology,
but each mock is an independent realization. These mocks have the same
angular and radial selection function as the NGC DR12 CMASS
sample. There are 84 mocks in total.  The N-body simulations from
which these cut-sky mocks were created use the high-resolution code
GADGET2 \citep{springel:2005}, using input parameters to ensure
sufficient mass and spatial resolution to resolve the halos that BOSS
galaxies occupy.  The second is a heterogeneous set in which different
galaxy bias models are built upon the same simulation. Thus these
mocks have not just the same underlying cosmology, but also the same
large scale structure. But the galaxy bias varies at the $\pm 5$ per
cent level. These mocks are built on periodic cubes of $\sim 2.5
h^{-1}\,{\rm Gpc}$ per side---roughly 4 times the volume of the DR12
CMASS sample. We use three different bias models in the second set of
simulations. The second set of mocks are based on the Big MultiDark
simulation (\citealt{riebe_etal:13}).

The first set of high resolution mocks tests quantifies the accuracy
of the methods, including all aspects of the cut-sky analysis, while
the second tests for possible theoretical systematics associated with
the complexities of galaxy bias. Although the second set of mocks does
not span the full possible range of galaxy bias models, they provide
confidence that the methods are accurately recovering $f\sigma_8$ for
the conventional space of cosmologies and galaxy evolution models.  We
use the quadrature sum of the errors on $f\sigma_8$ from these two sets
of mocks as an estimate of the systematic error from the
high-resolution mocks.  As we will show, in some cases the error from
the high-resolution mocks is smaller than the error obtained from the
MD-Patchy mocks described in the previous section (the `$\Delta
f\sigma_8$' column in Table \ref{tab:RSDcompmocks}). To be
conservative, we adopt the larger of the errors obtained from the high
resolution mocks and the MD-Patchy mocks as our final systematic error
for a given redshift bin.

\begin{figure*}
%\begin{minipage}{7in}
\includegraphics[width=0.8\textwidth]{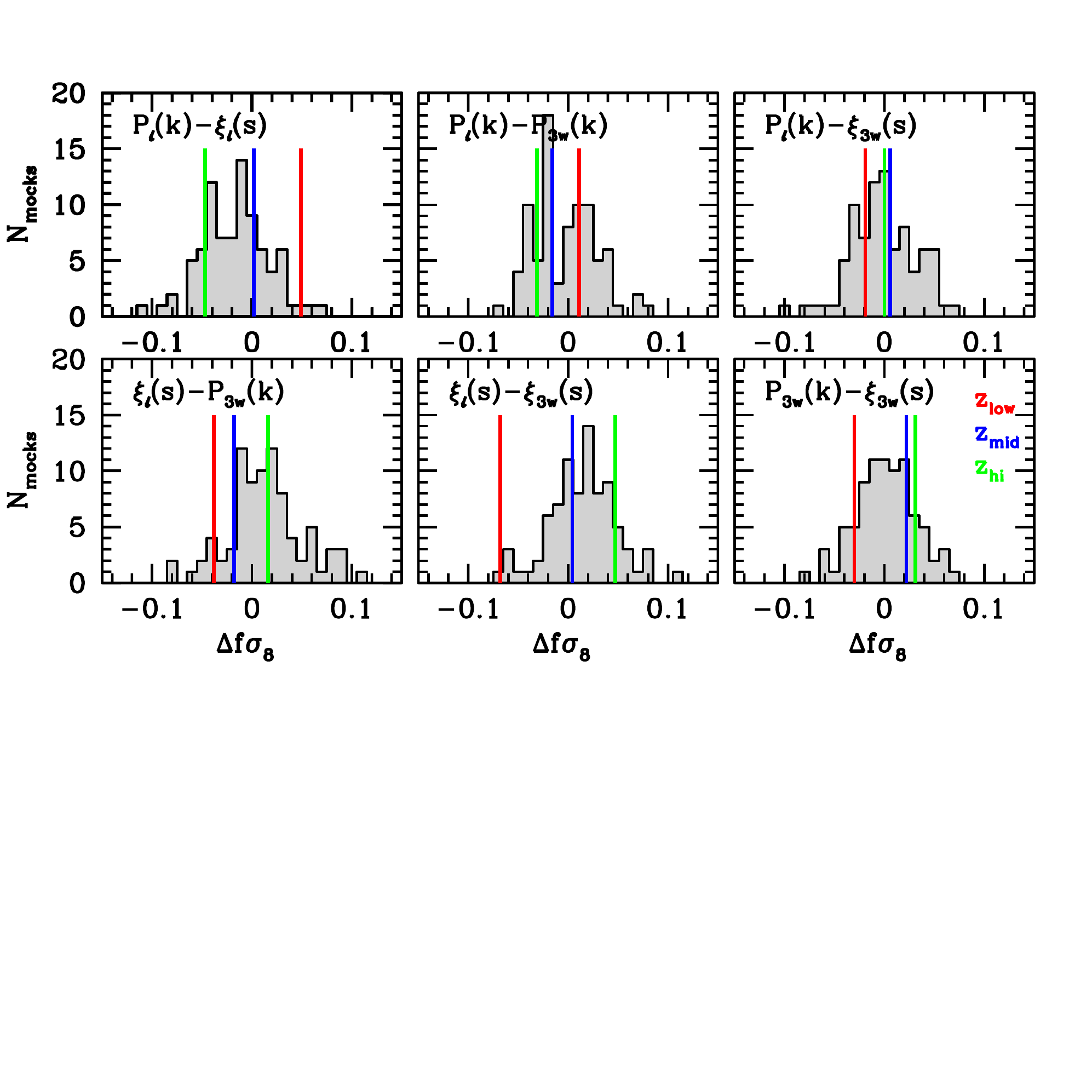}
\vspace{-5.5cm}
\caption{ The gray shaded histograms show the distribution of
  differences in $f\sigma_8$ between pairs of RSD methods (four
  methods, hence six pairs) applied to the 84 high-resolution cut-sky
  mocks described in \S \ref{sec:mock_challenge}. The cut-sky mocks
  are at $z=0.5$. The vertical coloured lines indicate the differences
  in $f\sigma_8$ when each pair of methods is applied to the DR12
  combined sample, as listed in Table \ref{tab:all_RSD}. Different
  coloured lines indicate different redshift bins. The pairwise
  differences found for the data are typical of those found in
  application to the cut-sky mocks.}
  \label{fig:fs8_nseries_diff}
%  \end{minipage}
\end{figure*}

Each of the four RSD methods used in our consensus results were
applied to 84 cut-sky mocks. Figure \ref{fig:fs8_nseries_diff} shows 
how the differences among all four methods in the $f\sigma_8$ values 
derived from the DR12 data
compare to the expectations from the cut-sky mocks. Each panel shows
$\Delta f\sigma_8$ between two methods for each redshift bin, and the
distribution of $\Delta f\sigma_8$ from the 84 mocks. The differences
between the four methods applied to BOSS data, listed in Table
\ref{tab:all_RSD}, are in line with expectations from the differences
between methods applied to the same mock survey.

\begin{figure}
%\begin{minipage}{7in}
\vspace{-1cm}
\includegraphics[width=3.5in]{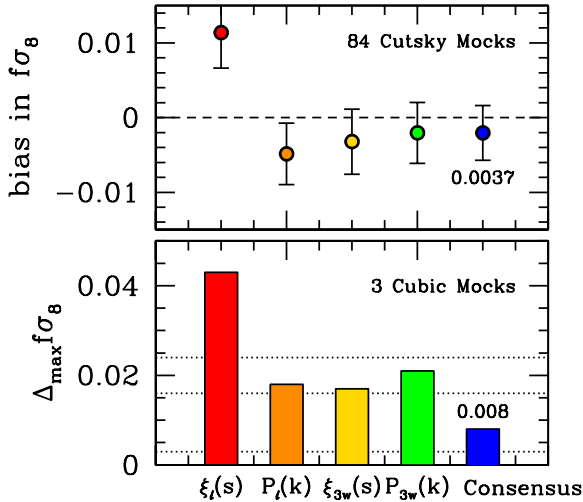}
\vspace{-3.5cm}
\caption{
  \label{fig:fs8_challenge_errors} {\it Top Panel:} The bias in
  $f\sigma_8$ in each full-shape analysis, including the consensus
  value, when applied to the 84 cut-sky mock galaxy catalogues. Here,
  bias is defined as the difference between the mean $f\sigma_8$ value
  from all mocks and the expected value given the input cosmology. The
  error on each point is the standard error in the mean. The bias in
  the consensus $f\sigma_8$ is smaller than the error in the mean,
  0.0037, so this value is adopted as the bias in the consensus
  $f\sigma_8$ value. {\it Bottom Panel:} The systematic variance of
  the $f\sigma_8$ in each full-shape analysis, including the consensus
  value, for three different galaxy bias models imprinted on the same
  N-body simulation (and thus the same intrinsic value of
  $f\sigma_8$). The $y$-axis is the maximal difference among the
  three values of $f\sigma_8$ obtained. For the consensus value, this
  is 0.008. The total systematic error on $f\sigma_8$ from the
  high-resolution mocks is the quadrature sum of the values in the top
  and bottom panels. See the text for more details. The three dashed
  lines represent the bias in $f\sigma_8$ compared to the MD-Patchy
  mocks for the three redshift bins. The low, middle, and high lines
  represent the high, middle and low redshift bins, respectively.}
%  \end{minipage}
\end{figure}

The top panel in Figure \ref{fig:fs8_challenge_errors} shows the bias in
each full shape method when applied to the cut-sky mocks. The error
bars represent the standard error in the mean.  For each cut-sky mock,
the results of the four RSD methods were combined in the same manner
as our consensus results.  Averaging over all 84 mocks, we find only a
modest mean bias in the measured value of $f\sigma_8$ of 0.0018.  This
value is smaller than the statistical precision of the mean
$f\sigma_8$ derived from 84 mocks, which is 0.0037, and so is not
statistically significant.  We adopt 0.0037 as an estimate of the
potential bias of $f\sigma_8$ based on these mocks.  To quantify a
systematic variance in our RSD methods, we also applied the same
analysis to the three cubic mocks with different bias models. Because
these mocks are built on the same N-body simulation, there is little
statistical significance in the comparison between the derived
$f\sigma_8$ and the expected value. However, given that the mocks are
built on the same large scale structure, any differences in the
derived $f\sigma_8$ values from mock-to-mock represent systematic
variations in the accuracy of the methods under different galaxy bias
models. Thus, we use the maximal difference in $f\sigma_8$ between the
three mocks as our systematic error from this test. The bottom panel
of Figure \ref{fig:fs8_challenge_errors} shows the quantity for all four
full-shape methods as well as the consensus value. For the consensus
value, we find the range in $f\sigma_8$ values to be 0.008. We then
place a total systematic error on $f\sigma_8$ from the high-resolution
mocks by adding 0.0037 and 0.008 in quadrature, yielding a value of
0.009 rms; however, we note that more exotic galaxy formation models
might produce larger effects. 

In principle, we can use these same high resolution mocks to quantify
a systematic error on $\epsilon$ from the full shape analyses. Using
the same procedure described above, where the cut-sky mocks define a
bias and the cubic mocks estimate a systematic variance, we find a
total error in $\epsilon$ of 0.0021. We will discuss this further in
the following subsection.

The bottom panel of Figure \ref{fig:fs8_challenge_errors} also shows the
bias in the consensus $f\sigma_8$ values with respect to the MD-Patchy
mocks. These bias values are shown with the horizontal dotted
lines. From top to bottom, respectively, they represent the low
redshift bin, the middle redshift bin, and the high redshift bin. The
error from the MD-Patchy mocks is larger than that derived from the
high-resolution mocks for the low and middle redshift bins (see the
exact values in Table \ref{tab:RSDcompmocks}). Thus, for the
systematic error in $f\sigma_8$, we use the
values from the MD-Patchy mocks for those two redshift bins, and we use
the value from the high resolution mocks for the high redshift bin.

\subsection{Tests of BAO Fitting Methodologies}
\label{sec:sys_fitting}

\citet{Anderson2014b} provides an extensive discussion of systematic
errors for the BAO measurements from the reconstructed density field.
The adopted estimate was 0.003 in $\alpha$ for systematics in the clustering
measurements and fitting systematics, 0.003 in $\alpha$ for
astrophysical systematics involving galaxy bias, and 0.005 in
$\epsilon$ for additional clustering and fitting systematics, all
taken in quadrature. \citet{Anderson2014b} does not identify any
dominant systematic of this type, arguing that all known effects were
plausibly below 0.001. 
The clustering and fitting systematics in the monopole include effects
such as mismatching of the power spectrum template and
averaging across a finite redshift
range. 
The clustering and fitting systematics in the quadrupole were
additionally due to uncertainties in the redshift distortion modelling,
possible small effects due to averaging over the finite redshift
range, and a persistent small bias in the estimation of $\epsilon$ in
our mock catalogues.  The astrophysical systematics were due to
potentially uncorrected shifts due to galaxy bias, despite past
experience that reconstruction tends to null these shifts.

For the present work, we
believe that several aspects of the results have improved.  We have
used several different fitting codes among the various methods and
found good performance with all.  \citet{Thep15} further limits the
template errors, in particularly finding only small shifts with
changes in the extra relativistic species $N_{\rm rel}$, a parameter
that in any case has been more sharply limited by recent Planck
results \citep{Planck2015}.  Most notably, the measurement of 
$\epsilon$ is less biased than before, giving us substantially 
more confidence in the anisotropic measurement.  

\cite{VargasMagana16} provides an additional exploration of potential
sources of theoretical systematic uncertainty in the anisotropic BAO
analysis of the completed BOSS galaxy samples. 
This paper also incorporates
results from previous systematic error analyses
\citep{VargasMaganaEtAl14,Cuesta16,VargasMaganaEtAl15,Ross16,BeutlerBAO16} 
to obtain a fuller accounting of potential systematic uncertainties. 
\citet{VargasMagana16} explores
the anisotropic BAO methodology in a step-by-step manner,
studying concerns such as (but not limited to): the effect of using
different 2-point estimators in configuration space; the effect of
using a finite sample of random catalogues; the manner in which the
covariance matrix is produced; and the fiducial cosmology assumed in
the analysis.  Most variations were found to be very small, below 0.0005.
The more important terms are variations induced by changes in
reconstruction smoothing length, by changes in covariance matrix
estimation, and by variations in the fiducial cosmology for the
distance-redshift relation.  However, these are all modest in size,
around 0.001.

Adding the various terms in quadrature, \citet{VargasMagana16} estimates 
the systematic errors due to analysis and fitting procedures to be
0.002 for $\alpha$ and 0.003 for $\epsilon$.  

\subsection{Summary Model of Systematic Errors}
\label{sec:sys_err}

Having discussed these tests of our methodologies,
we now bring the results together to estimate a
systematic error budget for the consensus results.
We note that while we have tested a wide range of 
variations in our analysis procedures, as well as 
several sets of mock catalogues, this necessarily 
depends on some extrapolation to the unknown.
As such, we opt to round up to an estimate of 
0.003 in $\alpha$ and 0.005 in $\epsilon$ for systematics
in the clustering measurements and fitting methodologies.
This is the same as the estimate in \citet{Anderson2014b}.
We treat these two errors as uncorrelated.

As stated in the previous subsection, we believe that the systematic
control on BAO fitting has continued to improve relative to \citet{Anderson2014b}.
However, these 
improvements in the BAO analysis must be balanced against the fact that the
reconstructed results are now being combined with fits to the full
anisotropic clustering of the unreconstructed density field.
In particular, while much of the $\epsilon$ information comes from
the BAO, some results from the Alcock-Paczynski signal from the
broadband clustering, which is partially degenerate with the
RSD anisotropies.  Therefore, errors in the RSD modelling could 
create systematic biases in $\epsilon$.  We therefore opt to keep
the $\epsilon$ error at 0.005, even though the BAO fitting studies
themselves do not indicate this much uncertainty.

We also continue to estimate an additional systematic error on $\alpha$ of 0.003,
to be added in quadrature so as to yield a total of 0.0042,
resulting from astrophysical systematics involving galaxy bias,
following \citet{Anderson2014b}.  We note that our fits to mock
catalogues continue to return smaller shifts than this, despite variations
in the physical models.  However, more extensive work with N-body
simulations and more complicated halo occupation models is needed
to confidently shrink this error term.

As in \citet{Anderson2014b}, we explicitly exclude from the systematic
error budget the possible shift in the acoustic peak due to a
coupling of the low-redshift galaxy density field to the small
relative velocity between baryons and cold dark matter at high
redshift \citep{TH10,Dalal10,Yoo11,SlepianSV,Blazek16,Schmidt16}.
Whether this effect exists at a measurable level remains speculative,
but observational work from BOSS argues that it is subdominant as a
systematic error. \citet{Yoo2013} first investigated relative
velocities in the power spectrum of BOSS DR9 galaxies, placing an
upper limit on their impact. \citet{Beutler16} sought the effect
in the cross-correlation of the WiggleZ and BOSS survey, again
finding no detection. More recently, \citet{Slepian16RV} search for
the distinctive acoustic-scale signature of this coupling in the
three-point correlation function of DR12 CMASS galaxies.  They find
no detection and use the results of \citet{Blazek16} to place an
0.3\% rms limit on the bias on the BAO-inferred distance scale 
resulting from the relative velocities.
We have also tested whether including these velocities in the BAO
template, following \citet{Blazek16} and marginalizing over a free
amplitude, alters the best fit to our pre-reconstruction measurements.
We find at most a 0.3$\sigma$ shift in $\alpha$, consistent with
the results of \citet{Slepian16RV},
indicating no preference for relative velocities
in the two-point clustering. \citet{Schmidt16} argues for further
acoustic-scale imprints of the relative variations of the baryon
and dark matter density fields, highlighting the possibility that
galaxy bias could depend on the small large-scale variations in the
baryon fraction \citep{Barkana11}. This remains an open topic ---
see \citet{Soumagnac16} for a novel but inconclusive search for a
related effect.

Turning now to the RSD fits, Fig.~\ref{fig:Patchyfits} compares the
two values for the systematic error on $f\sigma_8$ from the MD-Patchy
mock comparison and the high-resolution challenge mocks described in
sections \ref{sec:mockcomp} and \ref{sec:mock_challenge}. The
MD-Patchy bias is larger for the low- and intermediate-redshift bins,
while the error from the high resolution mocks is larger for the high
redshift bin.  We follow a conservative approach and define the
systematic errors for each redshift bin as the maximum of these two
estimates. We therefore adopt systematic errors on $f\sigma_8$ given
by 0.024, 0.015 and 0.009 for our low, intermediate and high redshift
bins, respectively.  We note that the middle bin differs (negligibly)
from the 0.016 in Section \ref{sec:mock_challenge} due to slight
evolution late in the development of the paper.

In principle, this systematic error in $f\sigma_8$ would be correlated
with the systematic error in $\epsilon$.  However, the correlations in
the statistical errors of these parameters are not particularly large
($\approx-0.6$ for all redshift bins) due to the sizeable role of the
BAO, which is less degenerate with modulations of the quadrupole
amplitude caused by the RSD. Taking the slope of the statistical error
correlation as an indication of the coupling in the quadrupole between
the broadband Alcock-Paczynski effect and the RSD, our systematic
errors on $f\sigma_8$ would map to errors on $\epsilon$ of 0.0033,
0.0018, and 0.0012 for each of the redshift bins. These are smaller
than the 0.005 rms error from \citet{Anderson2014b}, as is the error
estimated directly from the high resolution mocks in section
\ref{sec:mock_challenge}, 0.0021. Coupled with the improved fitting of
$\epsilon$, we opt to keep the systematic error on $\epsilon$ at 0.005
rms. We neglect the correlations of this with the $f\sigma_8$
systematic error for simplicity and assume the same systematic errors
in $\alpha$ and $\epsilon$ for out BAO-only, full-shape and final
BAO+FS constraints.

Having specified our estimate of independent systematic uncertainty in
$\alpha$, $\epsilon$, and $f\sigma_8$, we also need to specify how
these might correlate between our three redshift bins.  Declaring the
systematic errors to be independent between the redshift bins would be
over-optimistic as regards redshift-independent shifts if in fact the
errors in the three bins are highly correlated.  We do expect
substantial correlations across redshift: our fitting methodologies
are the same at each redshift, and the galaxies in the three samples
are rather similar, all red galaxies with rather little change in
luminosity or clustering amplitude.  We see little reason, for
example, that astrophysical shifts of the acoustic scale due to galaxy
bias would differ much between $z=0.6$ LRGs and those at $z=0.3$.  On
the other hand, treating the errors as fully correlated is also an
extreme, as it excludes mild systematic variation in redshift.  We
therefore adopt an intermediate ansatz by introducing off-diagonal
redshift couplings of 0.75 in the reduced covariance matrix
representing our systematic errors on all parameters.  This
corresponds to superposing a fully correlated error that is 0.87 of
the total with additional independent errors per redshift bin that are
0.50 of the total.  Alternatively stated, in this ansatz, the variance
of the common mode is 10-fold larger than the variance of the two
other modes.

We note, however, that because the systematic error for $f\sigma_8$
changes with redshift (unlike for $\alpha$ and $\epsilon$), the common
mode favoured by this ansatz is not redshift independent.  A
redshift-independent shift in $f\sigma_8$ is constrained to have 0.008
rms in our model.  For comparison, had we chosen the three bins to be
fully independent, the constraints on the redshift-independent shift
would have been 6\% stronger.  A correlation coefficient around 0.45
maximizes the error for this shift, but at a level only 12\% worse
than our model.  Noting that the systematic errors are subdominant to
statistical errors in all cases, we conclude that the choice of
correlation coefficients for $\alpha$, $\epsilon$, or $f\sigma_8$ does
not substantially impact our cosmological conclusions.

\section{Results from the BOSS combined sample}
%\section{Combining measurements and likelihoods}
\label{sec:combining_measurements}
\subsection{Combining measurements and likelihoods}\label{sec:consensus_method}

\begin{table*}
\centering
 \caption{Final consensus constraints on 
 $\DM\left(\rdfid/\rd\right)$,
 $H\left(\rd/\rdfid\right)$, and
 $f(z)\sigma_8(z)$
 for the BAO-only, full-shape and joint (BAO+FS) measurements . 
 Note that BAO-only results are post-reconstruction while full-shape
 results are pre-reconstruction, and the (strong) covariance between
 them is accounted for when combining to obtain the BAO+FS column.
 In each column,
 the first error corresponds to the statistical uncertainty derived 
 from the combination of the posterior distributions, while the 
 second value represents the systematic error assigned to these 
 results as described in Section~\ref{sec:errors}. In our fiducial cosmology, 
 $\rdfid = 147.78\,{\rm Mpc}$. The cosmological analysis 
 presented in Section~\ref{sec:cosmology} is based on these values. 
}
\begin{tabular}{ccccc}
\hline\hline
 Measurement & redshift &  BAO-only & Full-shape &  BAO+FS \\  \hline
$\DM\left(r_{\rm d,fid}/r_{\rm d}\right)$ [Mpc] & $z=0.38 $ & $1512 \pm 22 \pm 11 $ & $1529 \pm 24 \pm 11$ &  $1518 \pm 20 \pm 11$ \\ 
$\DM\left(r_{\rm d,fid}/r_{\rm d}\right)$ [Mpc] & $z=0.51 $ & $1975 \pm 27 \pm 14 $ & $2007 \pm 29 \pm 15$ &  $1977 \pm 23 \pm 14$ \\ 
$\DM\left(r_{\rm d,fid}/r_{\rm d}\right)$ [Mpc] & $z=0.61 $ & $2307 \pm 33 \pm 17 $ & $2274 \pm 36 \pm 17$ &  $2283 \pm 28 \pm 16$  \\
$H\left(r_{\rm d}/r_{\rm d,fid}\right)\,[{\rm km}\,{\rm s}^{-1}{\rm Mpc}^{-1}]$ & $z=0.38 $ & $81.2 \pm 2.2 \pm 1.0$ & $81.2 \pm 2.0 \pm 1.0$ & $81.5 \pm 1.7 \pm 0.9$  \\ 
$H\left(r_{\rm d}/r_{\rm d,fid}\right)\,[{\rm km}\,{\rm s}^{-1}{\rm Mpc}^{-1}]$ & $z=0.51 $ & $90.9 \pm 2.1 \pm 1.1$ & $88.3 \pm 2.1 \pm 1.0$ & $90.5 \pm 1.7 \pm 1.0$ \\ 
$H\left(r_{\rm d}/r_{\rm d,fid}\right)\,[{\rm km}\,{\rm s}^{-1}{\rm Mpc}^{-1}]$ & $z=0.61 $ & $99.0 \pm 2.2 \pm 1.2$ & $95.6 \pm 2.4 \pm 1.1$ & $97.3 \pm 1.8 \pm 1.1$  \\
$f\sigma_8$ 							& $z=0.38 $ & - & $0.502 \pm 0.041 \pm 0.024$ & $0.497 \pm 0.039 \pm 0.024$ \\ 
$f\sigma_8$ 							& $z=0.51 $ & - & $0.459 \pm 0.037 \pm 0.015$ & $0.458 \pm 0.035 \pm 0.015$  \\ 
$f\sigma_8$ 							& $z=0.61 $ & - & $0.419 \pm 0.036 \pm 0.009$ & $0.436 \pm 0.034 \pm 0.009$ \\ \hline\hline
\end{tabular}
\label{tab:combined_all}
\end{table*}

\begin{table*}
\caption{The covariance matrix and precision matrix of the BAO+FS consensus constraints,
including systematic errors.  The matrices $c_{ij}$ and $f_{ij}$ are the reduced covariance
and reduced precision matrix, multiplied by $10^4$ for conciseness.  $c_{ij}$ is in the lower
triangle; $f_{ij}$ is the upper triangle.  $\sigma_i$ is square root of the diagonal of the covariance
matrix; $s_i$ is the square root of the diagonal of the precision matrix.  Hence, the full
matrices would be $\sigma_i \sigma_j c_{ij}$ and $s_i s_j f_{ij}$.  The row labels
omit factors of $(\rd/\rdfid)$ and the units for conciseness; these are 
supplied in Table \ref{tab:combined_all}.  The on-line
files have the full numerical precision, which we recommend for parameter fits.}
\label{tab:covariance}
\begin{center}
\begin{tabular}{rccrrrrrrrrrc}
\hline\hline
& Mean & $\sigma_i$ & \multicolumn{9}{c}{$10^4c_{ij}$ (lower) or $10^4f_{ij}$ (upper)} & $1/s_i$ \\
\hline
$\DM(0.38)$ & 1518 & 22 & 10000 &  -750 & -3675 & -4686 &   239 &  1781 &   495 &  -166 &   -85 & 18 \\ 
$H(0.38)$ & 81.5 &  1.9 &  2280 & 10000 & -2904 &   126 & -4426 &  1196 &  -380 &   465 &   -73 &  1.6 \\ 
$f\sigma_8(0.38)$ &  0.497 &  0.045 &  3882 &  3249 & 10000 &  1764 &  1588 & -4669 &   299 &   -79 &   625 &  0.034 \\ 
$\DM(0.51)$ & 1977 & 27 &  4970 &  1536 &  1639 & 10000 &  -737 & -3662 & -4764 &   375 &  1922 & 18 \\ 
$H(0.51)$ & 90.4 &  1.9 &  1117 &  4873 &  1060 &  2326 & 10000 & -2855 &   253 & -5140 &  1452 &  1.4 \\ 
$f\sigma_8(0.51)$ &  0.458 &  0.038 &  1797 &  1726 &  4773 &  3891 &  3039 & 10000 &  1733 &  1631 & -4990 &  0.025 \\ 
$\DM(0.61)$ & 2283 & 32 &  1991 &   984 &   237 &  5120 &  1571 &  2046 & 10000 &  -906 & -4042 & 24 \\ 
$H(0.61)$ & 97.3 &  2.1 &   520 &  2307 &   108 &  1211 &  5449 &  1231 &  2408 & 10000 & -2565 &  1.7 \\ 
$f\sigma_8(0.61)$ &  0.436 &  0.034 &   567 &   725 &  1704 &  1992 &  1584 &  5103 &  4358 &  2971 & 10000 &  0.026 \\ 
\hline\hline
\end{tabular}
\end{center}
\end{table*}

\begin{figure*}
\begin{minipage}{7in}
\includegraphics[width=7in]{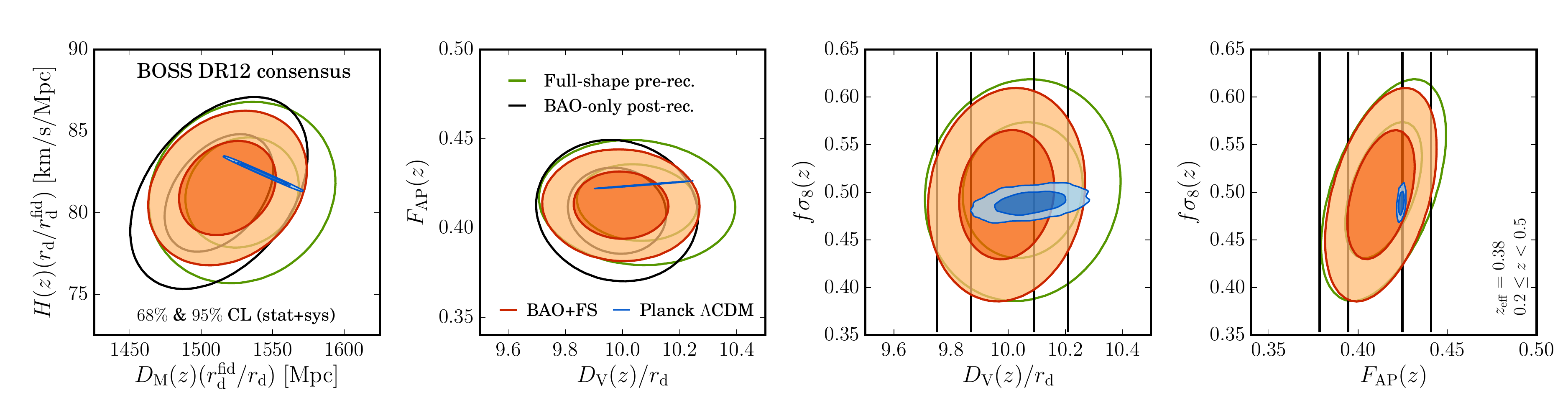}
\includegraphics[width=7in]{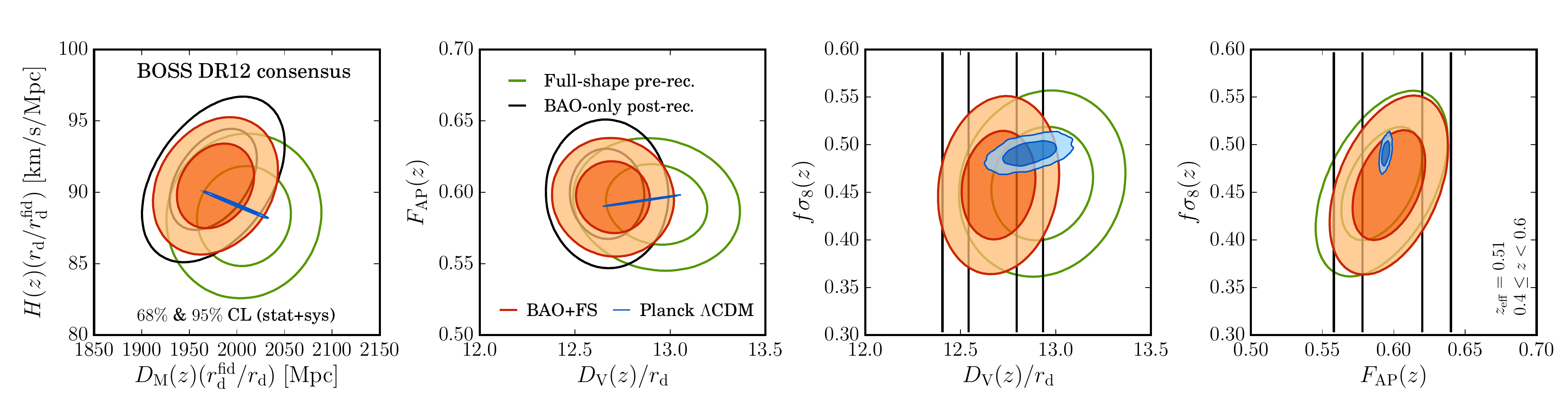}
\includegraphics[width=7in]{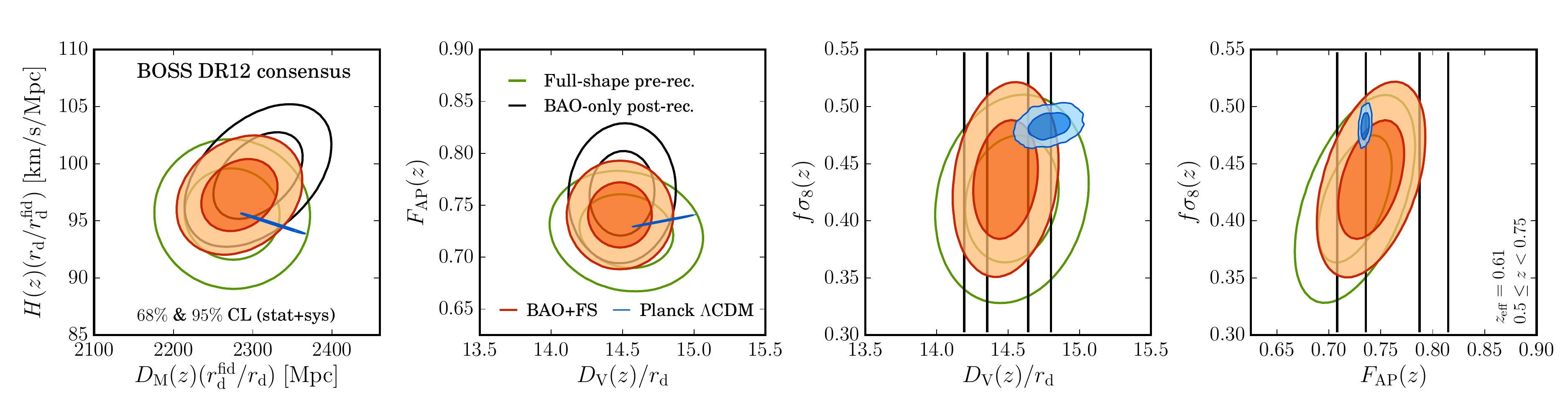}
  \caption{Likelihood contours, showing the 68 per cent and 95 per cent confidence intervals for various combinations of parameters 
  in our three redshift bins. 
  From left to right we show the constraints on: $H(z)(\rd/\rdfid)$ and 
  $\DM(z)(\rdfid/\rd)$, 
  $F_{\rm AP}(z)$ and $\DV(z)/\rd$, $f\sigma_8(z)$ and $\DV(z)/\rd$, and finally $f\sigma_8(z)$ and $F_{\rm AP}(z)$. 
  The black contours show the constraints from post-reconstruction BAO only, the green contours show the constraints from the 
  pre-reconstruction full-shape measurements, and the red filled contours show our final BAO+FS combined constraints. 
  These contours include of the systematic error bars quoted in
  Section~\ref{sec:errors}. The blue solid lines correspond to the constraints inferred from the Planck CMB measurements under the assumption of a $\Lambda$CDM model. }
  \label{fig:combined_contours}
  \end{minipage}
\end{figure*}

\begin{figure*}
\begin{minipage}{7in}
\includegraphics[width=7in]{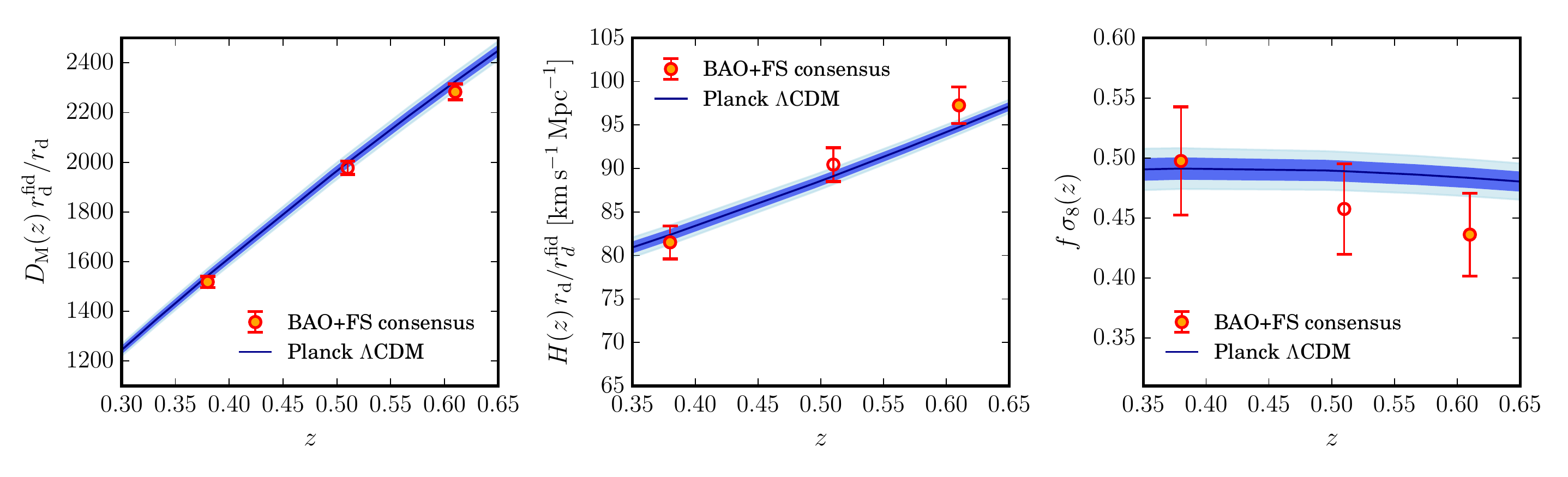}
  \caption{Final consensus constraints on $D_{\rm M}(z)$, $H(z)$ and $f\sigma_8(z)$, shown against the 
  $\Lambda$CDM predictions from the Planck observations of the CMB temperature and polarization.
  The error bars correspond to the total error including statistical variations and systematics.
  The results from the middle redshift bin are shown as a open symbol as a reminder that this bin overlaps with the other 
  two. This figure represents the values presented in the last column of Table~\ref{tab:combined_all}, and the error bars 
  shown include both the statistical and systematic error. }
  \label{fig:combined_redshift}
  \end{minipage}
\end{figure*}

As described in the previous sections, we have computed the parameter combinations 
$\DM(z)/\rd$, $H(z) \rd$ and $f\sigma_8(z)$ in three overlapping redshift slices 
using multiple clustering statistics and modelling assumptions. 
Although they are of course covariant, these
estimates do not contain the same information nor are they affected by noise in the same way. This implies 
that their combination can have a higher constraining power than each individual measurement.
With this in mind, we combine the posterior distributions obtained from our BAO-only and full-shape 
measurements into sets of consensus constraints that optimally capture all of the information
they provide. To do this we follow the method of \citet{SanchezStat16}, which we summarize below.

We wish to combine the results of $m$ different statistical analyses applied to a given 
data set, each leading to an estimate of the same set of  $p$ parameters. 
If the posteriors of these parameters are well described by a multivariate Gaussian distribution, 
the results of any given method $i$ can be represented by an array of $p$ measurements
${\bf D}_i$ and their corresponding $p\times p$ covariance matrix $\mathbfss{C}_{ii}$. 
The full set of measurements obtained from the $m$ different methods corresponds to a set of 
$m\times p$ highly correlated measurements.
As shown in \citet{SanchezStat16}, it is possible to compress the full information contained in these 
measurements into a single set of $p$ consensus values, ${\bf D}_{\rm c}$, with its corresponding 
$p\times p$ covariance matrix, $\mathbfss{C}_{\rm c}$.
A crucial ingredient for this combination is the %$N_{\rm tot}\times N_{\rm tot}$ 
total covariance matrix
\begin{equation}
\mathbfss{C}_{\rm tot} = \left( \begin{array}{ccc}
\mathbfss{C}_{11} & \cdots & \mathbfss{C}_{1m} \\
\vdots & \ddots & \vdots \\
\mathbfss{C}_{m1} & \cdots & \mathbfss{C}_{mm} \\ \end{array} \right),
\end{equation}
where each off-diagonal block $\mathbfss{C}_{ij}$ represents the cross-covariance matrix between the results 
of methods $i$ and $j$. In order to write down the explicit solutions for ${\bf D}_{\rm c}$ and
$\mathbfss{C}_{\rm c}$ we first define a total precision matrix as
\begin{equation}
\mathbf{\Psi}_{\rm tot} \equiv \mathbfss{C}_{\rm tot}^{-1},
\end{equation}
which we divide in blocks of size $p\times p$ as
\begin{equation}
{\bf \Psi}_{\rm tot} = \left( \begin{array}{ccc}
{\bf \Psi}_{11} & \cdots & {\bf \Psi}_{1m} \\
\vdots & \ddots & \vdots \\
{\bf \Psi}_{m1} & \cdots & {\bf \Psi}_{mm} \\ \end{array} \right).
\end{equation}
The general expression for $\mathbfss{C}_{\rm c}$ can then be written as 
\begin{equation}
\mathbfss{C}_{\rm c} \equiv \Psi_{\rm c}^{-1} \equiv \left(\sum_{i=1}^m\sum_{j=1}^m \Psi_{ij}\right)^{-1},
\label{eq:C_comb}
\end{equation}
while ${\bf D}_{\rm c}$ is given by
\begin{equation}
{\bf D}_{\rm c} = \mathbf{\Psi}_{\rm c}^{-1} \sum_{i=1}^{m}\left(\sum_{j=1}^{m} \Psi_{ji}\right){\bf D}_i.
\label{eq:D_comb}
\end{equation}

This methodology can also be used to combine posterior distributions with different number of parameters. 
In this case the final consensus constraints will correspond to the parameter space defined by the union of
those of the individual measurements.
In particular, in order to combine our BAO-only and full-shape constraints 
we are interested in the case in which a given method $i$ gives constraints
on the first $p-1$ parameters only, with an associated 
$(p-1)\times(p-1)$ covariance matrix $\tilde{\mathbfss{C}}_{ii}$.
These results can be considered as including a constraint on the remaining parameter,
but with an infinite uncertainty, that is 
\begin{equation}
\mathbfss{C}_{ii}=\left( \begin{array}{cc}
\tilde{\mathbfss{C}}_{ii} & 0 \\
0 & \infty \\ \end{array} \right).
\end{equation}
In the total covariance matrix $\mathbfss{C}_{\rm tot}$ the rows and
columns corresponding to the undetermined parameter will be zero. This structure will  
be inherited by ${\bf \Psi}_{\rm tot}$, where also the diagonal entry corresponding to this
parameter will cancel. It is then possible to apply equations (\ref{eq:C_comb}) and (\ref{eq:D_comb})
to derive the final consensus values that combine the information from all measurements.  

\citet{SanchezStat16} tested this technique by using it to combine the results obtained
from the application of the BAO-only and full-shape analyses described in Sections \ref{sec:measuring_BAO} and 
\ref{sec:full_shape} to a sub-set of 999 MD-Patchy mock catalogues described in Section \ref{sec:mocks}, 
showing that in all cases the obtained consensus constraints represent a reduction of the 
allowed region of the parameter space with respect to the results of each individual method. 

\subsection{Consensus constraints from BOSS}\label{sec:consensus_results}

As shown in our companion papers, the posterior distributions recovered from the different analysis 
methodologies applied to BOSS are well described by Gaussian multivariate distributions, which means 
that we can apply the methodology described in the previous section
to obtain our consensus results. We will obtain three sets of
consensus results, from the following measurements: post-reconstruction
BAO  (denoted as BAO-only), pre-reconstruction full-shape
measurements (denoted as full-shape or FS), 
and finally, a final consensus set from combining 
post-reconstruction BAO with pre-reconstruction full-shape measurements, 
denoted as BAO+FS.  
For BAO-only measurements, which are only sensitive to the geometric quantities $\DM(z)/\rd$
and $H(z) \rd$, we have $p=2$, while for full-shape
fits, which can also constrain $f\sigma_8(z)$, $p=3$.

The application of equations (\ref{eq:C_comb}) and (\ref{eq:D_comb}) requires the knowledge of the
total covariance matrices $\mathbfss{C}_{\rm tot}$ for each case. 
For the diagonal blocks $\mathbfss{C}_{ii}$, we use the covariance matrices derived from the posterior
distributions of each analysis method. We construct the off-diagonal blocks $\mathbfss{C}_{ij}$ using 
the cross-correlation coefficients derived by \citet{SanchezStat16} 
from the application of the different methods to the MD-Patchy mock catalogues.

The solid black contours in Figures~\ref{fig:2d_bao} and \ref{fig:2d_rsd} correspond to the BAO-only and 
full-shape consensus constraints, respectively, derived by 
applying equations (\ref{eq:C_comb}) and 
(\ref{eq:D_comb}) to the results of our companion papers.
The final covariance matrices of our consensus constraints are obtained by adding the matrices 
$\mathbfss{C}_{\rm c}$ derived from the combination of the posterior distributions, which represent 
the statistical uncertainties of our results, with that of the systematic errors described in 
Section~\ref{sec:sys_err}. The corresponding one-dimensional marginalized constraints are listed in 
the third and fourth columns of Table~\ref{tab:combined_all}, where the first error accompanying each 
value correspond to the statistical 68 per cent CL, and the second one represents the systematic error assigned to these results (see Section~\ref{sec:sys_err}).   

Fig.~\ref{fig:combined_contours} illustrates the principal observational
results of this paper in the form of confidence contours from the
BAO-only (black) and full-shape (green) consensus constraints
in each of our three redshift bins, for different pairwise combinations
of $\DM(z)\times(\rdfid/\rd)$, $H(z) \times (\rd/\rdfid)$, $\DV(z)/\rd$,
$f\sigma_8(z)$, and the Alcock-Paczynski parameter $F_{\rm AP}(z)$.
The filled contours represent the combination of these results into
the final set of BAO+FS consensus constraints representing
the full information obtained from our pre- and post-reconstruction
clustering measurements.
The corresponding one-dimensional constraints are quoted in the
last column of Table~\ref{tab:combined_all} and shown as a function
of redshift alongside the $\Lambda$CDM best-fit Planck prediction
in Fig.~\ref{fig:combined_redshift}.  The covariance and precision
matrices are in Table~\ref{tab:covariance}.

The statistical uncertainties
in $\DM(z)/\rd$, $H(z)\rd$, and $f\sigma_8(z)$
are all reduced in our final
consensus values, with respect to those in any individual method
or in the BAO-only and full-shape consensus constraints. The
improvement in the statistical uncertainty, with respect to the
{\it smallest} quoted uncertainty in each of the individual
measurements, is typically 15\% for $\DM(z)$, 20\% for
$H(z)$, and 10\% for $f\sigma_8$. 
These improvements are in
agreement with what is expected from tests on the mocks
\citep{SanchezStat16}. Figure \ref{fig:fs8_challenge_errors} further shows that,
on high-fidelity mocks, consensus results show a smaller systematic
bias than each individual method. It is this final set of consensus
values and derived likelihoods that we use in our cosmological
analysis in Section~\ref{sec:cosmology}.

When expressed in terms of the average distance $\DV(z)$, 
our final BAO+FS consensus constraints 
correspond to 
\begin{align}
\DV(0.38) &= \left(1477 \pm 16 {\rm\;Mpc}\right)\left(\rd\over\rdfid\right), \\
\DV(0.51) &= \left(1877 \pm 19 {\rm\;Mpc}\right)\left(\rd\over\rdfid\right), \\
\DV(0.61) &= \left(2140 \pm 22 {\rm\;Mpc}\right)\left(\rd\over\rdfid\right).
\end{align}
These values correspond to distance measurements of 1.1\% precision
for our low-redshift bin
and 1.0\% for the intermediate and high-redshift bins. 
The 0.2\% statistical error on $\rd$,
based on Planck 2015 CMB constraints assuming standard matter and radiation 
content, makes a negligible contribution when added in quadrature.
These $\DV$ values are covariant, though the first and third are only
weakly so; one should use the full likelihood for fitting models.

Although it is not appropriate for cosmological fits, 
it can be useful as a metric 
to compute the aggregate precision of the measurement by combining 
across the three redshift bins, including our systematic error estimates.  
Doing this, we find a precision of 1.0\% on the transverse
distance scale, 1.6\% on the radial distance scale, and  0.8\%
on the spherically averaged $\DV$.  We also find a 5.7\% aggregate 
precision on $f\sigma_8$.  
In all cases, these were computed as the error on a single rescaling
of the best-fit measurements in the three redshift bins for the chosen
parameter, holding the other six measurements fixed.  In the case of
$\DV$, we held $F_{\rm AP}$ fixed.  If instead
one marginalizes over the other six dimensions, the aggregate errors
degrade slightly by a factor of 1.1--1.2.  We note that the performance
on specific parametrized models, such as in Section \ref{sec:cosmology},
can be different than these values, as they correspond to other
weightings of the various measurements.

In Table~\ref{tab:combined_all}, the 1-dimensional errors on
$\DM(z)$ and $H(z)$ from the full-shape analyses are only slightly
worse than those of the BAO-only analyses even though the constraints
on these quantities come mainly from BAO and the BAO-only analyses
take advantage of precision gains from reconstruction.
Figure~\ref{fig:combined_contours} helps to resolve this conundrum.
The values of $\DM(z)$ and $H(z)$ are more strongly correlated
for the BAO-only analysis, so while the $\DV(z)$ constraints
from post-reconstruction BAO-only are appreciably tighter than those
from pre-reconstruction full-shape, the marginalized constraints
on $\DM(z)$ and $H(z)$ are not.  The constraints on $F_{\rm AP}(z)$
from sub-BAO scales in the full-shape analyses help to break
the degeneracy between $\DM$ and $H$, leading to rounder 
confidence contours and smaller errors on $F_{\rm AP}$.
The combined BAO+FS contours are able to take advantage of 
both the sharpening of the BAO feature by reconstruction and
the improved degeneracy breaking from the sub-BAO Alcock-Paczynksi
effect.  In all of the projections shown here, the 68\% CL
contour from our consensus constraints overlaps the 68\% CL
contour from the Planck 2015 CMB results assuming a 
$\Lambda$CDM cosmological model, demonstrating impressive success of this
model in reproducing the expansion history and rate of structure
growth over the redshift range $0.2 < z < 0.75$.
We provide more detailed assessment of the cosmological implications of
these measurements in \S\ref{sec:cosmology}.

\subsection{Comparison to Past Work}
\label{sec:comparison_to_previous}

\addtolength{\tabcolsep}{-3pt}
\begin{table*}
\centering
\caption{Comparison of BOSS BAO measurements from DR9, DR10, DR11, and DR12.
The new DR12 Combined Sample measurements (BAO-only) reported here for 
the low and high redshift bins have been extrapolated to $z=0.32$ 
and $z=0.57$ respectively, (assuming a $\Lambda$CDM model with 
$\Omega_m=0.31$) for direct comparison to previous measurements based
on the LOWZ and CMASS samples.
A fiducial sound horizon $r_{\rm d,fid} = 147.78\,{\rm Mpc}$ is assumed.
The last two lines under DR12 are combinations of the indicated
results listed earlier in the table, accounting for covariance.
}
\begin{tabular}{cccccccc}
\hline\hline
     & & $\DV\left(\rdfid/\rd\right)$ 
	   & $\DM\left(\rdfid/\rd\right)$ 
	   & $H\left(\rd/\rdfid\right)$ 
	   & $\DV\left(\rdfid/\rd\right)$ 
	   & $\DM\left(\rdfid/\rd\right)$ 
	   & $H\left(\rd/\rdfid\right)$ \\
     & & $z=0.32$ & $z=0.32$ & $z=0.32$ & $z=0.57$ & $z=0.57$ & $z=0.57$ \\
     & & [Mpc] & [Mpc] & $[{\rm km}\,{\rm s}^{-1}{\rm Mpc}^{-1}]$ 
	   & [Mpc] & [Mpc] & $[{\rm km}\,{\rm s}^{-1}{\rm Mpc}^{-1}]$ \\
\hline
DR9  & \cite{Anderson2012,Anderson2014a}  & --- & --- & --- & $2073 \pm 33$ & $2188 \pm 70$ & $93.8 \pm 7.9$ \\ \hline
DR10 & \cite{Anderson2014b}& $1262 \pm 36$ & --- & --- & $2034 \pm 28$ & $2154 \pm 40$ & $95.1 \pm 4.7$ \\
	 & \cite{TojeiroEtAl14}&  && &  &  & \\ \hline
DR11 & \cite{Anderson2014b} & $1251 \pm 25$ & --- & --- & $2035 \pm 20$ & $2209 \pm 31$ & $97.8 \pm 3.4$ \\ \hline
\multirow{11}{*}{{DR12}} & Chuang et al. 2016 & $1268\pm26$ & $1262\pm37$ & $75.0\pm4.0$ & $2050\pm22$ & $2204\pm36$ & $95.5\pm2.7$ \\ & Cuesta et al. 2016a & $1270\pm22$ & $1301\pm27$ & $78.8\pm5.6$ & $2037\pm21$ & $2210\pm33$ & $99.8\pm3.7$ \\ & Gil-Mar\'in et al. 2016a & $1274\pm22$ & $1299\pm31$ & $78.5\pm4.1$ & $2025\pm18$ & $2186\pm30$ & $98.5\pm2.5$ \\ & Gil-Mar\'in et al. 2016b & --- & $1239\pm37$ & $77.2\pm3.8$ & --- & $2186\pm35$ & $94.2\pm3.0$ \\ & Gil-Mar\'in et al. 2016c & --- & $1315\pm43$ & $79.5\pm3.7$ & --- & $2165\pm35$ & $93.2\pm1.9$ \\ & Pellejero-Iba\~nez et al. 2016 & --- & $1262\pm36$ & $79.1\pm3.3$ & --- & $2206\pm39$ & $96.7\pm3.1$ \\ & Slepian et al. 2016a & --- & --- & --- & $2025\pm35$ & --- & --- \\ & Wang et al. 2016 & --- & $1229 \pm 46$ & $74.3\pm5.7$ & --- & $2159\pm 56$ & $92.7 \pm 4.0$ \\ & Zhao et al. 2016 & --- & $1229\pm 52$ & $78.3\pm4.1$ & --- & $2153\pm 36$ & $94.2\pm 3.6$ \\ &  Cuesta + G-M 2016 a & $1272\pm22$ & $1301\pm29$ & $78.7\pm4.7$ & $2030\pm19$ & $2197\pm28$ & $99.3\pm2.8$ \\ &  G-M et al. 2016 (a+b+c) & --- & $1287\pm25$ & $78.2\pm2.6$ & --- & $2179\pm23$ & $94.9\pm1.5$ \\ \hline
Final & This work  & $1270 \pm 14$ & $1294 \pm 21$ & $78.4 \pm 2.3$ & $2033 \pm 21$           & $2179 \pm 35$ & $96.6 \pm 2.4$ \\
\hline
\hline
\end{tabular}
\label{tab:drcomparison}
\end{table*}
\addtolength{\tabcolsep}{3pt}

\begin{figure*}
\begin{minipage}{7in}
\includegraphics[width=3.5in]{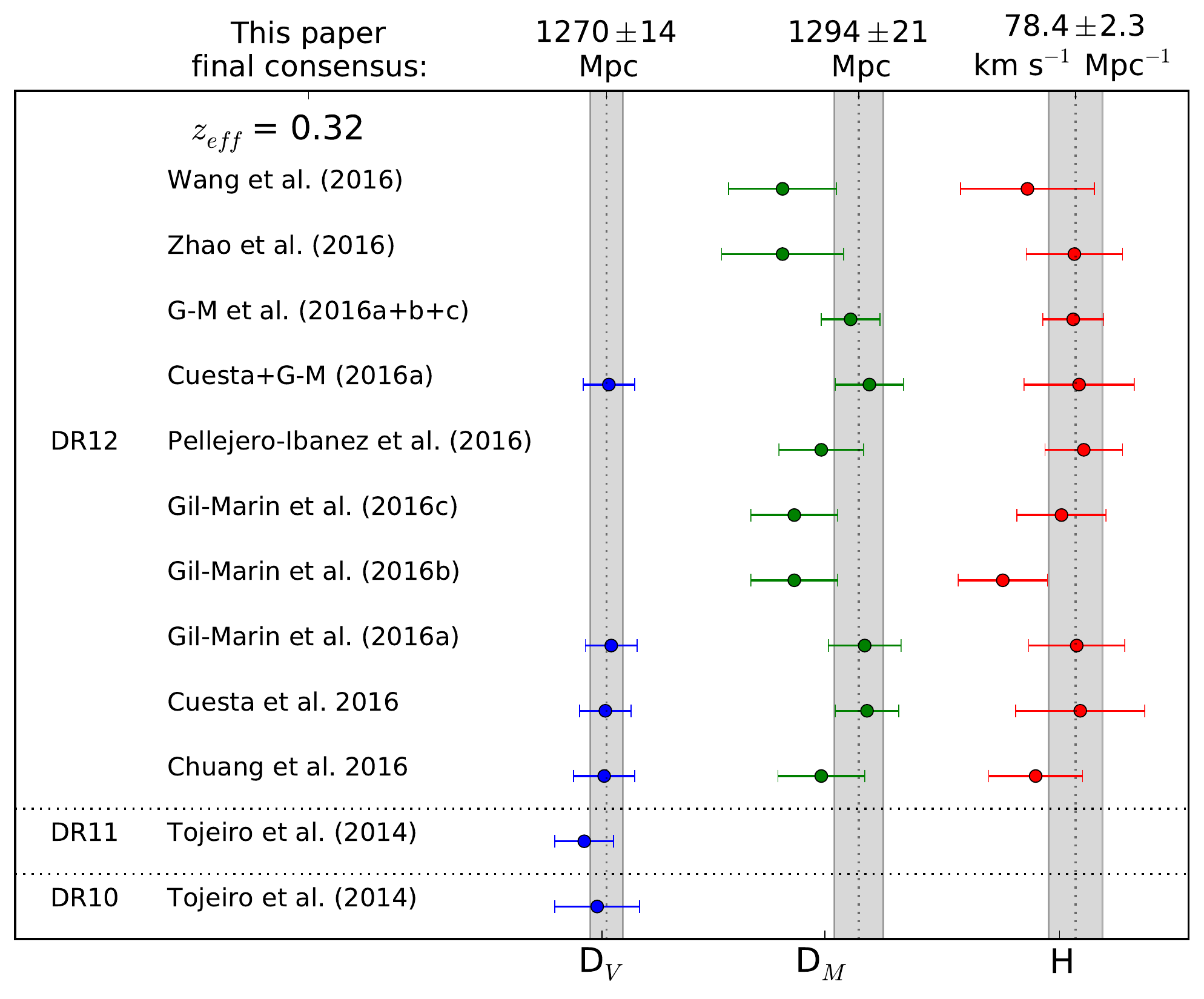}
\includegraphics[width=3.5in]{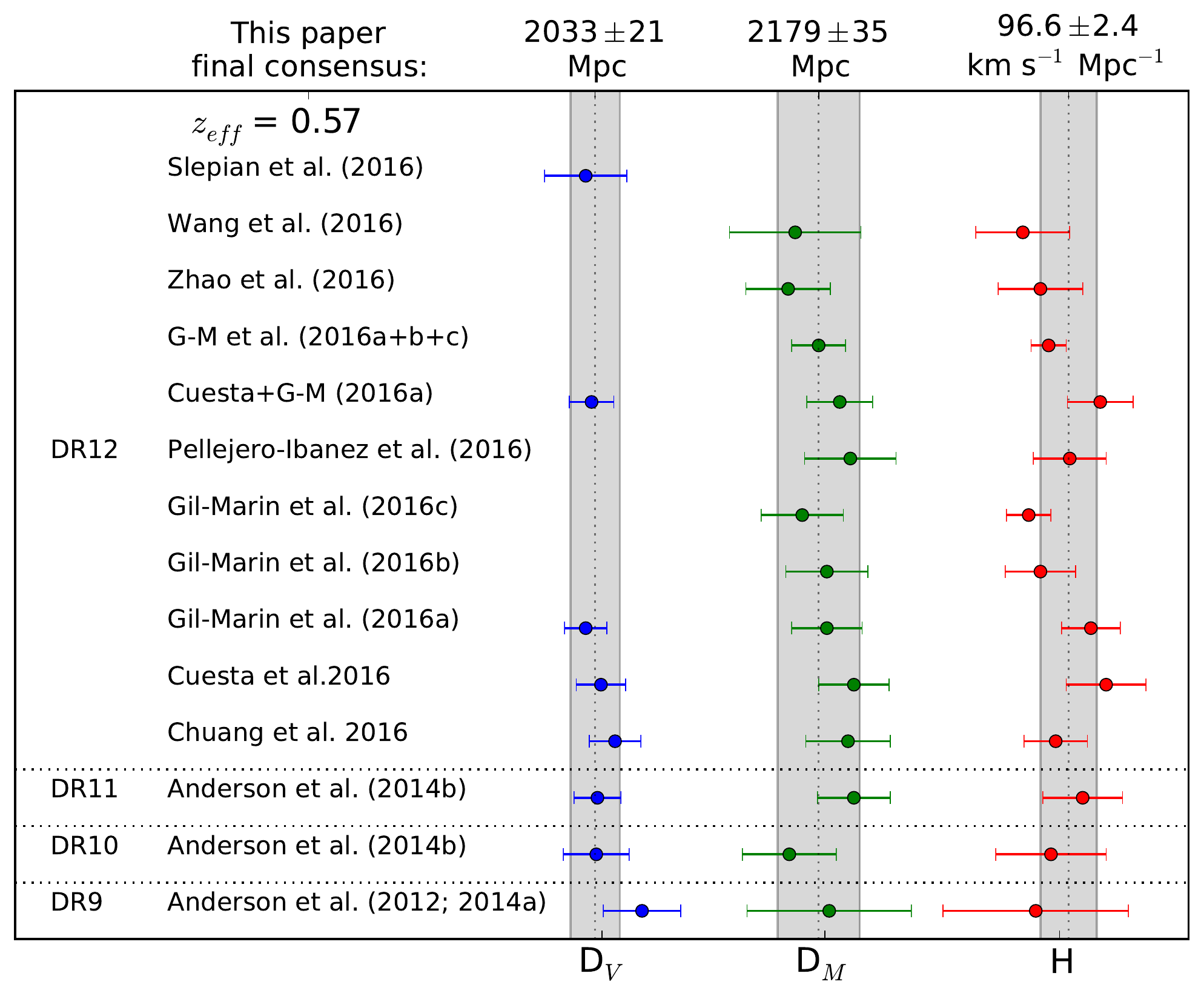}
  \caption{A summary and evolution of BOSS measurements since DR9. The left hand panel shows measurements done at low-redshift and the right-hand panel shows measurements done at high-redshift. All values are presented in Table~\ref{tab:drcomparison}. Error bars are $1\sigma$ and the grey band shows the results of this paper.}
  \label{fig:drcomparison}
  \end{minipage}
\end{figure*}

In this section, we compare our results to previous work.
We begin by summarizing BAO-distance measurements made by the BOSS team 
since DR9, which we collect in Table~\ref{tab:drcomparison} and 
Fig.~\ref{fig:drcomparison}. We quote values at $z_{\rm eff} = 0.32$ 
and $z_{\rm eff} = 0.57$, corresponding to the effective redshifts of the 
LOWZ and CMASS samples. To put the results of this paper in the same context, 
we extrapolate our distance measurements to the above values of 
effective redshift, assuming a flat $\Lambda$CDM cosmology ($\Omega_m = 0.31$). 
The DR9, DR10, and DR11 measurements use 2-point statistics in configuration
and Fourier space, and the improvement of precision with the growing BOSS
footprint is evident in the statistical error bars, while the agreement between the measurements is reassuring. 
The DR12 measurements come from a variety of methods including
3-point  statistics in both configuration and Fourier space  as well as different
approaches to redshift binning and quantifying anisotropy to separate
$\DM$ and $H$.  Some of these analyses use full-shape information
and others use BAO only.  
Most of the DR12 analyses listed in Table~\ref{tab:drcomparison} use the
LOWZ and CMASS catalogues, while this paper and its supporting papers
use a combined sample that is optimized to have better statistical power.
The \citet{Zhao16} and \citet{Wang16} results listed in Table~\ref{tab:drcomparison}
also use the combined sample.

We briefly review the different approaches of the previous
DR12 papers.
Because the underlying galaxy data
are the same, we expect consistency at the 1$\sigma$ level or better,
but the robustness of distance-scale inferences across such a wide
range of analysis methods is reassuring nonetheless.
\citet{Chuang2016} use 2-point functions in configuration space and seek to achieve the best systematic free measurement of distances and growth rate measurements by marginalizing over several nuisance terms in their analysis. 
\citet{Cuesta16} follows exactly the same methodology as in our DR10 and DR11 analysis \citep{Anderson2014b} with the same type of catalogues (LOWZ and CMASS). We expect the increase of statistical power of the derived parameters from \citet{Anderson2014b} to \citet{Cuesta16} to be purely due to the increase in our survey volume. 
\citet{Gil15BAO}  and \citet{Gil15RSD} used line of sight power-spectrum to measure the BAO position and growth rate respectively. \citet{Slepian16BAO} uses the
3-point function in configuration space to measure the BAO position. 
\citet{Gil16RSDbispect} uses full-shape measurement of the 3-point
statistics and 2-point statistics in Fourier space to constrain the
BAO and RSD parameters. Our analysis presented in this paper does
not involve using any 3-point function statistics. In principle,
there is additional information that can come from the
higher order correlation function (see discussions of information
correlation between 2- and 3-point functions in \citet{Slepian16BAO}
and \citet{Gil16RSDbispect}).
\citet{Wang16} and \citet{Zhao16} analyzed the BAO distances in nine redshift bins instead of the three in our analysis in both configuration space and Fourier space. 
\citet{Pellejero-Ibanez16} analyzed the sample with minimal assumptions of cosmological priors and found consistent results as our analysis.

A comparison with \citet{Cuesta16} and \citet{Gil15BAO} 
is of particular interest, as those papers present 
similar configuration and Fourier space analyses to the ones used here,
for the same BOSS data set, but breaking the samples by the LOWZ 
and CMASS target selections rather than the finer redshift binning
adopted in this paper. In the following discussion we will focus on their consensus results, obtained from combining the likelihoods derived from the correlation and power spectrum. Those consensus results are presented in \citet{Gil15BAO}. 
The performance of our updated methodology can be tested against the above consensus results by comparing the precision in cosmic distance measurements. %Even though our choice of redshift bins does not exactly match the redshift ranges spanned by the LOWZ and CMASS samples, 
We make an approximate comparison by equating LOWZ to our 
low redshift bin, and CMASS to our high redshift bin. Note 
that our low redshift bin has a larger effective volume than the LOWZ 
sample $V_{\rm eff,low}/V_{\rm eff,LOWZ}=1.7$, and our high redshift bin has a smaller effective volume than the CMASS sample, $V_{\rm eff,high}/V_{\rm eff,CMASS}=0.8$. There is a trade-off in the precision of the low redshift bin, at the expense of having less precision in the high redshift bin, motivated by the redshift boundary being shifted from $z=0.43$ to $z=0.50$. To clarify the comparison, we will rescale in the following discussion the LOWZ uncertainties by a factor of $\sqrt{V_{\rm eff,LOWZ}/V_{\rm eff,low}}=0.77$ and the CMASS uncertainties by a factor of $\sqrt{V_{\rm eff,CMASS}/V_{\rm eff,high}}=1.12$, so the reader should assume this factor implicitly in all text throughout this section. However, Fig. \ref{fig:drcomparison} and Table \ref{tab:drcomparison} have no such corrections applied to them.

For comparison, we focus on the $\DV$ constraints, as these provide the most information from the post-reconstruction BAO analysis and we regard the LOWZ volume as too small to obtain robust $H(z)$ likelihoods (the LOWZ $\DV$ likelihood is what was used in the \citealt{Cuesta16} cosmological analysis). The consensus precision on $\DV$ from the combination of the \citet{Cuesta16} and \citet{Gil15BAO} results is 1.3 per cent for LOWZ and 1.0 per cent for CMASS, 
after the above scaling by $\sqrt{V_{\rm eff}}$.
The consensus $\DV$ precision we obtain (see Section \ref{sec:consensus_results}) is 20 per cent better at low redshift and the same at high redshift, and these $\DV$ constraints come almost entirely from the post-reconstruction BAO analysis (see the second column of Fig. \ref{fig:combined_contours}). Our improvement at low redshift is compatible with the fact that our error in $\DV$ is smaller than the standard deviation of the mock samples (see Table 5) by 20 per cent, while the results presented in \cite{Cuesta16} obtained slightly worse precision than the equivalent quantity from the mocks. Such fluctuations in precision are consistent with those found in our mock samples. In terms of the standard deviation, the consensus mock results for $\DV$ in \cite{Cuesta16} agree with the consensus results presented in Table \ref{tab:Pkxicompmocks}, at the number of significant digits we quote. Thus, results from this comparison are consistent with the expectation from the tests in mock catalogues described in Section~\ref{sec:combined_sample}.

\begin{figure*}
\begin{minipage}{7in}
\includegraphics[width=6in]{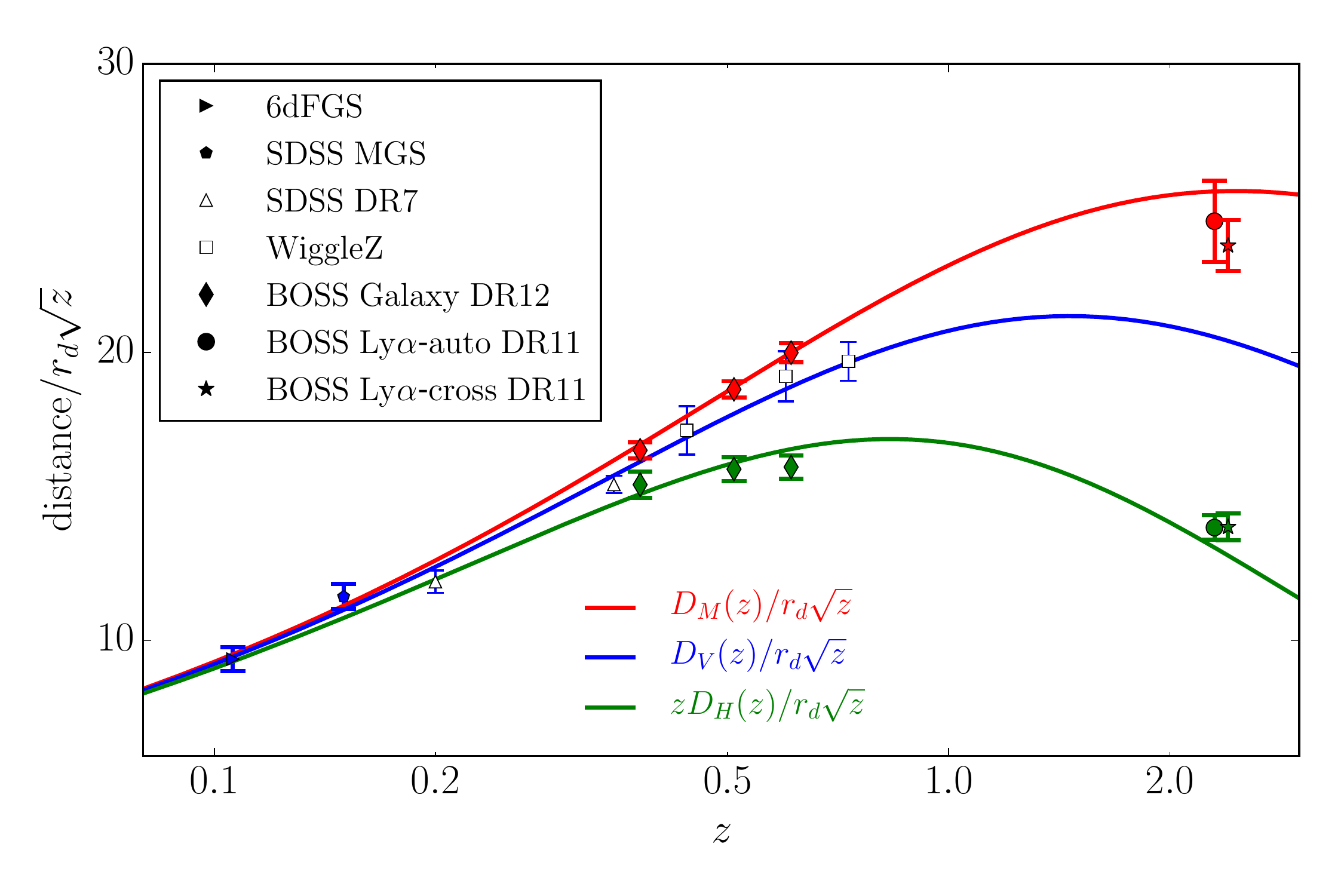}
  \caption{The ``Hubble diagram'' from the world collection of spectroscopic BAO detections.
  Blue, red, and green points show BAO measurements
    of $\DV/r_d$, $\DM/r_d$, and $\DHub/r_d$, respectively, from the
    sources indicated in the legend.
    These can be compared to the correspondingly coloured lines,
    which represents predictions of the fiducial Planck
    $\Lambda$CDM model (with $\Omega_m=0.3156$, $h=0.6727$).
    The scaling by $\sqrt{z}$ is arbitrary, chosen to compress the
    dynamic range sufficiently to make error bars visible on the plot.
    For visual clarity, the Ly$\alpha$ cross-correlation points
    have been shifted slightly in redshift; auto-correlation points are 
    plotted at the correct effective redshift.
	Measurements shown by open points are not incorporated in our
	cosmological parameter analysis because they are not independent
	of the BOSS measurements.
 }
  \label{fig:distancescale}
  \end{minipage}
\end{figure*}

Figure~\ref{fig:distancescale} plots our
BAO-only results in the wider context of other surveys 
and higher redshift measurements from the BOSS \lya\ forest.
Blue, green, and red curves/points show $\DV(z)$, $\DM(z)$,
and $\DHub(z) \equiv c/H(z)$, divided by $\rd$ and with redshift 
scalings that fit all three curves on the same plot with visible error bars.
The three lines show the predictions of a $\Lambda$CDM model
with the Planck 2015 parameters.  
Symbols show BAO measurements from
$z\approx 0.1$ to $z\approx 2.2$ collected from 
6dFGS \citep{Beutler11}, 
SDSS-I/II \citep{Percival2010,Ross15MGS},
WiggleZ \citep{Blake11,Blake11b},
and the BOSS \lya\ forest auto- and cross-correlations
(\citealt{Delubac} and \citealt{Font14}, respectively),
in addition to the BOSS galaxy measurements described here.
The \cite{Percival2010} analysis includes SDSS LRGs and overlaps
significantly with BOSS, while the main galaxy sample (MGS) 
analyzed, with reconstruction, by \cite{Ross15MGS} is essentially
independent.  The WiggleZ survey volume also overlaps BOSS, 
but 6dFGS is again independent.  
We find consistency across all galaxy BAO measurements. 
Moderate tension with the \lya\ forest BAO measurements remains,
as discussed in detail by \cite{Delubac} and \cite{Aubourg}.
BAO analyses of the DR12 \lya\ forest data set are in process
(J.\ Bautista et al., in prep.).

\begin{figure*}
 \includegraphics[width=3in]{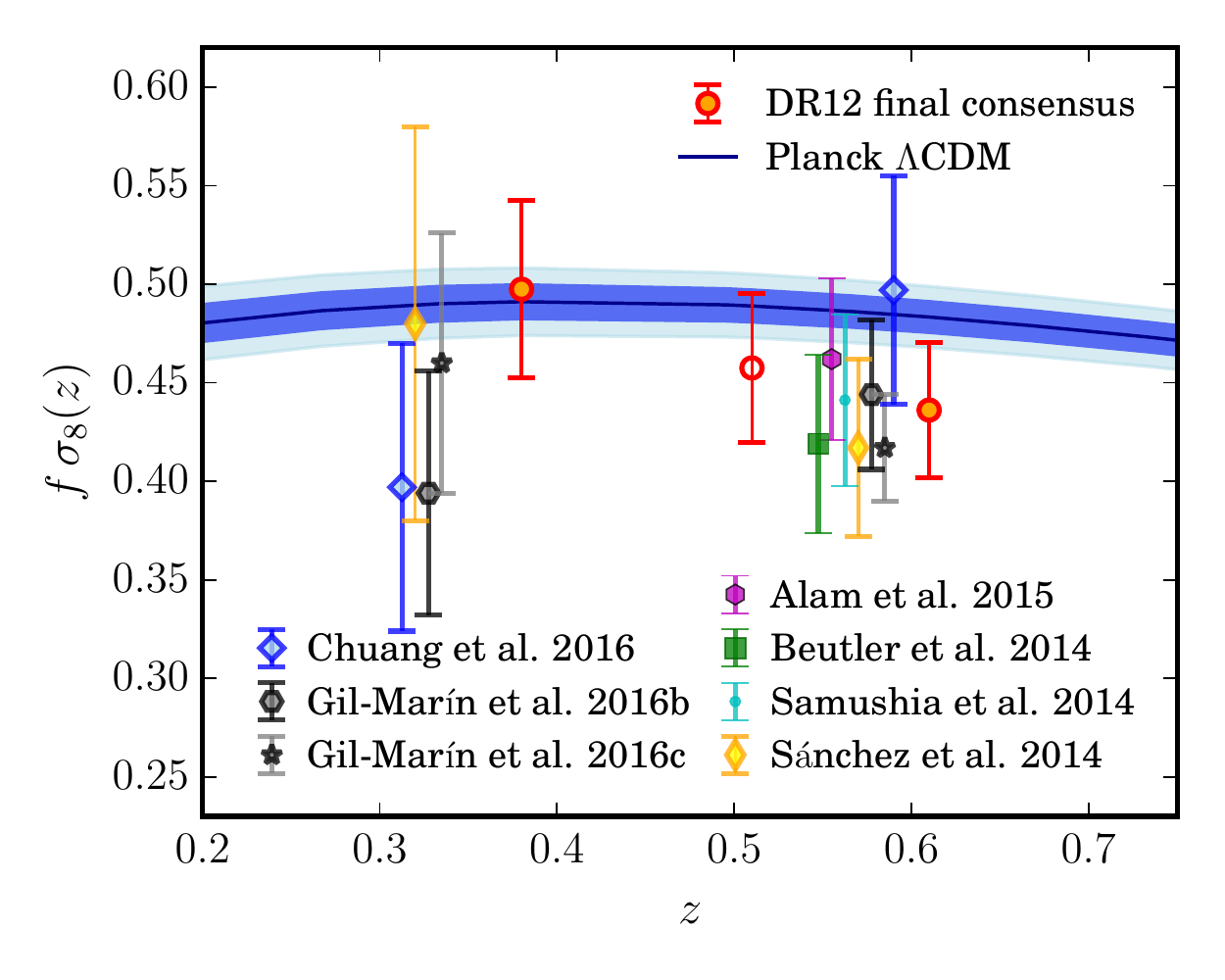}
\includegraphics[width=3in]{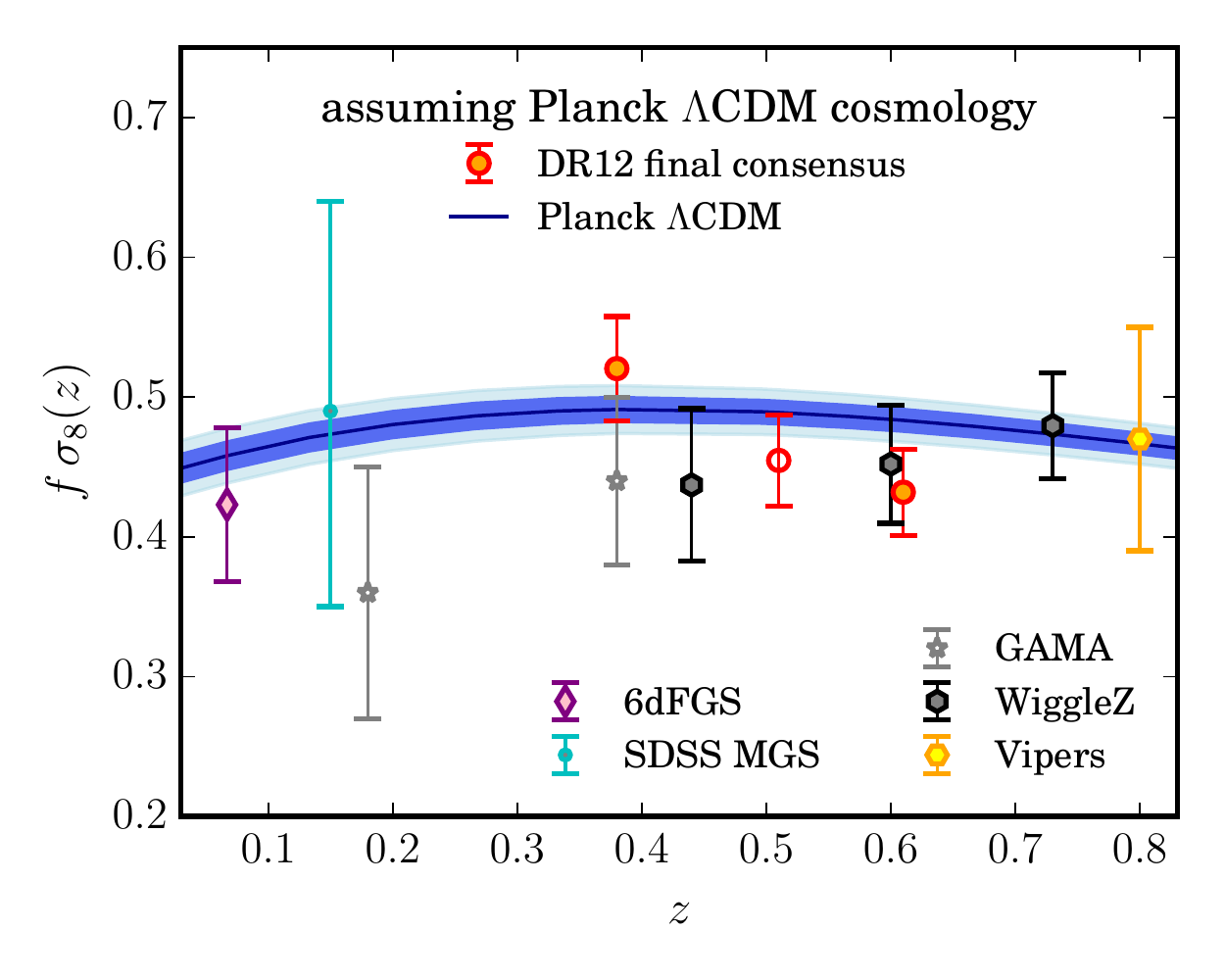}
 \caption{Left-hand panel:  Comparison of $f \sigma_8(z)$ measurements across previous BOSS measurements in DR11 \citep{AlamRSDDR112015,Beutler14,Samushia14,Sanchez14}
  and DR12 \citep{Gil15RSD, Gil16RSDbispect, Chuang2016} samples. 
   Right-hand panel: The $f \sigma_8(z)$ results from this work compared with the measurements of
  the 2dfGRS \citep{Percival04} and 6dFGS \citep{Beutler12},
  the GAMA \citep{Blake13}, the WiggleZ \citep{Blake12}, the VVDS \citep{Guzzo08}, and the VIPERS \citep{delaTorre2013} surveys,
  as well as the measurements from the SDSS-I and -II main galaxy sample \citep[MGS]{Howlett2015} and
  the SDSS-II LRG sample \citep[DR7]{Oka14}. We have plotted conditional constraints on  $f\sigma_8$ assuming a Planck $\Lambda$CDM background cosmology. This is one of the best evidence of how growth rate measurements from BOSS again reaffirm the validity of General Relativity in large scales.
 }
\label{fig:fsig8_other}
\end{figure*}

Next we compare our $f \sigma_8$ results to those from the literature. As before, we begin by collecting the work done by the BOSS team, which we summarize on the left-hand side of Fig.~\ref{fig:fsig8_other}. We include measurements and quoted uncertainties from DR11 studies \citep{AlamRSDDR112015,Beutler14,Samushia14,Sanchez14} and DR12 \citep{Gil15RSD,Chuang2016}. The 
improved precision at low redshift in the present analysis
greatly helps to test the predictions of structure growth in the universe,
showing consistency with $\Lambda$CDM and GR. 
We find excellent consistency among different methods and data releases.
Given the small area increase between DR11 and DR12,
the differences seen in Figure~\ref{fig:fsig8_other} are likely a
consequence of different redshift binning and analysis/modelling methods.
A more detailed study of the impact of different methodologies on $f\sigma_8$ measurements, using high-fidelity mocks, can be found in \cite{tinker_etal:2016} for DR12 measurements.

The right panel of Figure ~\ref{fig:fsig8_other} compares
our measurements of $f\sigma_8$ results those from other surveys: 
2dfGRS \citep{Percival04}, 6dFGS \citep{Beutler12}, GAMA \citep{Blake13}, 
WiggleZ \citep{Blake12}, VVDS \citep{Guzzo08}, and 
VIPERS \citep{delaTorre2013}, 
as well as the measurements from the SDSS-I and -II main galaxy 
sample \citep[MGS]{Howlett2015} and the SDSS-II LRG sample \citep[DR7]{Oka14}. 
The measurements plotted are conditional constraints on $f\sigma_8$ based on the Planck 2015 $\Lambda$CDM cosmological model. This can be seen as a direct test of General Relativity. We find that our results confirm the validity of General Relativity. 
We also find reassuring consistency between our measurements and 
those by different surveys.

It is also interesting to compare this paper's full-shape results
(Table~\ref{tab:combined_all}) with the 
full-shape analysis of the DR12 LOWZ and CMASS samples, 
done in Fourier space by \citet{Gil15RSD}
(scaled again by $\sqrt{V_{\rm eff}}$ factors).
Approximating LOWZ to our low redshift bin and CMASS to our high redshift bin, we 
find a $\DM$ measurement of 1.7\% in the low redshift bin 
and 1.8\% in the high redshift bin, which compares to 2.3\%
and 1.8\% in \citet{Gil15RSD}, respectively. 
 Regarding $H(z)$, our measurement of 2.8\% in
both the low and high redshift bins compares to 3.8\% and 3.6\% 
in \citet{Gil15RSD}, again showing a clear improvement in the 
precision when using our new methodology. Finally our $f\sigma_8$ 
constraint of 9.5\% and 8.9\% in the low and high redshift 
bin compares to the LOWZ constraint of 12.1\% and 9.6\%
in \citet{Gil15RSD}, which similarly to $\DM$ and $H$, 
shows a clear improvement in the low redshift bin. 

Additionally, we 
display the results based on the combination of the pre-reconstructed 
power spectrum, bispectrum and post-reconstruction BAO (from \citealt{Gil15BAO, Gil15RSD, Gil16RSDbispect}), which is presented in Table \ref{tab:drcomparison} 
and denoted as G-M et al. (2016 a+b+c). The combination of these three sets of results is presented at the end of \cite{Gil16RSDbispect}.
As before, this case is compared to our full-shape column of Table~\ref{tab:combined_all},
approximating LOWZ to our low redshift bin and CMASS to our high 
redshift bin, where the volume difference factor has been taken into account. Our $\DM$ measurement of 1.7\% in the low 
redshift bin and 1.8\% in the high redshift bin compares to 
1.5\% and 1.1\%, respectively, in Gil-Mar\'in 2016 a+b+c. 
Regarding $H(z)$, our measurement of 2.8\% in both the low and high redshift bins 
compares to 2.5\% and 1.8\% in Gil-Mar\'in 2016 a+b+c. 
Finally our $f\sigma_8$ constraint of 9.5\% and 8.9\%
in the low and high redshift bin compares to the LOWZ and CMASS measurements of 
9.2\% and 6.0\% by Gil-Marin 2016a+b+c. 
One can attribute the improvement in Gil-Mar\'in 2016a+b+c when compared to our measurement 
to the use of the bispectrum, which has not been used in our analysis.

%\section{Results from the BOSS combined sample}
%\label{sec:measurements}
%\input{measurements}

\section{Cosmological parameters}
\label{sec:cosmology}
\subsection{Data sets}

We now turn to cosmological interpretation of our results.  We will
use the consensus measurements, including our estimated systematic
error contribution to the covariance matrix,
from the BAO-only and BAO+FS columns of
Table \ref{tab:combined_BAO}.  In our subsequent figures
and tables, the former case is simply labeled ``BAO.''

Following \citet{Aubourg},
we include the 6dFGS and SDSS MGS BAO measurements and 
the BOSS DR11 \lya\ forest BAO measurements
(see Fig.~\ref{fig:distancescale} and \S\ref{sec:comparison_to_previous}).
These 
are largely independent and have utilized similar methodologies.
We opt not to include other BAO measurements, notably those
from photometric clustering and from the WiggleZ
survey \citep{Blake11,Blake12}, as the volumes partially overlap BOSS  and
the errors are sufficiently large that a proper inclusion would not
substantially affect the results.
As shown in \citet{Aubourg}, these 
measurements are in good agreement with those from BOSS.  
We note in particular the good match to the WiggleZ results, 
as this was a sample
of strongly star-forming galaxies in marked contrast to the 
red massive galaxies used in BOSS.  The dual-tracer opportunity was studied extensively
with a joint analysis of the overlap region of WiggleZ 
and BOSS \citep{Beutler16}.

We further opt not to include other RSD measurements beyond BOSS, 
as they come from a variety of analysis and modelling approaches.
One can see from Figure \ref{fig:fsig8_other} that the measurements from
other surveys are consistent with those from BOSS within their
quoted errors, and the error bars in all cases are large enough
that there are potential gains from combining multiple measurements.
However, in contrast to BAO measurements, systematic errors associated
with non-linear clustering and galaxy bias are a major component
of the error budget in any RSD analysis, and these systematics
may well be covariant from one analysis to another in a way that
is difficult to quantify.  Because of systematic error contributions,
we do not consider it feasible to carry out a robust joint RSD
analysis with other measurements.

In all cases, we combine with CMB anisotropy data from the Planck
2015 release \citep{Planck2015}.  
We use the power spectra for both temperature and polarization; in detail,
we use the likelihoods 
plik\_dx11dr2\_HM\_v18\_TTTEEE and lowTEB for the high and low multipoles, respectively.
We do not include the information from the lensing of the CMB in the
4-point correlations of the CMB temperature anisotropies.  We will
discuss the impact of the recent \citep{Planck2016} large-angle polarization 
results in \S \ref{sec:growth}.

We note that there is some mild tension between the Planck 2015
results and those from combining WMAP, SPT, and ACT \citep{Calabrese13,Spergel15,Bennett16}.
The Planck data set yields a mildly higher matter density $\Omega_mh^2$, which for
$\Lambda$CDM implies a higher $\Omega_m$ and $\sigma_8$ and a lower $H_0$.
As in the DR11 results, our BOSS results for $\Lambda$CDM fall in 
between these two and therefore do not prefer either CMB option.
We have presented non-Planck results in \citet{Anderson2014b} and \citet{Aubourg}
and do not repeat that here, as the sense of the differences has not changed.

Finally, for some cases, we utilize measurements of the distance-redshift
relation
from Type Ia supernovae (SNe) from the Joint Lightcurve Analysis 
\citep[JLA, ][]{JLA}, which combined
SNe from the SDSS-II Supernova Survey \citep{Sako14}
and the Supernova Legacy Survey 3-year data set \citep{Conley11}
together with local and high-$z$ data sets.
The combination of SN measurements with BAO is
particularly powerful for constraining the low-redshift distance
scale \citep[e.g.,][]{Mehta12,Anderson2014b}.  The SNe provide a higher precision measurement
of relative distance at lower redshift where the BAO is limited by
cosmic volume, but the BAO provides an absolute scale that connects
to higher redshift and particularly to the CMB acoustic scale at
$z=1000$.
The combination of BAO and SN data also allows an ``inverse distance ladder''
measurement of $H_0$ that uses the CMB-based calibration of $\rd$ but
is almost entirely insensitive to the dark energy model and space
curvature over the range allowed by observations \citep{Aubourg}.

\subsection{Cosmological Parameter Results: Dark Energy and Curvature}

\begin{figure*}
\begin{minipage}{7in}
\includegraphics[width=2.3in]{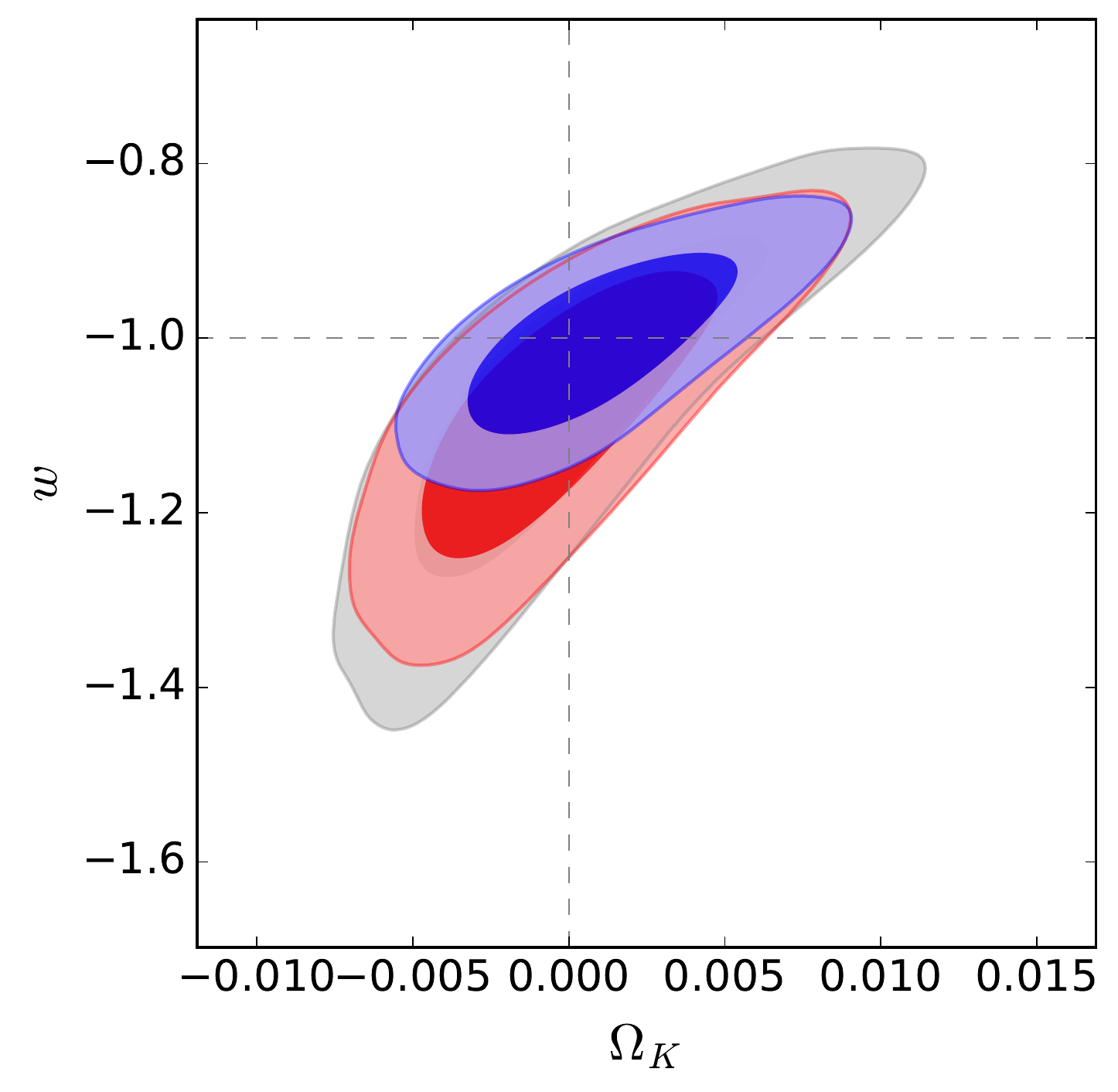}
\includegraphics[width=2.2in]{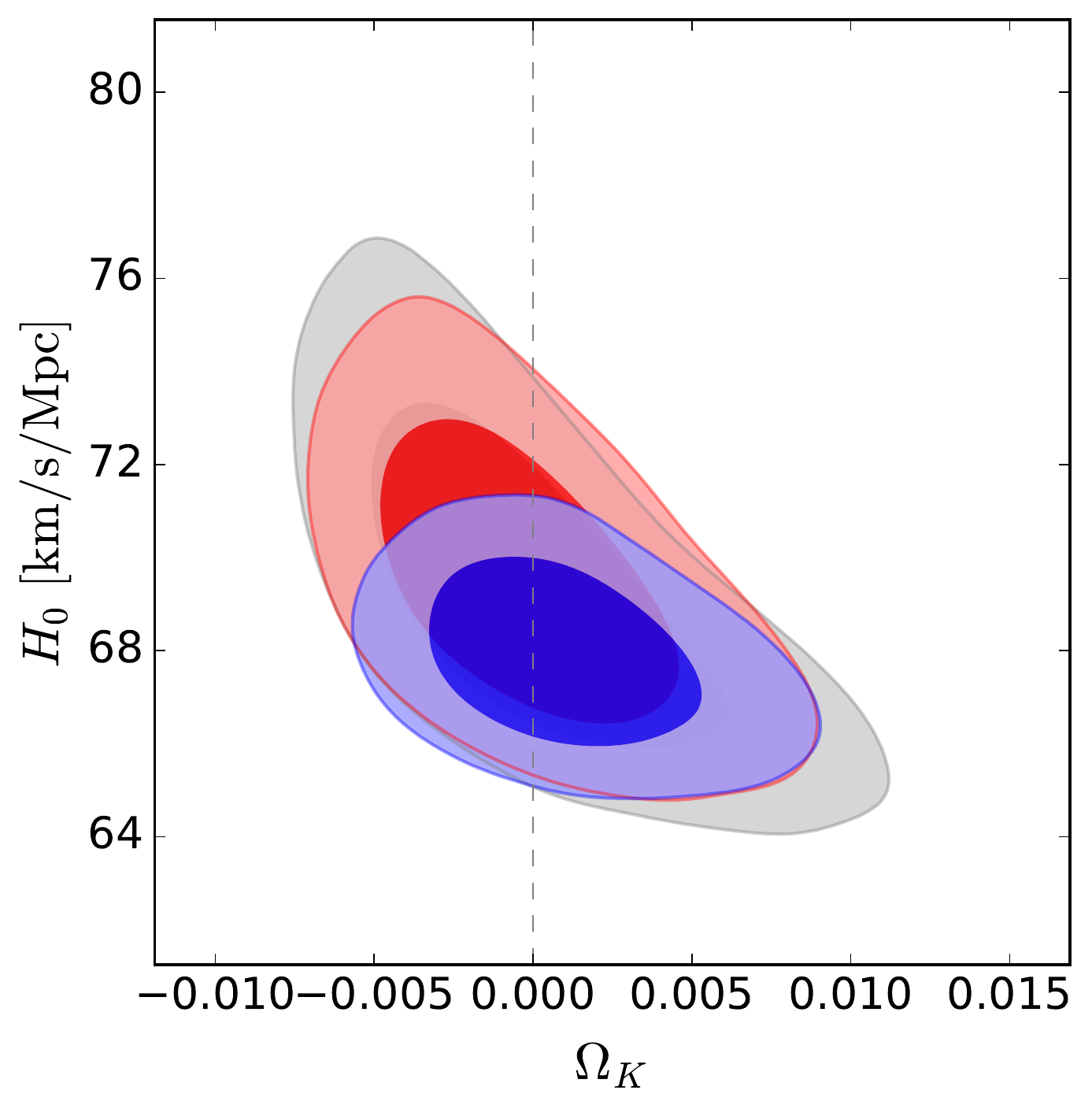}
\includegraphics[width=2.2in]{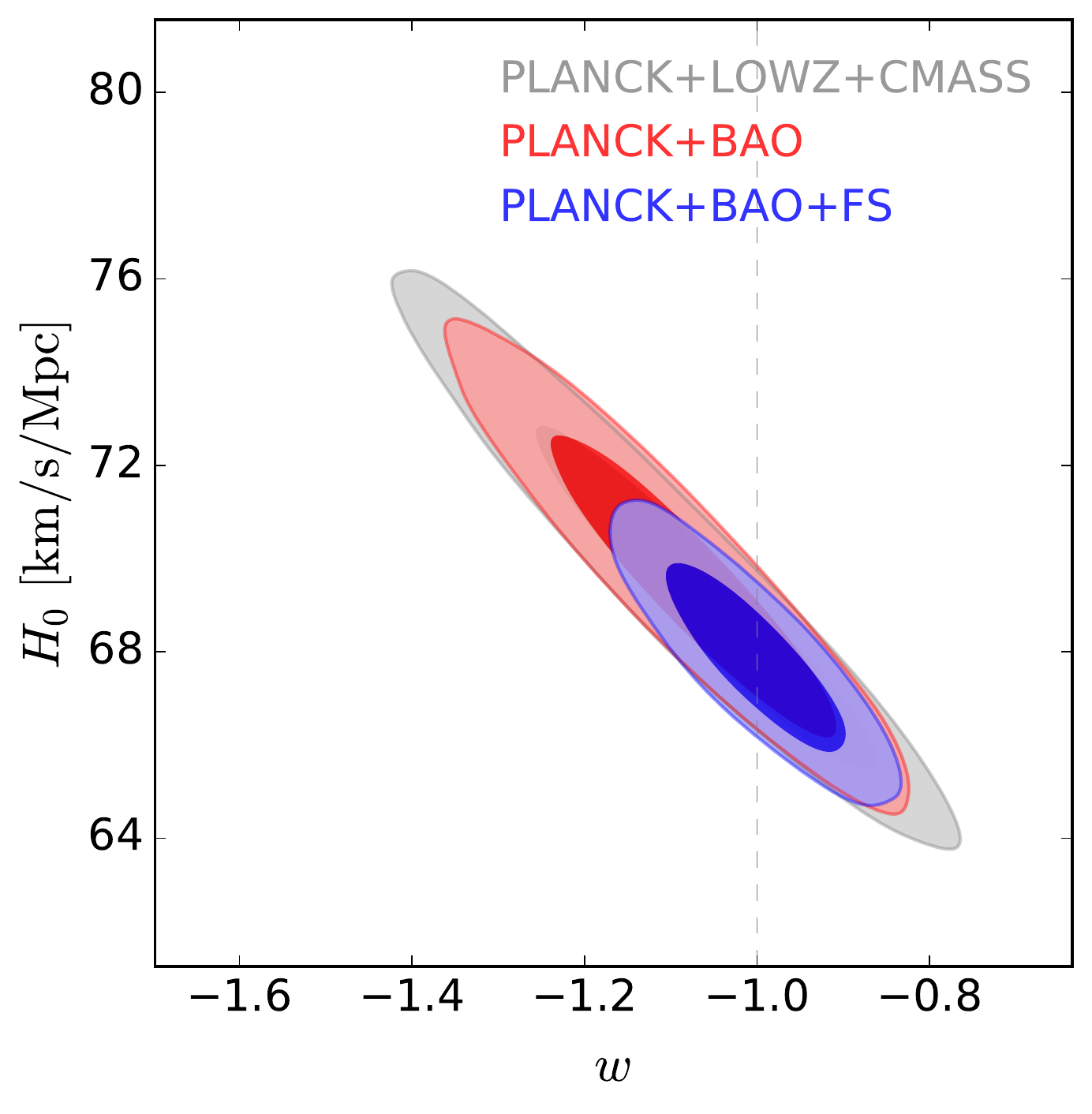}
  \caption{Parameter constraints for the $ow$CDM cosmological model, comparing the BAO and
  BAO+FS results from this paper as well as the DR12 LOWZ+CMASS results from \citet{Cuesta16}.
  One sees that adding a 3rd redshift bin has improved the constraints somewhat, but full-shape information, especially the constraint on $H(z)\DM(z)$ from
  the Alcock-Paczynski effect on sub-BAO scales, sharpens
  constraints substantially. }
  \label{fig:comparison_lowz}
  \end{minipage}
\end{figure*}

\begin{figure*}
\begin{minipage}{7in}
\includegraphics[width=3in]{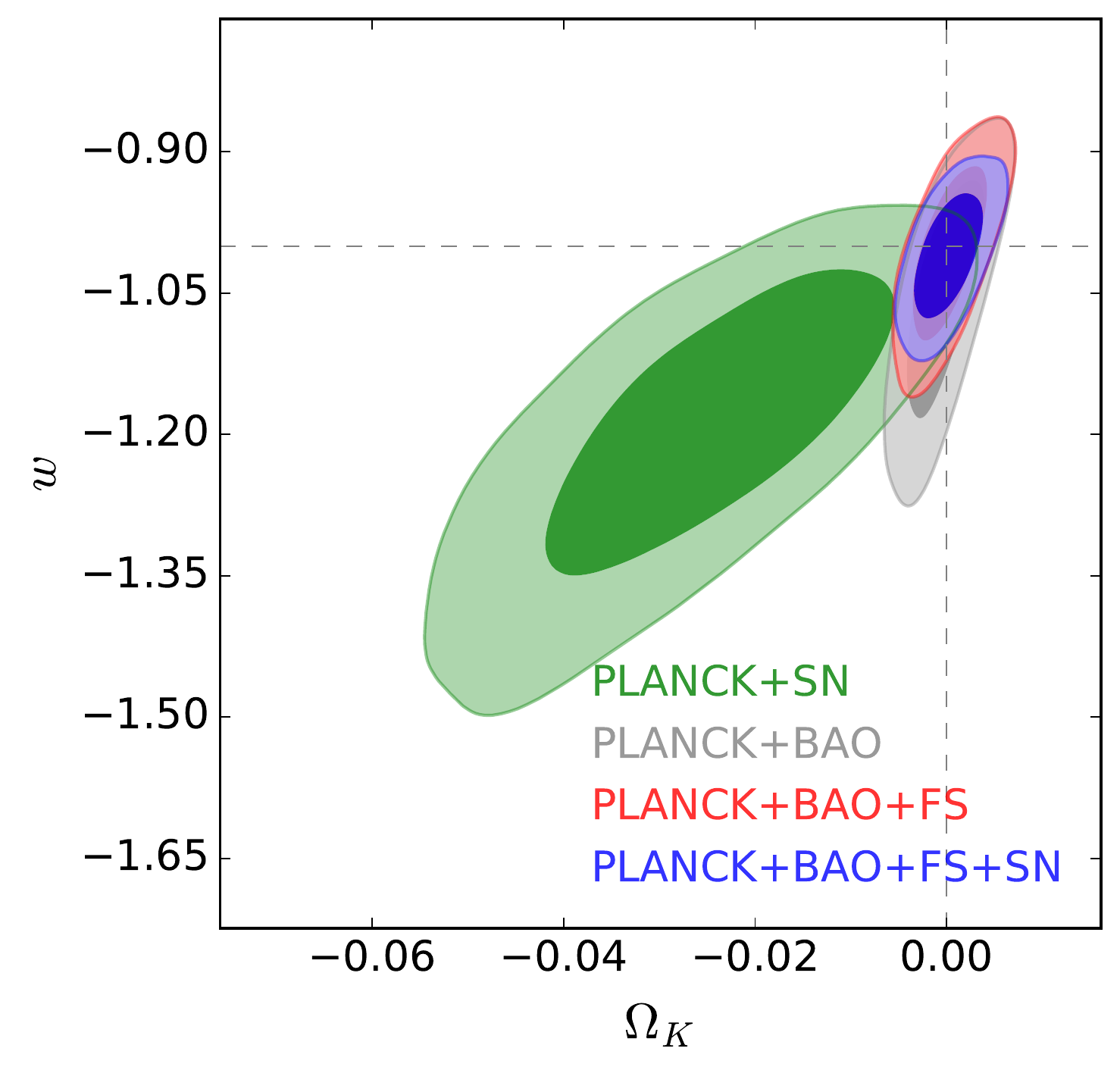}
\includegraphics[width=3in]{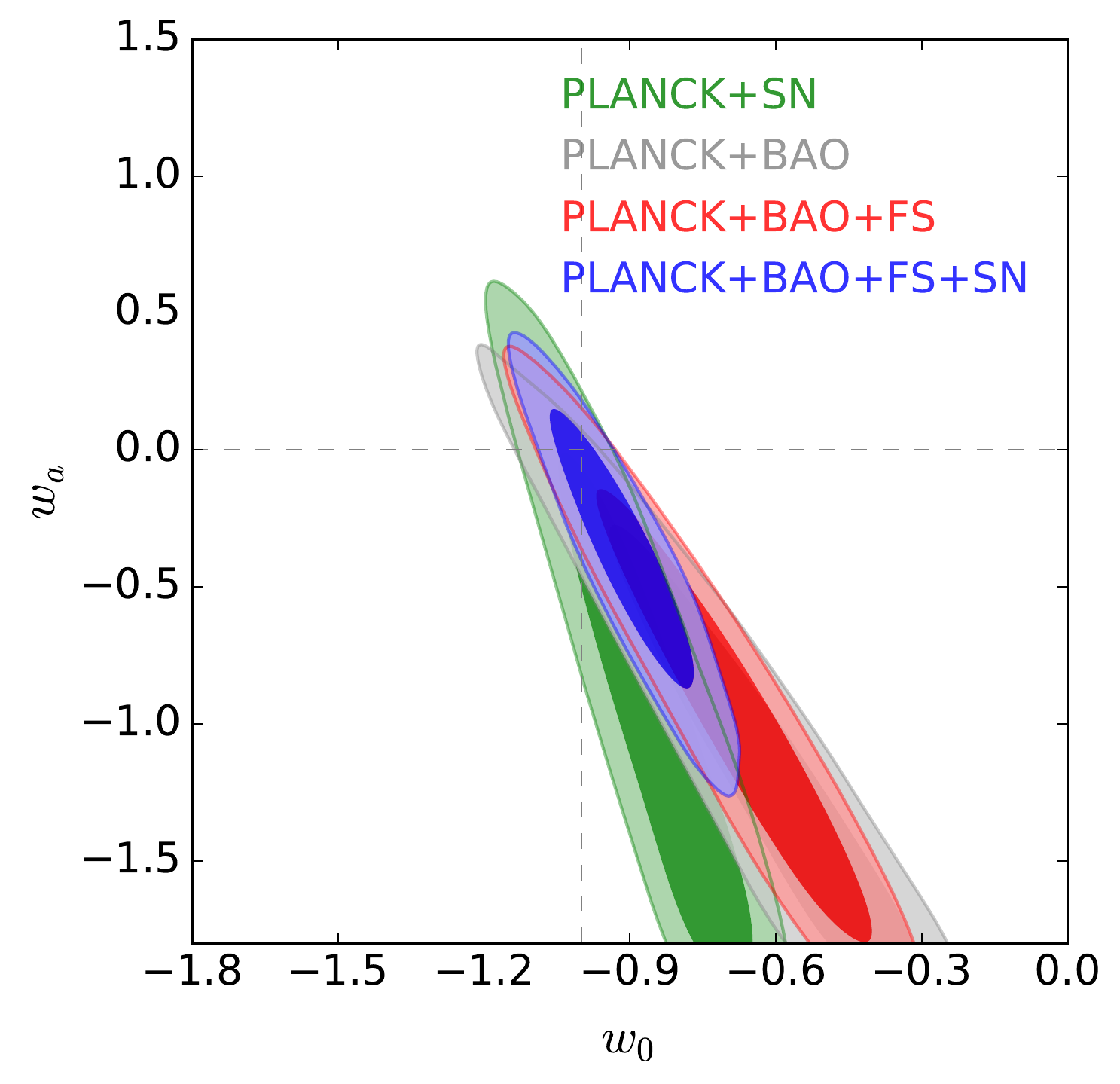}
  \caption{Parameter constraints for the $ow$CDM (left) and $w_0w_a$CDM (right) cosmological models,
comparing the results from BAO and BAO+FS to those with JLA SNe.  One sees that the galaxy
clustering results are particularly strong in the $\Omega_K$--$w$ space and
are comparable to the SNe in the $w_0$--$w_a$ space.
  }
  \label{fig:sncomp}
  \end{minipage}
\end{figure*}

\addtolength{\tabcolsep}{-2pt}
\begin{table*}
\centering
\begin{tabular}{p{1.800000cm}p{3.300000cm}p{1.600000cm}p{1.500000cm}p{1.600000cm}p{1.900000cm}p{1.600000cm}p{1.600000cm}}
\hline
Cosmological  & Data Sets & $\Omega_{\rm m} h^{2}$&$\Omega_{\rm m}$&$H_0$&$\Omega_{\rm K}$&$w_0$&$w_a$ \\
Model         &           & &&km/s/Mpc&&& \\
\hline
$\Lambda$CDM&Planck&$0.1429\ (14)$&$0.317\ (9)$&$67.2\ (7)$&...&...&...                     \\
$\Lambda$CDM&Planck + BAO&$0.1418\ (10)$&$0.309\ (6)$&$67.7\ (5)$&...&...&...                     \\
$\Lambda$CDM&Planck + BAO + FS&$0.1419\ (10)$&$0.311\ (6)$&$67.6\ (5)$&...&...&...                     \\
$\Lambda$CDM&Planck + BAO + FS + SN&$0.1419\ (10)$&$0.310\ (6)$&$67.6\ (5)$&...&...&...                     \\
\hline
$o$CDM&Planck + BAO&$0.1422\ (14)$&$0.309\ (7)$&$67.9\ (7)$&$+0.0007\ (20)$&...&...                     \\
$o$CDM&Planck + BAO + FS&$0.1422\ (14)$&$0.310\ (6)$&$67.7\ (6)$&$+0.0004\ (20)$&...&...                     \\
$o$CDM&Planck + BAO + FS + SN&$0.1421\ (14)$&$0.310\ (6)$&$67.8\ (6)$&$+0.0005\ (20)$&...&...                     \\
\hline
$w$CDM&Planck + BAO&$0.1424\ (13)$&$0.302\ (12)$&$68.8\ (14)$&...&$-1.05\ (6)$&...                     \\
$w$CDM&Planck + BAO + FS&$0.1421\ (11)$&$0.309\ (10)$&$67.9\ (12)$&...&$-1.01\ (5)$&...                     \\
$w$CDM&Planck + BAO + FS + SN&$0.1420\ (11)$&$0.308\ (9)$&$67.9\ (9)$&...&$-1.01\ (4)$&...                     \\
\hline
$ow$CDM&Planck + SN&$0.1418\ (14)$&$0.379\ (37)$&$61.4\ (31)$&$-0.0252\ (121)$&$-1.19\ (11)$&...                     \\
$ow$CDM&Planck + BAO&$0.1423\ (14)$&$0.301\ (14)$&$68.8\ (16)$&$-0.0003\ (27)$&$-1.05\ (8)$&...                     \\
$ow$CDM&Planck + BAO + FS&$0.1421\ (14)$&$0.310\ (11)$&$67.8\ (12)$&$+0.0003\ (26)$&$-1.01\ (6)$&...                     \\
$ow$CDM&Planck + BAO + FS + SN&$0.1421\ (14)$&$0.309\ (9)$&$67.9\ (9)$&$+0.0002\ (23)$&$-1.01\ (4)$&...                     \\
\hline
$w_0w_a$CDM&Planck + SN&$0.1428\ (14)$&$0.294\ (16)$&$69.8\ (18)$&...&$-0.85\ (13)$&$-0.99\ (63)$                     \\
$w_0w_a$CDM&Planck + BAO&$0.1427\ (11)$&$0.336\ (21)$&$65.2\ (21)$&...&$-0.63\ (20)$&$-1.16\ (55)$                     \\
$w_0w_a$CDM&Planck + BAO + FS&$0.1427\ (11)$&$0.334\ (18)$&$65.5\ (17)$&...&$-0.68\ (18)$&$-0.98\ (53)$                     \\
$w_0w_a$CDM&Planck + BAO + FS + SN&$0.1426\ (11)$&$0.313\ (9)$&$67.5\ (10)$&...&$-0.91\ (10)$&$-0.39\ (34)$                     \\
\hline
$ow_0w_a$CDM&Planck + BAO&$0.1422\ (14)$&$0.331\ (21)$&$65.6\ (21)$&$-0.0022\ (30)$&$-0.66\ (19)$&$-1.22\ (53)$                     \\
$ow_0w_a$CDM&Planck + BAO + FS&$0.1422\ (14)$&$0.333\ (16)$&$65.4\ (16)$&$-0.0020\ (28)$&$-0.67\ (18)$&$-1.12\ (59)$                     \\
$ow_0w_a$CDM&Planck + BAO + FS + SN&$0.1420\ (14)$&$0.314\ (10)$&$67.3\ (10)$&$-0.0023\ (28)$&$-0.87\ (11)$&$-0.63\ (45)$                     \\
\hline
\hline
\end{tabular}

\addtolength{\tabcolsep}{2pt}
\caption{Cosmological constraints for models varying the expansion history because
of spatial curvature and/or evolving dark energy.  $o$CDM varies the spatial curvature,
$w$CDM allows a constant equation of state of dark energy, 
and $w_0w_a$CDM allows a time-evolving $w(a) = w_0+(1-a)w_a$.
The models $ow$CDM and $ow_0w_a$CDM combine these factors.  All errors are 1$\sigma$ rms from
our Markov chains.
} 
\label{tab:DR12}
\end{table*}

We now use these results to constrain parametrized cosmological
models.  We will do this using Markov Chain Monte Carlo, following 
procedures similar to those described in \citet{Aubourg}, but due to
use of the full power spectrum shape data we do not run any chains
using that paper's simplified ``background evolution only'' code. Instead, we
calculate all our chains using the July 2015 version of the workhorse
\textsc{CosmoMC} code \citep{cosmomc}. The code was minimally
modified to add the latest galaxy data points and their covariance, the
\lya\ BAO datasets, and two optional
$A_{f\sigma_8}$ and $B_{f\sigma_8}$ parameters described later in the
text.  We use a minimal neutrino sector, with one species with a mass
of 0.06 eV/$c^2$ and two massless, corresponding to the lightest
possible sum of neutrino masses consistent with atmospheric and solar
oscillation experiments \citep{T2K,MINOS,KAMLAND}, unless otherwise mentioned.

We first consider models that vary the cosmological distance scale with 
spatial curvature or parametrizations of the dark energy equation of
state via $w(a) = w_0 + w_a(1-a)$ \citep{Chevallier01,Linder03}.  
These results are 
shown in Table \ref{tab:DR12} for various combinations of measurements.
In all cases, the table shows the mean and 1$\sigma$ error, marginalized
over other parameters.  Of course, some parameters are covariant,
as illustrated by contours in some of 
our figures.  Our model spaces always 
include variations in the matter density $\Omega_m h^2$, 
the baryon density $\Omega_b h^2$, the 
amplitude and spectral index of the primordial spectrum, and the 
optical depth to recombination.  However, we do not show results for
these parameters as they are heavily dominated by the CMB and are not
the focus of our low-redshift investigations.

We begin with the standard cosmology, the $\Lambda$CDM model, which
includes a flat Universe with a cosmological constant and cold dark
matter.  As is well known, CMB anisotropy data alone can constrain this
model well: the acoustic peaks imply the baryon and matter density, and
thereby the sound horizon, allowing the acoustic peak to determine the
angular diameter distance to recombination, which in turn breaks the 
degeneracy between $\Omega_m$ and $H_0$ \citep[e.g.,][]{WMAP}.  The Planck 2015
measurements do this exquisitely well, 
yielding $\Omega_mh^2 = 0.1429\pm 0.0014$,
$\Omega_m = 0.317 \pm 0.009$, 
and $H_0 = 67.2\pm 0.7\hubunits$ \citep{Planck2015}.

As shown already in Figures \ref{fig:combined_contours} and
\ref{fig:combined_redshift}, our BOSS measurements are fully
consistent with the Planck $\Lambda$CDM model results.  The
$\Lambda$CDM predictions from the Planck model fits for our distance
and growth observables match our measurements well, typically within
1$\sigma$. The combined Lyman-$\alpha$ data do deviate at 
the $2-2.5 \sigma$ level, 
which has been extensively discussed in literature 
\citep{Aubourg,Delubac,Font14,Sahni14}, 
but the overall $\chi^2$ is consistent with a minimal model.  As such, the BOSS
data do not require more complicated cosmologies.

As was seen in \citet{Anderson2014b} and \citet{Planck2015}, the addition 
of BOSS clustering data to the Planck results for the minimal $\Lambda$CDM model
does further improve the constraints on cosmological parameters.
In particular, we find $\Omega_mh^2 = 0.1419 \pm 0.0010$ (0.6 per cent), 
$\Omega_m = 0.311\pm0.006$,
and $H_0 = 67.6 \pm 0.5\hubunits$ (0.6 per cent).  
Adding the JLA SNe data does not
further improve the errors.

We next turn to extensions that affect the distance scale, notably spatial
curvature and variations in the dark energy density.  In these cases, the
most precise aspects of the CMB data sets suffer from a geometrical degeneracy:
the CMB determines the angular diameter distance to recombination very accurately,
but models that trade off low-redshift behaviour while holding this quantity
fixed are more difficult to distinguish.  The latest CMB data, such as from
Planck, does offer ways to break the geometrical degeneracy, most effectively
with gravitational lensing of the CMB, but the measurement of the 
low-redshift distance
scale with BAO and the Alcock-Paczynski effect offers a more direct route.

As reported in Table \ref{tab:DR12}, the BOSS data do this very well.
Combining Planck and BOSS for the non-flat model with a cosmological
constant yields a spatial curvature measurement of \upd{$\Omega_K = 0.0004
\pm 0.0020$}, confirming flatness at the $10^{-3}$ level.
Similarly, for the flat model with a constant dark energy equation of
state, we measure a value 
$w= -1.01 \pm 0.05$, 
highly consistent with
the cosmological constant.  Opening both of these parameters yields a
joint measurement of \upd{$\Omega_K = 0.0003\pm 0.0026$ and $w=-1.01\pm
0.06$}.  Focusing only on BAO, excluding the full-shape information,
degrades this to \upd{$\Omega_K = -0.0003\pm 0.0027$ and $w=-1.05 \pm
0.08$}.  

We stress that this consistency is a stringent test 
of the cosmological standard model.
The clustering of galaxies is based on the same underlying physics as that of
the CMB anisotropies.  This is most obvious for the acoustic scale, but it is
also true of the broadband power.  We are now measuring the imprints of this physics
over a wide range of redshifts, including at recombination, and finding a cosmic
distance scale that returns the simple, flat, cosmological constant model while 
opening not one but two new degrees of freedom.  
The 6 per cent measurement of $w$ 
is a compelling demonstration of the power of galaxy clustering to measure dark
energy and is an excellent counterpart to the 
dark energy evidence from supernovae.

Beyond this, one can consider more complicated dark energy models.  
However, current
data do not constrain these tightly.
We use here the common $w(a) = w_0 + (1-a) w_a$ model.
For the case with non-flat curvature, this is the fitting space for the 
Dark Energy Task Force Figure of Merit \citep{Albrecht06}.  
We continue to find superb agreement
with a flat Universe, with only \upd{0.0028} 
errors on $\Omega_K$ with or without
inclusion of SNe.  
Including the SNe data does sharpen the dark energy constraints,
and we find a \upd{0.45} error on $w_a$.  
While this constraint still allows order unity change in $w$ over
$\Delta z \approx 1$,
it is one of the strongest limits (perhaps the strongest) yet obtained
on equation-of-state evolution.

Because of the permissive limits on evolution, the errors on $w$ at
$z=0$ in these models are correspondingly worse.  However, there is an
intermediate ``pivot'' redshift $z_p$ where the errors on $w$ are
minimized and where the covariance between $w(z)$ and $w_a$ disappears
\citep{Albrecht06}.  For the combination Planck+BAO+FS+SN in the
$w_0w_a$CDM model, we find that the pivot redshift \upd{$z_p= 0.37$ and
$w(z_p) = -1.05 \pm 0.05$}.  For $ow_0w_a$CDM, we find \upd{$z_p=0.29$ and
$w(z_p) = -1.05 \pm 0.06$}.  We conclude that the current combination
of data is able to say that Universe is flat at the $10^{-3}$ level and
that the dark energy was within $\Delta w\sim 10^{-1}$ 
of a cosmological constant at some epoch in the fairly recent past,
but our knowledge of $w(z)$ remains limited.

Using our constraints 
to compute the Dark Energy Task Force Figure of Merit 
\citep{Albrecht06}, we find a result of 32.6 with SNe and 22.9 without SNe
for $[\sigma(w(z_p))\sigma(w_a)]^{-1}$.
For comparison, the BAO-only analysis of \citet{Cuesta16} found 
a Figure of Merit of 24.3 with Planck+BAO+SN and only 8.3 without SNe.
These comparisons show that
the present analysis with 3 redshift bins and including
the full-shape fits has notably improved the Figure of Merit.
If we construct the Figure of Merit while assuming flatness (and thereby 
different from the Dark Energy Task Force), we find 75.4 with SNe and 44.6 
without.

Our results are consistent with the distance-redshift relation from
the JLA SNe.  For example, adding the SNe does not significantly alter the best-fit
model parameters.  But the errors on a constant $w$ do continue to improve,
to \upd{0.04} in both the flat case and in the joint fit with
curvature.  Figure \ref{fig:sncomp} shows a comparison for both $ow$CDM and
$w_0w_a$CDM for galaxy clustering and SN results separately as well as the combination.
We see that in $ow$CDM, the dark energy constraints even without SNe are now very
tight, but the SN results are consistent and decrease the errors.
It is notable that the two data sets have sharply different degeneracy directions and
therefore will continue to be good partners in our cosmological constraints.
For $w_0w_a$CDM, the two data sets are of more comparable power, again with different
degeneracy directions, so that the combination is substantially tighter.

\subsection{Cosmological Parameter Results: Dark Radiation}

\begin{table*}
\centering
\begin{tabular}{p{2.000000cm}p{3.600000cm}p{1.600000cm}p{1.600000cm}p{1.600000cm}p{1.600000cm}p{1.600000cm}p{1.600000cm}}
\hline
Cosmological  & Data Sets & $\Omega_{\rm m} h^{2}$&$\Omega_{\rm m}$&$H_0$&$\Omega_{\rm K}$&$w_0$&$N_{\rm eff}$ \\
Model         &           & &&km/s/Mpc&&& \\
\hline
$\Lambda$CDM + $N_{\rm eff}$&Planck&$0.1418\ (32)$&$0.320\ (12)$&$66.6\ (16)$&...&...&$2.97\ (20)$                     \\
$\Lambda$CDM + $N_{\rm eff}$&Planck + BAO + FS&$0.1418\ (32)$&$0.311\ (7)$&$67.5\ (12)$&...&...&$3.03\ (18)$                     \\
$\Lambda$CDM + $N_{\rm eff}$&Planck + BAO + FS + $H_0$&$0.1452\ (28)$&$0.302\ (6)$&$69.3\ (10)$&...&...&$3.28\ (16)$                     \\
\hline
$ow$CDM + $N_{\rm eff}$&Planck + BAO + FS&$0.1418\ (34)$&$0.311\ (11)$&$67.6\ (15)$&$+0.0006\ (30)$&$-1.00\ (6)$&$3.02\ (23)$                     \\
$ow$CDM + $N_{\rm eff}$&Planck + BAO + FS + SN&$0.1417\ (32)$&$0.308\ (9)$&$67.8\ (12)$&$+0.0003\ (23)$&$-1.01\ (5)$&$3.02\ (21)$                     \\
$ow$CDM + $N_{\rm eff}$&Planck + BAO + FS + SN + $H_0$&$0.1446\ (30)$&$0.299\ (8)$&$69.5\ (11)$&$-0.0003\ (23)$&$-1.03\ (4)$&$3.22\ (19)$                     \\
\hline
\hline
\end{tabular}

\caption{Cosmological constraints for models varying amount of relativistic energy
density, as parametrized by the effective number of neutrino species $N_{\rm eff}$.
We consider both the $\Lambda$CDM and $ow$CDM case.
All errors are 1$\sigma$ rms from our Markov chains.
}
\label{tab:DR12b}
\end{table*}

\begin{table*}
\centering
\begin{tabular}{p{3.300000cm}p{2.900000cm}p{1.500000cm}p{1.500000cm}p{1.500000cm}p{1.500000cm}p{1.500000cm}p{1.500000cm}}
\hline
Cosmological  & Data Sets & $\sum m_{\nu}$~[eV/$c^2$]&$A_L$&$A_{f\sigma_8}$&$B_{f\sigma_8}$&$\sigma_8$&$\Omega_{\rm m}^{0.5}~\sigma_8$ \\
Model         &           & 95\% limit&&&&& \\
\hline
$\Lambda$CDM + $m_{\nu}$&Planck + BAO + FS&$<$ 0.16&...&...&...&$0.829\ (16)$&$0.462\ (9)$                     \\
$\Lambda$CDM + $m_{\nu}$ + $A_L$&Planck + BAO + FS&$<$ 0.23&$1.19\ (8)$&...&...&$0.795\ (22)$&$0.441\ (12)$                     \\
$\Lambda$CDM + $m_{\nu}$ + $A_{f\sigma_8}$&Planck + BAO + FS&$<$ 0.15&...&$0.96\ (6)$&...&$0.833\ (16)$&$0.464\ (9)$                     \\
$\Lambda$CDM + $m_{\nu}$ + $A_L$ + $A_{f\sigma_8}$&Planck + BAO + FS&$<$ 0.25&$1.19\ (8)$&$1.00\ (7)$&...&$0.793\ (25)$&$0.440\ (14)$                     \\
\hline
$\Lambda$CDM + $A_{f\sigma_8}$&Planck + BAO + FS&...&...&$0.96\ (6)$&...&$0.833\ (13)$&$0.464\ (9)$                     \\
$\Lambda$CDM + $A_{f\sigma_8}$ + $B_{f\sigma_8}$&Planck + BAO + FS&...&...&$0.97\ (6)$&$-0.62\ (40)$&$0.832\ (13)$&$0.463\ (9)$                     \\
\hline
$ow$CDM + $m_{\nu}$&Planck + BAO + FS + SN&$<$ 0.31&...&...&...&$0.826\ (21)$&$0.459\ (11)$                     \\
$ow$CDM + $A_{f\sigma_8}$ + $B_{f\sigma_8}$&Planck + BAO + FS + SN&...&...&$0.96\ (6)$&$-0.60\ (39)$&$0.840\ (18)$&$0.464\ (9)$                     \\
\hline
\hline
\end{tabular}

\caption{Cosmological constraints for models varying the neutrino mass or
allowing a modification of the growth rate.  The parameters $A_L$, $A_{f\sigma_8}$,
and $B_{f\sigma_8}$ are described in the text; the notation +$A_L$ means
that this parameter has been varied, which means that the information from
CMB lensing has been decoupled from the rest of the cosmological parameter 
inference.  The model $+B_{f\sigma_8}$ also allows $A_{f\sigma_8}$ to vary.
All errors are 1$\sigma$ rms from our Markov chains, save that the neutrino
masses are given as 95 per cent upper limits.  We include 
$\Omega_m^{0.5}\sigma_8$ (evaluated at $z=0$) as this is a well-constrained parameter combination in
cluster abundance and lensing studies.
}
\label{tab:DR12c}
\end{table*}

\begin{figure}
%%% \begin{minipage}{7in}
\includegraphics[width=3in]{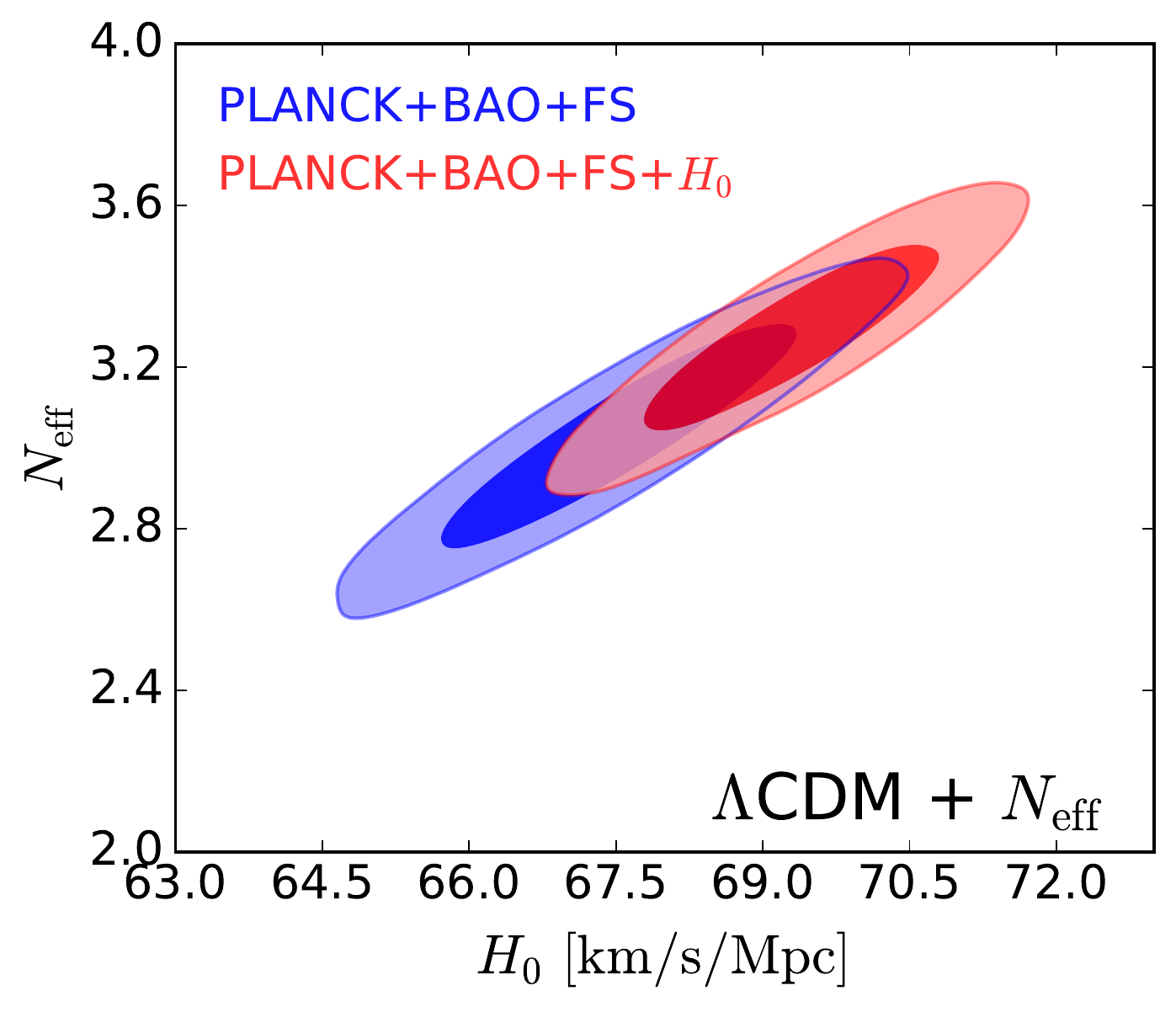}
\includegraphics[width=3in]{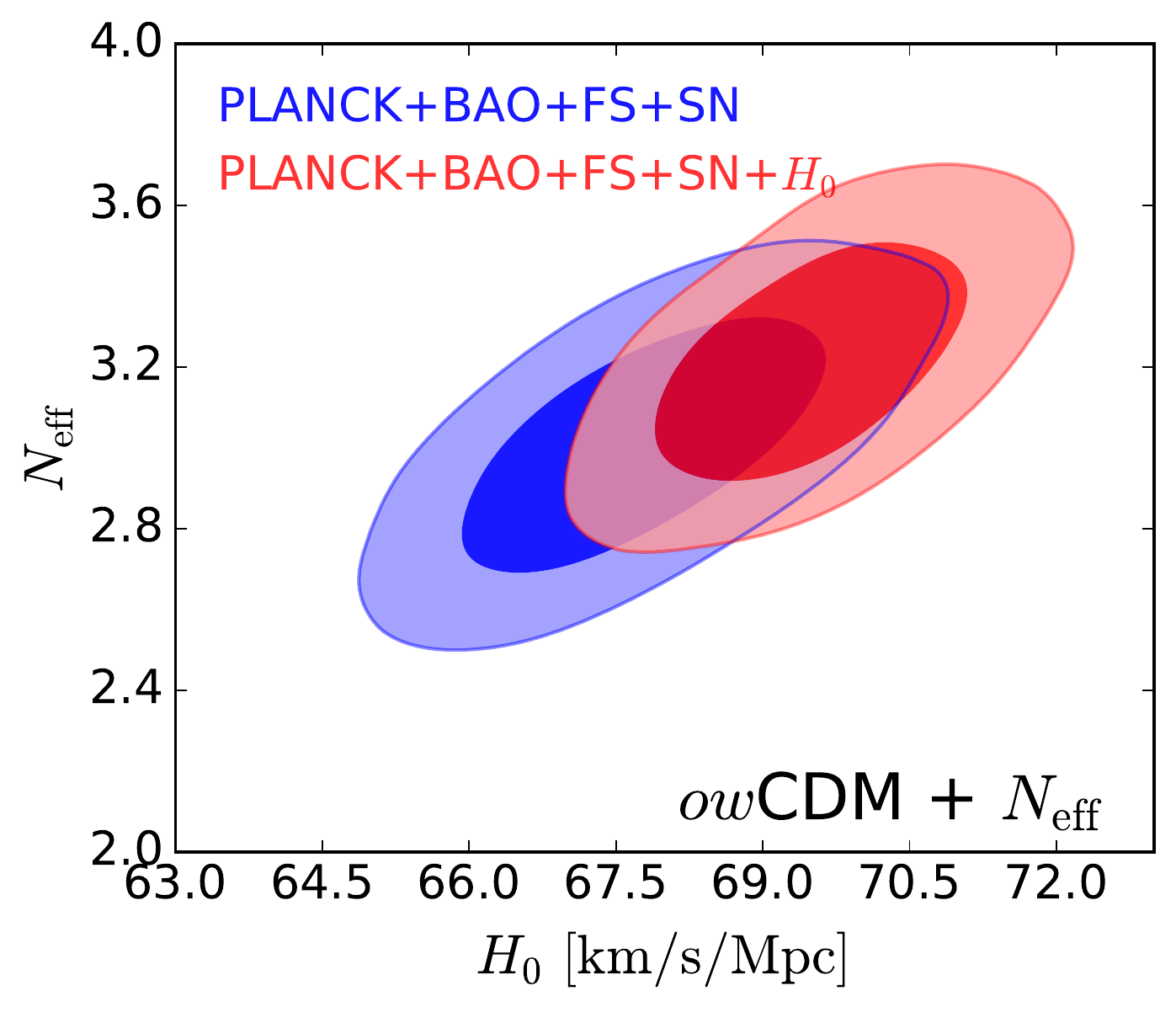}
  \caption{The constraints on $H_0$ and the relativistic energy density, parametrized
by $N_{\rm eff}$.  
(Top) Constraints for the $\Lambda$CDM parameter space using 
Planck+BAO+FS, with and without direct $H_0$ measurements.
(Bottom) Constraints for the $ow$CDM parameter space using 
Planck+BAO+FS+SNe, with and without direct $H_0$ measurements.
In both cases, the combination with the $H_0=73.0\pm1.8\hubunits$ measurement
of \citet{Riess16} causes a shift toward higher $N_{\rm eff}$ and higher $H_0$,
with $\chi^2$ rising by 8.
%%%  Some Chisq values.  $\Lambda$CDM$+N_{\rm eff}$: 12963.46 and 12971.48 (+HST); 
%%%  $ow$CDM$+N_{\rm eff}$: 13658.76 and  13666.82 (+HST).
 }
  \label{fig:darkradiation}
%%%   \end{minipage}
\end{figure}

We next consider models with variations in the relativistic energy density.
These are parametrized by $N_{\rm eff}$, the effective number of neutrino species.
Any new density above the 3.046 expected from standard model neutrino decoupling 
\citep{Mangano05}
is assumed to be a massless species, 
sometimes referred to as ``dark radiation''
(e.g., \citealt{Archidiacono11,Calabrese11}),
which may or may not result from the neutrino
sector
\citep{Steigman77,Seljak06,Ichikawa07,Mangano07}.  
Such models are important in BAO studies because the extra density in 
the early Universe results in a higher Hubble parameter before recombination,
which in turn produces a smaller sound horizon $r_d$.  Since the BAO method 
actually measures the ratio of distance to the sound horizon, this results in 
smaller inferred distances and larger low-redshift Hubble parameters 
\citep{Eisenstein04,Archidiacono11,Mehta12,Anderson2014b,Aubourg}.  This is
of substantial current interest because several high-precision direct measurements
of $H_0$ yield values about 10 per cent higher than that inferred from combinations of
Planck and BOSS BAO data \citep{riess11,Freedman12,Riess16}.

However, the Planck 2015 results appear to largely close the window 
for altering the sound horizon enough to reconcile the BAO+SN
``inverse distance ladder'' $H_0$ with these higher direct measurements.
\citep{Planck2015}.  The physics underlying this constraint is that
Silk damping \citep{Silk68} is a diffusion process whose lengthscale
depends on the square root of time, while the 
sound horizon depends linearly on time.  The amplitude of the small-angle
CMB fluctuations, when standardized by the angular acoustic scale, thereby
measures the Hubble parameter at recombination and thus constrains
$N_{\rm eff}$.

Table \ref{tab:DR12b} shows our parameter results for models 
with free $N_{\rm eff}$, for several model and data combinations.
Like \citet{Planck2015}, our chains for $\Lambda$CDM find tight constraints,
\upd{$N_{\rm eff} = 3.03\pm 0.18$}.  As this central value matches that of 
the 
standard model, the central values of $\Omega_m$ and $H_0$ move negligibly;
however, the error on $H_0$ with free $N_{\rm eff}$
increases from \upd{0.5 to $1.2\hubunits$}.  The error
on $\Omega_m$ increases only from \upd{0.006 to 0.007}, 
indicating that $N_{\rm eff}$
is primarily degenerate with $H_0$, not $\Omega_m$.  
Figure \ref{fig:darkradiation}
shows the covariance between $H_0$ and $N_{\rm eff}$.
If we add $N_{\rm eff}$ as a degree of freedom to the $ow$CDM model,
then constraints on $\Omega_K$ and $w$ are not
substantially affected, as one can see by comparing the $ow$CDM lines
in Tables~\ref{tab:DR12} and~\ref{tab:DR12b}.
If SNe are added as an observational constraint, then $ow$CDM
constraints on $N_{\rm eff}$ and $H_0$ remain tight, with
\upd{$N_{\rm eff}=3.02\pm0.21$ and $H_0 = 67.8\pm 1.2 \hubunits$}
(see Fig.~\ref{fig:darkradiation}, right).

\citet{Riess16} present a measurement of $H_0$ of 
$73.0\pm 1.8\hubunits$ (2.4 per cent), 
while \citet{Freedman12} find $74.3\pm2.1 \hubunits$ (2.8 per cent).
If we include the \cite{Riess16} measurement as a constraint in our
fits, then the preferred values of $H_0$ and $N_{\rm eff}$ shift
upward (see Table~\ref{tab:DR12b} and Fig.~\ref{fig:darkradiation}),
as one would expect given the disagreement between the
direct $H_0$ and the value inferred from Planck+BAO+FS+SN.
Addition of this one observation increases $\chi^2$ of
the best-fit model by 8.

In the most flexible dark energy model that we consider,
$ow_0w_a$CDM, with $N_{\rm eff}= 3.046$ we find
$H_0 = 67.3 \pm 1.0\hubunits$.  This can be taken as the
updated value of the ``inverse distance ladder'' $H_0$ measurement
assuming standard matter and radiation content
from \cite{Aubourg}. They obtained the same central value and a
$1.1\hubunits$ error bar using a flexible polynomial description of the low
redshift energy density with Planck 2013, DR11 BAO, and JLA SN data.

Our inference of $H_0$ rests on (1) the inference of the matter density
from the CMB, (2) the inference of the sound horizon from the CMB,
(3) the measurement of the BAO peak in the galaxy distribution, and
(4) the tracking of the expansion history from $z\sim 0$ to $z\sim 0.6$
with SNe.  The good agreement between Planck 2013 and 2015 parameter 
determinations argues that ingredient (1) is robust.
Pre-Planck CMB data implied somewhat lower values of the matter
density \citep{Calabrese13}, which would go in the direction of
reconciliation \citep{Bennett16}, but even this shift is small
if one includes BAO information in addition to CMB \citep{Anderson2014b}.
A substantial change in ingredient (2) appears less likely 
with the improved $N_{\rm eff}$ constraints from Planck 2015
discussed above.  Continued improvement in the measurement of
the CMB damping tail from larger aperture ground-based experiments
should clarify any remaining systematic concerns and tighten
the sound horizon constraints.
Regarding (3), as discussed throughout the paper,
we do not see a plausible way to systematically
shift the BAO measurement of $\DM/\rd$ at the several per cent level
that would be needed to substantially reduce the tension with 
the direct $H_0$ measurements.
There are some rather contrived possibilities, which would be
physically interesting in themselves, such as 
a well-tuned admixture of isocurvature perturbations that 
remains undetected in the CMB and yet affects fitting templates
enough to distort our distance measurements,
or a very large coupling of late-time galaxy bias to the
the relative baryon-CDM velocity field at high redshift, which 
has escaped our searches due to cancellation with a second 
unexpected effect.
Ingredient (4) is what makes our $H_0$ inference insensitive to the
assumed dark energy model, since the SNe provide an empirical 
measurement of the distance ratios needed to transfer our precise
BAO measurements at $z \sim 0.5$ down to $z=0$.
Our analysis includes the systematic error contributions to the
covariance matrix estimated by \cite{JLA}, and \cite{Aubourg}
show that similar results are obtained using the 
Union 2 SN compilation of \cite{Suzuki12}.
Our modeling adopts flexible but smooth parametric forms for
the evolution of dark energy density, and it is possible that
a model with more rapid low-redshift changes could shift
the value of $H_0$ while remaining consistent with the SN data.

It is also possible that systematic errors in the direct $H_0$
measurement are larger than estimated by \cite{Riess16}.
For example, \citet{Efstathiou13} presents an
alternative analysis of the local data, arguing for a lower value of
$70.6\pm3.3$ or $72.5\pm2.5 \hubunits$, depending on the choice
of primary standards. 
\citet{Rigault15} argue that the dependence of the supernova luminosity 
after correction for light-curve fitting on the host galaxy 
star-formation rate causes a net calibration offset between the SNe in 
the Hubble flow and those with nearby Cepheid measurements; they
find that this reduces $H_0$ by 3.3\% (but see discussion by 
\citealt{Riess16}).
It is also possible that everyone's error estimates are correct and we are
simply being unlucky, e.g., if the cosmologically inferred $H_0$ is $2\sigma$
low and the direct measurement is $2\sigma$ high.
For now, we continue to see this tension as provocative, but not conclusive.
Further work that tightens the statistical errors and examines
systematic uncertainties in direct $H_0$ measurements is clearly desirable,
as this tantalizing tension could yet reveal either astrophysical
or cosmological exotica.

\subsection{Cosmological Parameter Results: Growth of Structure}
\label{sec:growth}

\begin{figure}
%\begin{minipage}{7in}
\includegraphics[width=3in]{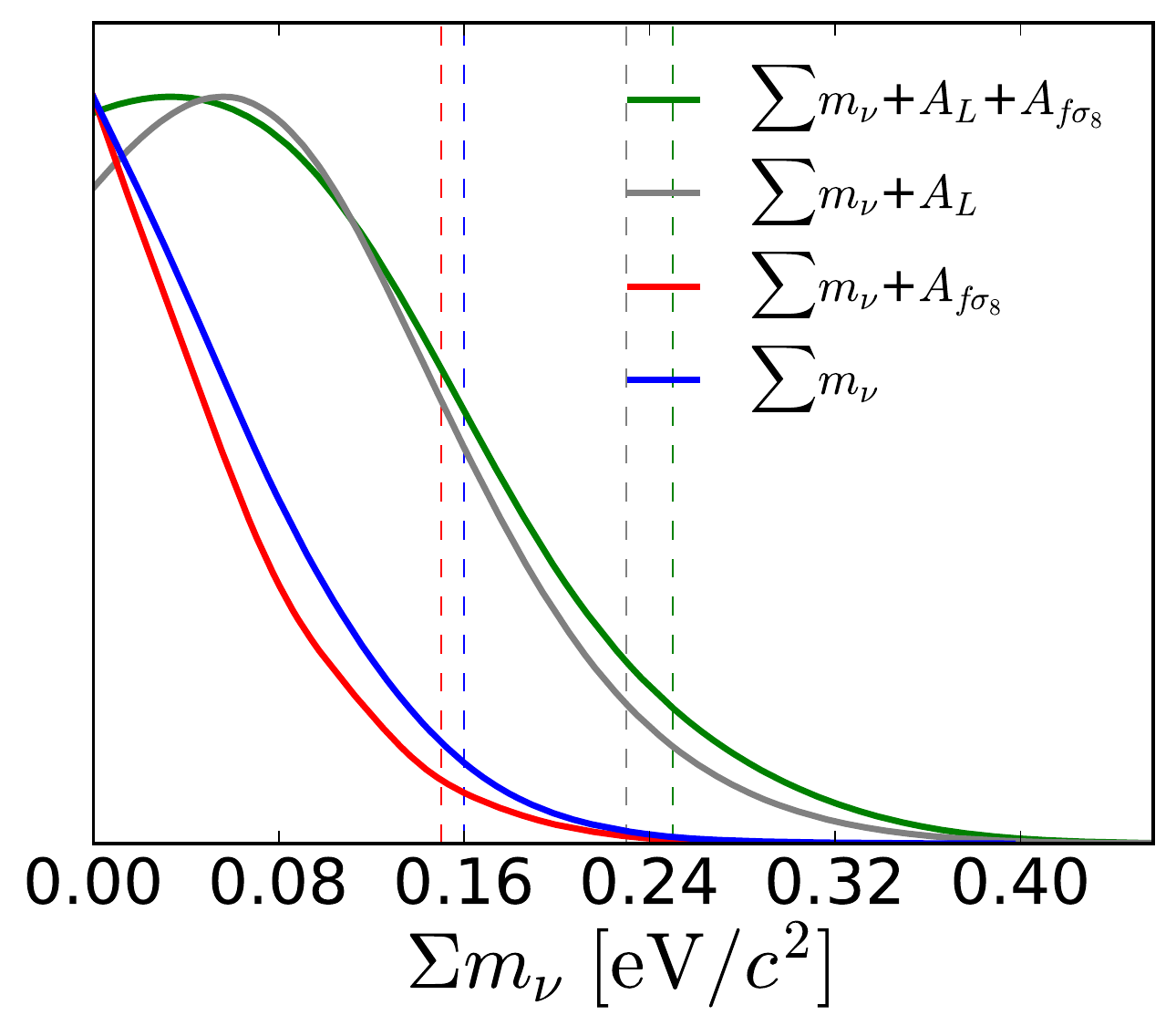}
  \caption{Posterior distribution for the sum of the mass of neutrinos in the 
$\Lambda$CDM cosmological model.  The blue curve includes the growth measurement
from the lensing impacts on the CMB power spectrum and from the BOSS RSD measurement
of $f\sigma_8$.  The green curve exclude both of these constraints; one still gets
constraint on the neutrino mass from the impact on the distance scale.  Red and grey
curves relax one of the growth measurements at a time; showing that most of the extra information
comes from the CMB lensing.  The vertical dashed lines indicate the 95\% upper limits
corresponding to each distribution.
}
  \label{fig:neutrinos}
% \end{minipage}
\end{figure}

\begin{figure}
%%% \begin{minipage}{7in}
\includegraphics[width=3.2in]{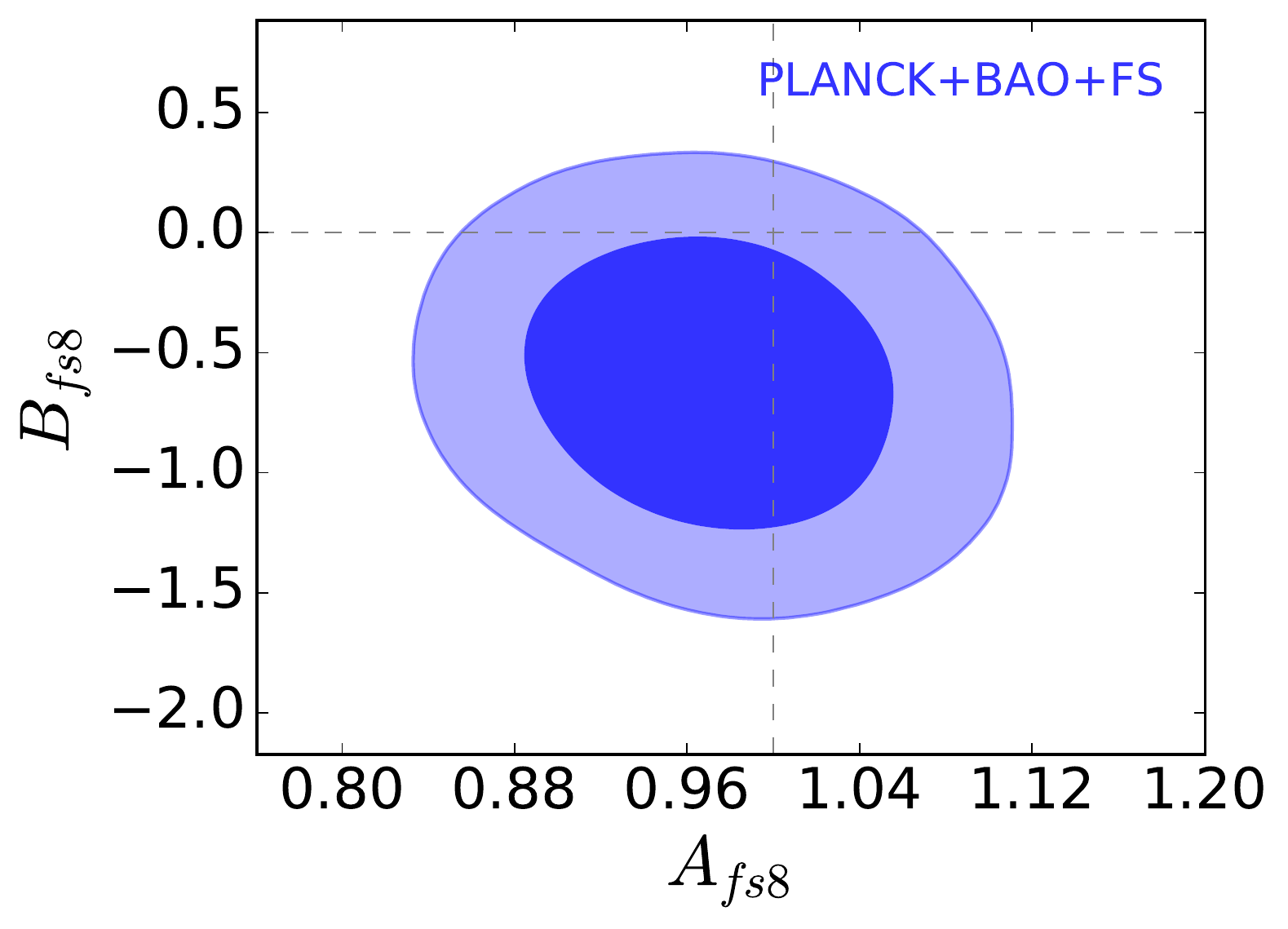}
  \caption{Results for modification of the growth function in the
    $\Lambda$CDM cosmological model.    The results are
    consistent with the predictions of General Relativity: $A_{f\sigma_8}=1$,
    $B_{f\sigma_8}=0$. }
  \label{fig:mg_Afs8}
%%%   \end{minipage}
\end{figure}

We next turn to models that assume a simpler distance scale but consider parameters
to vary the growth of structure, notably through massive neutrinos or 
modifications of the growth rates predicted by General Relativity.
These results are found in Table \ref{tab:DR12c}.

We start with $\Lambda$CDM models that include an unknown total mass of the 
three neutrino species.  In detail, we assume that all of the mass is in only
one of the three weakly coupled species, but 
the difference between this assumption and three nearly degenerate
species of the same total mass is small for our purposes.
Neutrinos of sub-eV mass serve as a sub-dominant admixture
of hot dark matter.  Because of their substantial velocity, they fail to 
fall into small-scale structure at low redshift, thereby suppressing the 
growth of structure from recombination until today \citep{Bond83,Hu98}.
The measurement of the amplitude of the CMB anisotropy power spectrum 
and the optical depth to recombination $\tau$ implies the amplitude of the
matter power spectrum at $z\approx 1000$.  The measurement of the expansion
history along with the assumptions of GR and minimal neutrino
mass then determines the amplitude
of the matter power spectrum at $z=0$, typically reported as $\sigma_8$.
Variations in the neutrino mass then cause the expected $\sigma_8$ to vary.

Measurements of the low-redshift amplitude of structure can therefore 
measure or limit the neutrino mass.  Here, we utilize two measurements:
the lensing effects on the Planck CMB anisotropy power spectrum and the BOSS RSD.  
Using these, we find a 95 per cent upper limit on the neutrino mass of 0.16 eV/$c^2$.  

We then consider how the constraints vary if one relaxes these
measurements, as shown in Figure \ref{fig:neutrinos}.  We include
additional nuisance parameters $A_L$ that scale the impact of the CMB
lensing and $A_{f\sigma_8}$ that scales the RSD following as
\begin{equation}
f\sigma_8 \rightarrow f\sigma_8
  \left[A_{f\sigma_8}+B_{f\sigma_8}(z-z_p)\right]
\end{equation}
with $z_p=0.51$ (chosen to be the central measurement redshift and
also close to actual redshift pivot point for these two
parameters). However, for the discussion of neutrinos, we keep $B_{f \sigma_8}=0$.
We note that $A_L$ is defined scaling the power spectrum of fluctuations, 
whereas $A_{f\sigma_8}$ varies the amplitude.  This means that errors
on $A_L$ will be double those on $A_{f\sigma_8}$.

From this, we find
that the measured CMB lensing power spectrum
is about \upd{$19\pm8$} per cent stronger (so about 9.5 per cent on the 
amplitude of fluctuations) than what
the $\Lambda$CDM model would prefer, while the measured RSD is within
1$\sigma$ of the base level: \upd{$A_{f\sigma_8} = 0.96\pm0.06$}.  This
means that the RSD measured in BOSS is a 6 per cent test of the expected
amplitude of structure, with the central value of the measurement
being slightly lower than the $\Lambda$CDM prediction.

Interestingly, even with $A_L$ and $A_{f\sigma_8}$ varying and hence with no
low-redshift measurement of the growth of structure save for a weak contribution from
the Integrated Sachs Wolfe effect in the large-angle CMB anisotropies, 
we find a 95 per cent upper limit of \upd{$m_\nu<0.25$~eV/$c^2$}.  This comes from the impact
of the neutrino mass on the expansion history of the Universe \citep{Aubourg}.
Essentially, the CMB inference of the balance of matter and radiation at recombination
yields the density of baryons and CDM, while the measurements of the low-redshift
distance scale infer a matter density that now includes the massive neutrinos as
well.

Considering growth measurements one at a time, we find that including the
CMB lensing effect is primarily responsible for shrinking the 95 per cent
upper limit from \upd{0.25 eV/$c^2$ to 0.16 eV/$c^2$}.  The RSD measurement alone only
reaches $<$0.23 eV/$c^2$.  This is not surprising: a 1$\sigma$ variation of
order 0.13 eV/$c^2$ corresponds to a 1 per cent mass fraction of neutrinos, which
yields a roughly 4 per cent change in the small-scale growth function to low
redshift.  This is somewhat smaller than the 6 per cent rms measurement from
RSD.  But the relative improvements are also being impacted by the
central values of the RSD and CMB lensing inferences.  RSD prefers a
slightly lower normalization of small-scale power, thereby favouring a
larger neutrino mass.  Meanwhile, the CMB power spectrum appears to
indicate a larger $A_L$ and hence a higher normalization of
small-scale power, which pulls neutrino masses lower and makes the
upper limit stronger. 

As the distance scale itself is providing some constraint on the neutrino mass,
we also consider fits in the $ow$CDM model.  These degrade the limits to 0.31 eV/$c^2$.
While this is a moderate degradation, it demonstrates that the distance
scale data are sufficiently good that one can simultaneously fit for expansion
history and growth rate.

These limits on the neutrino mass are comparable to numerous other
recent measurements. The strongest bound so far, 0.12~eV/$c^2$ at 95 per cent,
is presented
in \cite{Palanque15} for the combination of Planck 2015
data and the one-dimensional flux power spectrum of the BOSS
Lyman-$\alpha$ forest in quasar absorption spectra.
Recent attempts to combine
the galaxy power spectrum with
with Planck 2015 data \citep{Giusarma16,Cuestaneut} produce
bounds between 0.25 and 0.30~eV/$c^2$, depending on the power spectrum datasets
used and the number of massive neutrino states assumed in the
analysis (or $\sim$0.20~eV/$c^2$ if a compilation of recent BAO data is
used instead of the power spectrum).  This can be brought further
down to $\sim$0.12~eV/$c^2$ if a Hubble constant prior from direct $H_0$
measurements is imposed additionally.  However, the combination of
cosmological datasets in tension with each other can drive a spurious
neutrino mass signal, so it is important to address these issues
before naively interpreting as a neutrino mass detection a signature
of systematic effects. For example \cite{Beutlerneut} showed that a 
somewhat large neutrino mass of $\sum {m_{\nu}} = 0.36 \pm 0.14\,$eV/$c^2$ is 
favoured when combining CMASS Data Release 11 with WMAP9 data. 
A similar result is confirmed when combining CMASS DR11 with Planck 2013 
if the $A_L$ lensing parameter is marginalized out. If $A_L$ is not 
marginalized over, this is not the case, as reported in the Planck paper. 
Finally, the identification and removal of systematic effects on large 
angular scales in the polarization data of Planck has resulted in a stronger 
bound on neutrino mass from CMB data alone, placing a limit of 0.59~eV/$c^2$ 
without polarization and 0.34~eV/$c^2$ with polarization \citep{Planck2016}.

Instead of explaining any variations in the amplitude of structure by
a non-minimal neutrino mass, one could instead view it as a test of
the growth rate of structure under General Relativity.  
In this sense these nuisance parameters can be regarded as a test similar
to that usually carried out using the phenomenological $\gamma$ parameter (\citealt{Wang98,Linder07}).
This has the advantage of being
independent of the model of structure formation, simple to interpret and directly measured by the
data, at the expense of not constraining any concrete theories of
modified gravity.  Again for $\Lambda$CDM, we find \upd{$A_{f\sigma_8} =0.96\pm 0.06$}; that
is, via the BOSS RSD measurement, we infer $f\sigma_8$ to be within
6 per cent of the $\Lambda$CDM prediction.  While this level of precision on
$\sigma_8$ can be achieved by several methods, such as cluster
abundances or weak lensing, the measurement of the time derivative $f$
of the growth function is harder to access with methods that measure
only the single-redshift amplitude of the power spectrum.  

Extending the model to include a redshift-dependent variation $B_{f\sigma_8}$,
we find \upd{$B_{f\sigma_8}=-0.62\pm0.40$}.  
This is a mild indication of
evolution, with the ratio of the measured to the predicted value
decreasing toward higher redshift.  
The results for LCDM are visualized in Figure \ref{fig:mg_Afs8}.
This is consistent with the trend 
from Figure \ref{fig:combined_redshift}.  As this slope is only
non-zero at 1.5$\sigma$, we do not regard this as a statistically
significant detection of this second parameter.  We conclude that
our RSD measurements indicate that structure is growing in a manner consistent with General
Relativity even in the epoch dominated by dark energy.  

Table \ref{tab:DR12c} further shows that the constraints on 
$A_{f\sigma_8}$ and $B_{f\sigma_8}$ change negligibly if we 
extend the expansion history to the $ow$CDM model.  This implies
that the distance scale information is setting the GR prediction for
$f\sigma_8$ to a level that is well better than we can measure it 
with RSD.

We note that the Planck collaboration has recently concluded \citep{Planck2016} that the 
optical depth to reionization inferred from large-angle $E$-mode polarization 
is $\tau=0.055\pm0.009$,
about 30 per cent less than the value of $\tau=0.078\pm0.019$ that results 
from the \citet{Planck2015} likelihood that we use here.
This has
the consequence of decreasing the amplitude of structure at recombination
by 2 per cent, which in turn reduces the prediction of $\sigma_8$ at low redshift
by the same amount.  This will not affect our errors on $A_{f\sigma_8}$,
but would increase the central value by 2 per cent.  It will push the neutrino
masses toward lower values, slightly reducing our upper limits (as 
well as any others based on \citealt{Planck2015}), as 
there is less room for a decrement of low-redshift power caused by 
hot dark matter.

RSD measurements are only one part of an active current debate about 
the amplitude of low-redshift structure.  Measurements of cosmic shear
and galaxy-galaxy lensing
\citep{Heymans13,Mandelbaum13,MacCrann15,Hildebrandt16} and of cluster abundances \citep{Vikhlinin09,Rozo10,Planck2014xx,Planck2015xxiv} have often yielded notably
lower amplitudes than the Planck predictions in $\Lambda$CDM or the Planck measurement
of the lensing of the CMB from 4-point correlations \citep{Planck2015}.
The tension can be up to 10 per cent in the amplitude and 2-3$\sigma$ per measurement, 
although there are 
cosmic shear measurements \citep{Jee13} and
cluster mass calibrations \citep{Mantz15} that argue for a higher amplitude.
The small-scale clustering of the Lyman $\alpha$ forest provides another 
data point, more in line with the higher Planck prediction.  Our BOSS RSD
measurement falls in the middle of the dispute, with $A_{f\sigma_8}=0.96\pm0.06$
being consistent with the Planck
prediction but also with the lower values.  
For example, our LCDM chain with varying $A_{f\sigma_8}$ finds 
$\Omega_m^{0.5} \sigma_8 = 0.464\pm0.009$, so the RSD measurement itself would favor a value 
$4\pm6$ per cent lower, e.g., $\Omega_m^{0.5} \sigma_8 = 0.445\pm0.03$.
This might be compared, for example, to the measurement 
$\Omega_m^{0.5} \sigma_8 = 0.408\pm0.02$
from \citet{Hildebrandt16}.
While this is not a provocative
position, we note that all of these routes to the low-redshift amplitude depend 
on controlling some thorny systematic or modelling issue.  It is therefore
fortunate
that there are multiple viable methods as we attempt to reach 
sub-percent precision.

\bigskip

\section{Conclusion}
\label{sec:discuss}
We have presented measurements of the cosmological distance-redshift
relation as well as the growth rate of large-scale structure
using an extensive analysis of the clustering of galaxies from the
completed SDSS-III Baryon Oscillation Spectroscopic Survey.
The final sample includes 1.2 million massive galaxies over 9382~deg$^2$
covering $0.2<z<0.75$, making it the largest spectroscopic galaxy
sample yet utilized for cosmology.  We split this sample into 
three partially overlapping redshift bins, each large enough for
clear detections of the baryon acoustic oscillations, so as to study
the expansion history and evolving structure formation of the
Universe.  These bins have effective redshifts of 0.38, 0.51, and 0.61.

The consensus results of this paper are the synthesis of the results of several
companion papers studying this sample 
with a variety of methods and with the support
of large suites of mock catalogues.  
\cite{BeutlerBAO16,Ross16,VargasMagana16}
have measured the distance scale by localizing the BAO feature and
estimated systematic uncertainties in these measurements, while
\cite{BeutlerRSD16,Grieb16,Sanchez16,Satpathy16} have modelled the
RSD signature in the full-shape of the clustering measurements to
add structure growth constraints and improve AP effect measurements.
Studies of high-resolution mock samples described in
\cite{tinker_etal:2016} have enabled the estimate of systematic
uncertainties of the structure growth measurements.
\cite{SanchezStat16} describes how the results of the independent analyses have been
combined into one 9-dimensional Gaussian likelihood that includes
the covariance between our 3 redshift slices and between our 
pre-reconstruction and post-reconstruction analyses.

These results represent the first instance in which post-reconstruction
BAO distance measurements have been combined with structure growth
measurements obtained by modelling the RSD signature.  We expect this
will be the standard in future analyses, e.g., with data from the
DESI experiment, and that methods will be further improved to
simultaneously model post-reconstruction BAO information and RSD
signatures \citep{White15}.

The consensus likelihood presented here is then used to measure
parametrized models of cosmology, including variations in 
dark energy, spatial curvature, neutrino masses, extra relativistic
density, and modifications of gravity.  In all cases, we combine our BOSS
measurements with those from the power spectra of CMB temperature
and polarization anisotropies from \citet{Planck2015}.  The common
physics and theoretical model underlying the phenomena of CMB
anisotropies and late-time large-scale structure make this an
extremely powerful cosmological probe.  We now have compelling
measurements of the baryon acoustic oscillations at a variety
of redshifts, including the exquisite detection at $z\approx1080$,
demonstrating the commonality of the physical basis for structure
formation from recombination to today.
The standard ruler provided by the BAO is a clear and robust marker 
of the distance-redshift relation.  Moreover, the expansion history
this implies is in excellent agreement with the inference of the 
matter-radiation equality from the CMB acoustic peak heights.
This is a remarkable qualitative success of modern cosmology.

Turning to our quantitative results, we highlight the following conclusions:

1) The results of the seven data analyses using a variety of
methodologies are demonstrated to be consistent at the level expected
based on analysis of mock galaxy samples.  Notably this includes
four different analyses of redshift-space
distortions and the Alcock-Paczynski effect, the models of which
were validated on a variety of N-body simulations and mock catalogues.
Our measurements include an estimate of systematic uncertainties, but 
we expect we are limited by our statistical uncertainties. 

2) We measure the Hubble parameter to better than 2.4 per cent and the
angular diameter distance to better than 1.5 per cent
accuracy in each of our redshift bins.  When combined, the measurements
represent a 1.0 per cent constraint on the transverse distance scale
and a 1.6 per cent constraint on the radial distance scale.

3) From the anisotropy of redshift-space clustering, we measure 
the amplitude of the peculiar velocity, parametrized as $f\sigma_8$, 
to 9.2 per cent or better than precision in each redshift bin.  
In total, we find a 6 per cent measurement of a bulk shift of $f\sigma_8$ 
relative to the flat $\Lambda$CDM model.

We find no tensions in
our combined measurements when they are compared to the predictions
of the Planck best-fit $\Lambda$CDM model.

4) Combining with the Planck 2015 power spectrum likelihood, we
find no preference for a model that includes additional parameters
beyond the vanilla spatially flat $\Lambda$CDM model.  This remains
true when combined with JLA SNe data.

5) In the simplest spatially flat $\Lambda$CDM model, our data
moderately tightens the errors from Planck alone, yielding
$\Omega_m=0.311\pm0.006$ and $H_0 = 67.6\pm0.5\hubunits$
Allowing extra relativistic density loosens the errors 
but does not notably shift the central value, yielding $\Omega_m =
0.311\pm0.007$, $H_0=67.5\pm1.2\hubunits$, and $N_{\rm eff} = 3.03\pm
0.18$.

6) Models simultaneously varying a constant dark energy equation
of state parameter and spatial curvature are tightly constrained.
Using Planck and BOSS data alone yields $\Omega_K=0.0003\pm0.0026$
and $w=-1.01\pm0.06$, in tight agreement with the flat $\Lambda$CDM
model despite having opened two new degrees of freedom.  Adding 
JLA SNe improves the dark energy result to $w=-1.01\pm0.04$ while
also yielding $H_0=67.9\pm0.9\hubunits$.

7) Flat models with a time-variable equation of state are less 
well constrained, with finding $w_a = -0.98\pm0.53$ without SNe
and $-0.39\pm0.34$ with SNe.  We do continue to find tight errors
on $w(z)$ at a pivot redshift, $w(0.37)=-1.05\pm0.05$.

8) We find tight and stable limits on $H_0$ for all cases.  For
example, for our most general $ow_0w_a$CDM model, we find
$H_0=67.3\pm1.0\hubunits$ with SNe.  We also find 
$H_0=67.8\pm1.2\hubunits$
for the $ow$CDM model with extra relativistic species.  As
such, our results do nothing to reduce the tension with the direct
measurements of $H_0$ that have found higher values, such as the
$73.0\pm1.8\hubunits$ of \citet{Riess16}.  Whether this remains
to be explained as some combination of statistical and systematic
errors or is an indication of a breakdown of the flat $\Lambda$CDM
model is an enticing open question; our results indicate that 
curvature, smooth evolution of dark energy at low redshift,
or extra pre-recombination energy as parameterized by $N_{\rm eff}$
are not enough to resolve the discrepancy.

9) We place strong constraints on the sum of the neutrino masses.
The 95 per cent upper limit is 0.16 eV; this can be compared to the
minimum of 0.06 eV.
Removing any growth of structure information
(i.e., $f\sigma_8$ information from our data set and CMB lensing
information from Planck), we find the upper limit increases to 0.25
eV, with the information coming primarily from the effect of the
neutrino mass on the expansion history.

10) Alternatively, if one interprets the measurement of $f\sigma_8$
as a test of the GR prediction for the growth rate of large-scale 
structure given the measured expansion history, we find a rescaling of
$A_L=0.96\pm0.06$, which is a 6 per cent measurement consistent with GR.
Testing for redshift evolution, we find a mild preference, 
about 1.5$\sigma$, for evolution compared to the value predicted
by GR.  We do not regard this preference as statistically significant.

This work represents the culmination of large-scale structure goals of the BOSS 
galaxy survey. The
survey fulfilled its experimental design and produced a 
three-dimensional map of the structure of the Universe
over a volume of 18.7 Gpc$^3$ 
with sufficient sampling to be dominated
by sample variance on scales modelled by cosmological linear theory.
BOSS showed that the BAO feature exists in the
distribution of galaxies to greater than 10$\sigma$ significance
and that the subsequent recovery of the acoustic scale allows robust and precise
measurements of angular diameter distance to and the expansion rate
at the redshift of the galaxies. 
These BAO distance measurements form a compelling low-redshift complement 
to the beautifully detailed view of early structure gained from CMB 
observations.  The ability to observe a single well-modelled physical
effect from recombination until today is a great boon for cosmology
and now underlies much of cosmological parameter estimation.
Further, our analyses have extended the use of the anisotropic
galaxy clustering signatures of RSD and the Alcock-Paczynski effect 
to the unprecedented size of the BOSS sample, producing
robust measurements of the expansion history and the rate of structure growth.  
We believe that BOSS has marked an important cosmological milestone,
combining precise clustering measurements of an enormous volume
with detailed modelling from cosmological simulations and extensive
observations of the primary CMB anisotropies to produce a 
persuasive jump in the quality of our cosmological inferences from large-scale structure
and a firm platform for the search for extensions to the standard
cosmological model.
We look forward to seeing this program extended with the coming decade
of large spectroscopic surveys.

\section*{Acknowledgements}

SSA, JNG, and AGS acknowledge support from the Trans-regional
Collaborative Research Centre TR33 `The Dark Universe' of the German
Research Foundation (DFG). FB acknowledges support from the UK Space
Agency through grant ST/N00180X/1. ASB acknowledges support from the
U.S. Department of Energy, Office of Science, Office of High Energy
Physics under Award Number DE-SC0010331. AJC is supported by the
European Research Council under the European Community's Seventh
Framework Programme FP7-IDEAS-Phys.LSS 240117 and the Spanish MINECO 
under projects AYA2014-58747-P and MDM-2014-0369 of ICCUB (Unidad de 
Excelencia 'Mar{\'\i}a de Maeztu'). KSD acknowledges
support from the U.S. Department of Energy, Office of Science, Office
of High Energy Physics under Award Number DE-SC0009959.  DJE
acknowledges support from the U.S. Department of Energy, Office of
Science, Office of High Energy Physics under Award Number
DE-SC0013718.  HGM acknowledges Labex ILP (reference ANR-10-LABX-63)
part of the Idex SUPER, and received financial state aid managed by
the Agence Nationale de la Recherche, as part of the programme
Investissements d'avenir under the reference ANR-11-IDEX-0004-02.  WJP
acknowledges support from the UK Science and Technology Facilities
Research Council through grants ST/M001709/1 and ST/N000668/1, the
European Research Council through grant 614030 Darksurvey, and the UK
Space Agency through grant ST/N00180X/1.  NPR acknowledges support
from the STFC and the Ernest Rutherford Fellowship scheme. GR
acknowledges support from the National Research Foundation of Korea
(NRF) through NRF-SGER 2014055950 funded by the Korean Ministry of
Education, Science and Technology (MoEST), and from the faculty
research fund of Sejong University. H-JS acknowledges support from the
U.S. Department of Energy, Office of Science, Office of High Energy
Physics under Award Number DE-SC0014329.  RT acknowledges support from
the Science and Technology Facilities Council via an Ernest Rutherford
Fellowship (grant number ST/K004719/1).

Funding for SDSS-III has been provided by the Alfred P. Sloan
Foundation, the Participating Institutions, the National Science
Foundation, and the U.S. Department of Energy Office of Science. The
SDSS-III web site is http://www.sdss3.org/.

SDSS-III is managed by the Astrophysical Research Consortium for the
Participating Institutions of the SDSS-III Collaboration including the
University of Arizona,
the Brazilian Participation Group,
Brookhaven National Laboratory,
University of Cambridge,
Carnegie Mellon University,
University of Florida,
the French Participation Group,
the German Participation Group,
Harvard University,
the Instituto de Astrofisica de Canarias,
the Michigan State/Notre Dame/JINA Participation Group,
Johns Hopkins University,
Lawrence Berkeley National Laboratory,
Max Planck Institute for Astrophysics,
Max Planck Institute for Extraterrestrial Physics,
New Mexico State University,
New York University,
Ohio State University,
Pennsylvania State University,
University of Portsmouth,
Princeton University,
the Spanish Participation Group,
University of Tokyo,
University of Utah,
Vanderbilt University,
University of Virginia,
University of Washington,
and Yale University.

Based on observations obtained with Planck (http://www.esa.int/Planck), an ESA science
mission with instruments and contributions directly funded by ESA Member States, NASA, and Canada.

This research used resources of the National Energy Research Scientific Computing Center, a DOE Office of Science User Facility supported by the Office of Science of the U.S. Department of Energy under Contract No. DE-AC02-05CH11231.

\appendix

\section{North-South discrepancy}
\label{app:NGC_vs_SGC}

\begin{figure*}
 \centering
\includegraphics[width=\columnwidth]{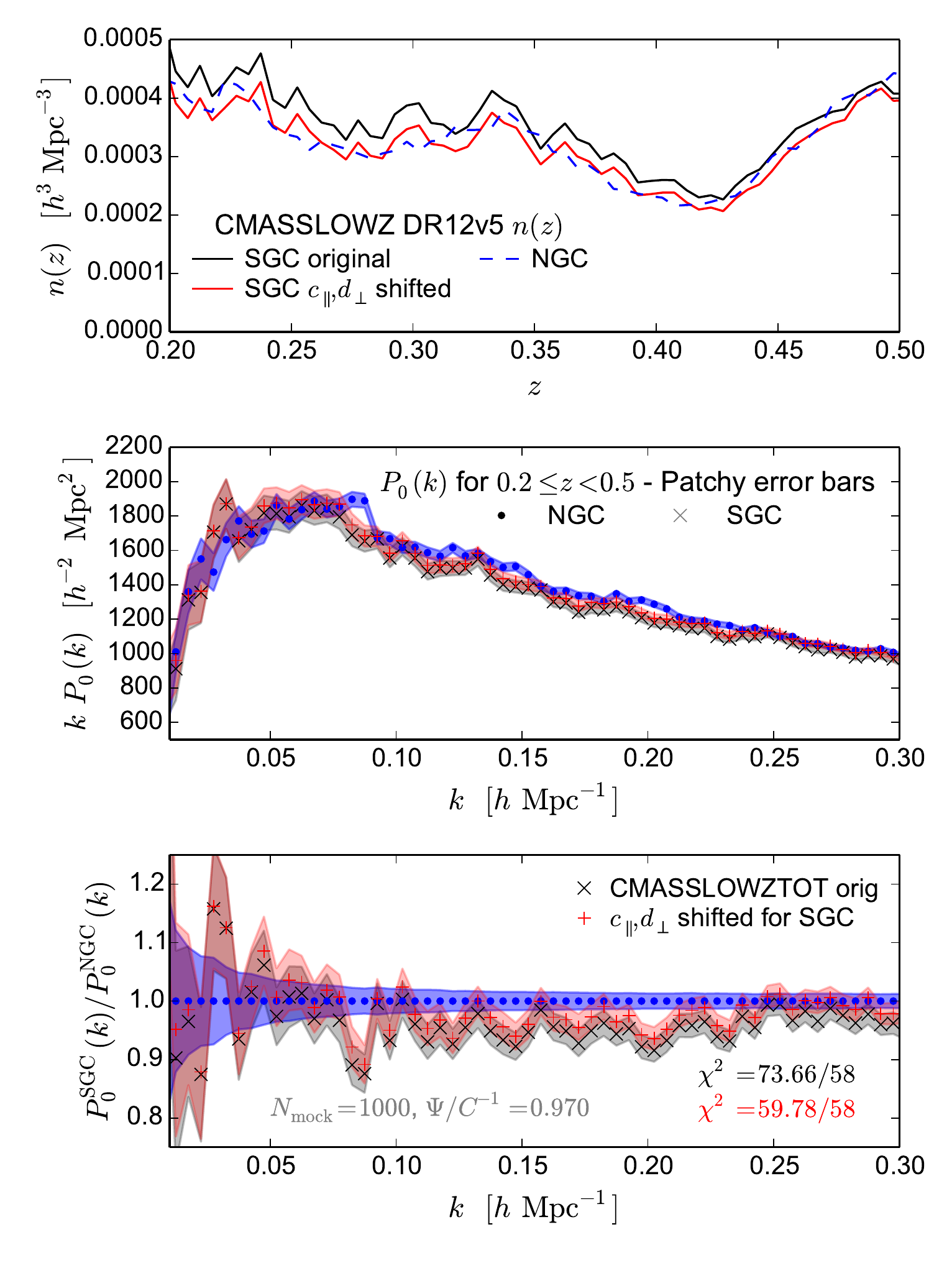}
\includegraphics[width=\columnwidth]{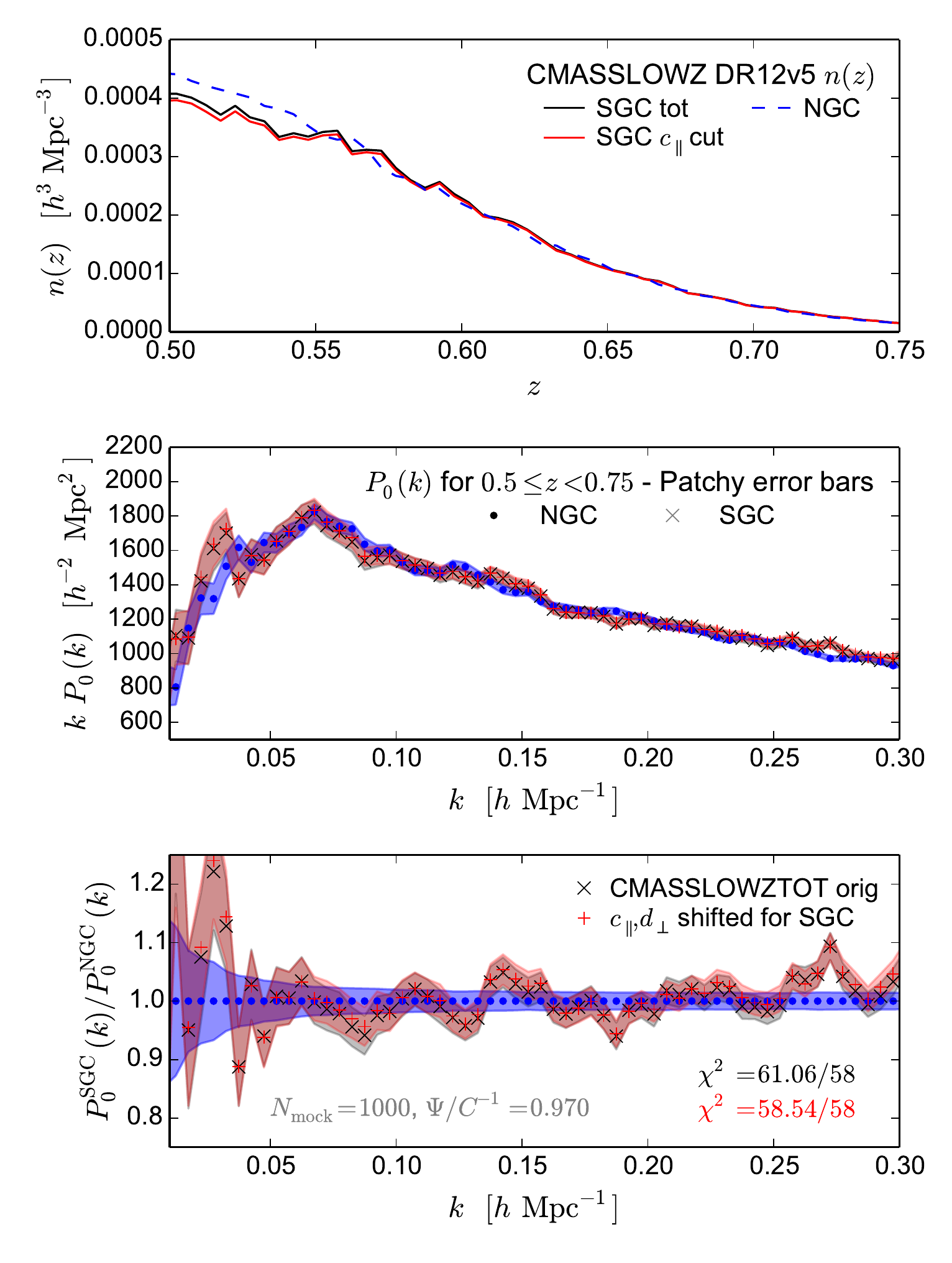}
 \caption{The selection function (upper panels) and power spectrum monopole (centre panels) for the NGC (blue) and SGC (black lines) subsamples of the combined sample in the low (left-hand panels) and high (right-hand panels) redshift bins.
  The error bars (shaded area) are given by the diagonal entries of the covariance matrix that is obtained from the MD-Patchy mock catalogues.
  The red line corresponds to a corrected SGC sample taking the colour shifts between SDSS photometry in the North and South into account \citep[for more details, see][]{Ross16}.
  The lower panels shows the $P_0(k)$ ratio to highlight the deviations between the samples in the two hemispheres.}
 \label{fig:NGC_vs_SGC_PatchyCov}
\end{figure*}

In Fig{.}~\ref{fig:NGC_vs_SGC_PatchyCov}, we show the power spectrum monopole for each subsample with error bars derived from the diagonal of the MD-Patchy covariance matrix (described in section~\ref{sec:mocks_cov}) for the low (left-hand panels) and high (right-hand panels) redshift bins.
The centre panels show the $P_0(k)$ up to a wavenumber of $k_\mathrm{max} = 0.3 \; h \, \mathrm{Mpc}^{-1}$, the lower panels show the ratio of SGC power spectrum to the NGC power spectrum.
The upper panel shows the $n(z)$ of the different subsamples, excluding the early LOWZ regions in the North.
The comparison of the measured power spectrum monopole, $P_0(k)$, of the NGC and SGC subsamples of the final catalogue with the predicted NGC--SGC dispersion from the MD-Patchy mocks shows significant tension for the low redshift bin, while the differences in the high redshift bin are consistent with the sample variance of the mocks.
In terms of the power spectrum monopole, the SGC clustering in the low redshift bin shows a 4 per cent amplitude mismatch.

As discussed in more detail in \citet{Ross16}, the discrepancy can be solved by taking into account the colour shifts between SDSS DR8 photometry \citep{DR8} in the North and South that have been identified by \citet{Schlafly:2010dz}.
These corrections affect the LOWZ SGC colour cut on $c_\parallel$ \citep[equation~9]{ReidEtAl15},
\begin{equation}
 r_\mathrm{cmod} < 13.5 + c_{\parallel,\mathrm{corr}} / 0.3,
\end{equation}
where $c_{\parallel,\mathrm{corr}} = c_\parallel - 0.015$, and the CMASS SGC colour cuts on $d_\perp$ \citep[equations~13 and 14]{ReidEtAl15},
\begin{align}
  \nonumber
  d_{\perp,\mathrm{corr}} &> 0.55 \quad \text{and} \\
  i_\mathrm{cmod} &< \min \left( 19.86 + 1.6 (d_{\perp,\mathrm{corr}} - 0.8), 19.9 \right),
\end{align}
where $d_{\perp,\mathrm{corr}} = d_\perp - 0.0064$.

The selection function and power spectrum monopole of the corrected SGC subsample are overplotted in Fig{.}~\ref{fig:NGC_vs_SGC_PatchyCov}.
The SGC $n(z)$ is reduced by 10 per cent at low redshifts.
The power spectrum monopole of the corrected SGC sample has a larger amplitude than the original sample, but is still lower than the one of NGC for most wavenumber bins.
As the window function induces a correlation between the measurement bins, only the analysis of the log likelihood $\chi^2$, can quantify the level of consistency.
We obtain the inverse covariance matrix from the inverse of the co-added NGC and SGC covariance matrices,
\begin{equation}
 {\boldsymbol \psi}_\mathrm{diff} = (1 + D) \left[ {\mathbfss C}^{P_0}_\mathrm{diff} \right]^{-1}, \quad \text{where} \quad {\mathbfss C}^{P_0}_\mathrm{diff} = {\mathbfss C}^{P_0}_\mathrm{NGC} + {\mathbfss C}^{P_0}_\mathrm{SGC},
\end{equation}
These covariance matrices of the subsamples were obtained from 1000 MD-Patchy realizations.
We correct the inverse of the co-added covariance matrix for sampling noise using the correction factor as proposed in \citet{Hartlap:2006kj}, $(1 + D)$, that is given in the figure.

The $\chi^2$ analysis shows that the amplitude mismatch in $P_0(k)$ is lowered to a level that is consistent with the North.
The low-redshift NGC--SGC difference in the corrected sample is of the order of what can be expected ($\chi^2 = 59.78$ instead of $\chi^2 = 73.66$ for $58$ bins) given the distribution of the mock catalogues.
Further, the high redshift bin also shows slightly increased consistency ($\chi^2 = 58.548$ for $58$ bins), even though it was already in good agreement in the original sample ($\chi^2 = 61.06$).

These results on the shifts of $n(z)$ and $P_0(k)$ are in good agreement with those obtained in the configuration space analysis of the DR9 CMASS  sample presented in \citet{RossEtAl12}.
In that work, no significant effect on the galaxy clustering was found correcting for the shifted photometry.
Further tests on the DR12 combined sample in configuration space \citep{Ross16} show a much better degree of consistency than what is seen in Fourier space.
The amplitude mismatch for the correlation function is not significant as the broad-band effect seen in Fourier space corresponds to scales smaller than those probed in the clustering analysis ($r \lesssim 20 \; h^{-1} \, \mathrm{Mpc}$).
Also, the relative errors bars are larger in configuration space.

Due to the significant deviations in $n(z)$ and $P_0(k)$ between the NGC and SGC subsamples, we see this analysis as good evidence that these two subsamples probe slightly different galaxy populations for redshifts lower than $z \leq 0.5$.

\label{lastpage}

\end{document}